\documentclass[11pt,letter]{article}
\usepackage[margin=1in]{geometry}
\usepackage{amsmath,amssymb,amsthm}
\usepackage{graphicx}
\usepackage{booktabs}
\usepackage{natbib}
\usepackage{microtype}
\usepackage{hyperref}
\usepackage{xcolor}
\hypersetup{colorlinks=true, citecolor=blue, linkcolor=blue, urlcolor=blue}
\usepackage{setspace}
\usepackage{ctable}
\usepackage{multirow}
\usepackage{caption}
\usepackage{subcaption}
\usepackage{placeins}
\usepackage{xr-hyper}
\title{\vspace{-3.5em}\Large\textbf{How an Economy Shrinks in Space: \\
  Concavity-on-Jobs and Upward Consolidation \\ under Demographic Decline}\thanks{\fontsize{8.5}{9.6}\selectfont This study is conducted as a part of the Project ``Sustainability of Cities and Regions in Japan under Population Decline'' undertaken at the Research Institute of Economy, Trade and Industry (RIETI). The draft of this paper was presented at the RIETI Discussion Paper Seminar; the authors would like to thank the participants for their helpful comments. The data of the \emph{Japanese Panel Study of Employment Dynamics} (JPSED), Recruit Works Institute, were provided by the Social Science Japan Data Archive, Center for Social Research and Data Archives, Institute of Social Science, The University of Tokyo. This work was supported by JSPS KAKENHI Grant Numbers 25H00543 and 24H00012, the Mitsubishi Foundation Humanities Research Grant, a research grant from the Kajima Foundation, and the Joint Usage/Research Center program of the Institute of Economic Research, Kyoto University. The authors used Anthropic's Claude (Fable 5) as a coding and writing assistant during the preparation of this paper, in particular for data-pipeline scripting in Python, regression-output formatting in LaTeX, and prose drafting and editing. All conceptual choices, empirical specifications, and substantive interpretations are the authors' own. The Online Appendix is available as an ancillary file of the arXiv version: \url{https://arxiv.org/abs/2609.09859}.}}

\author{Miki Ogawa\thanks{Corresponding author. Graduate School of Economics, Kyoto University, Yoshida-honmachi, Sakyo-ku, Kyoto 606-8501, Japan. Email: \href{mailto:ogawa.miki.26v@st.kyoto-u.ac.jp}{ogawa.miki.26v@st.kyoto-u.ac.jp}.} \and Tomoya Mori\thanks{Institute of Economic Research, Kyoto University, Yoshida-honmachi, Sakyo-ku, Kyoto 606-8501, Japan; and Research Institute of Economy, Trade and Industry (RIETI). Email: \href{mailto:mori@kier.kyoto-u.ac.jp}{mori@kier.kyoto-u.ac.jp}.}}

\date{\today}

\makeatletter
\renewcommand\paragraph{\@startsection{paragraph}{4}{\z@}%
    {1.2ex \@plus 0.3ex \@minus 0.2ex}%
    {-1em}%
    {\normalfont\normalsize\bfseries}}
\makeatother

\begin{document}
\maketitle
\thispagestyle{empty}
\vspace{-2.8em}
\begin{abstract}
\begin{spacing}{1.02}
\noindent When a country's population declines, the aggregate economy appears to contract on the intensive margin: industrial diversity intact, every industry a little smaller. At the regional level, contraction is uneven and takes the \emph{extensive} form: entire industries disappear, one after another. The relevant unit is the city: industries are nested by size --- the \emph{hierarchy property of industrial location} --- each viable only above a minimum population. Necessity industries' thresholds bunch at the low end, so a city's industry count --- and its jobs --- is sharply concave in size (\emph{concavity on jobs}). A modest loss pushes a small city below many thresholds at once; a large core sheds a few specialized industries, one at a time. Lost industries \emph{consolidate upward} to the next city large enough to host them; for the worker it means a step down to a lower-paid local job if one is left, and a fall to the bottom of the hierarchy if none is. To recover the lost wage, workers move up to the apex --- the only city hosting the full industry range. In Japan --- two decades ahead of the OECD, Tokyo at its apex --- a wage regression in real, housing-inclusive wages, on worker-level panel data for the young workers who carry the migration, finds a Tokyo-bound migration incentive that varies across origins exactly as concavity on jobs predicts.


\end{spacing}
\end{abstract}

\clearpage
\setcounter{page}{1}

\section{Introduction}\label{sec:intro}

Working-age decline is now one of the most consequential structural challenges facing developed economies: the United Nations projects that more than thirty countries will see their working-age population shrink by over $10\%$ by 2050, with Korea, Italy, Germany, and Japan contracting by more than $25\%$ \citep{UN-WPP-2024}. The consequences of population \emph{aging} --- for growth, productivity, and automation --- have been studied extensively \citep{Acemoglu-Restrepo-AER-PP-2017,Acemoglu-Restrepo-REStud2022,Maestas-Mullen-Powell-AEJM2023}; outright population \emph{decline} has become a subject in its own right only recently: \citet{Jones-AER2022} shows that a shrinking population can end growth by drying up the creation of new ideas, while \citet{Hopenhayn-Neira-Singhania-ECMA2022} and \citet{Karahan-Pugsley-Sahin-AER2024} trace slowing population growth to firm aging and the start-up deficit. What remains far less understood is how sustained contraction reshapes the \emph{spatial} organization of economic activity within a country.

At the national level, the reshaping is nearly invisible: every 3-digit industry present in Japan in 2009 was still present nationally in 2021 (Section~\ref{sec:descriptive_concavity}), so aggregate contraction looks like every industry shrinking a little --- an intensive-margin adjustment. This paper shows that at the city level, decline reorganizes the economy along its urban hierarchy. The smaller the city, the more its contraction takes the \emph{extensive} form: entire industries vanish, one after another; the displaced workers follow their industries up the hierarchy, and the process consolidates population in Tokyo, the terminal metropolis. We study its key consequence, the wage premium that rural contraction generates for migration into Tokyo. What the paper tests is the mechanism as a whole, not the effect of any single variable. The mechanism makes a distinctive joint pattern of predictions --- including where the premium must \emph{not} vary with origin size --- and the question is whether the data reproduce that pattern in full.

\begin{figure}[h!]
\centering
\includegraphics[width=0.85\textwidth]{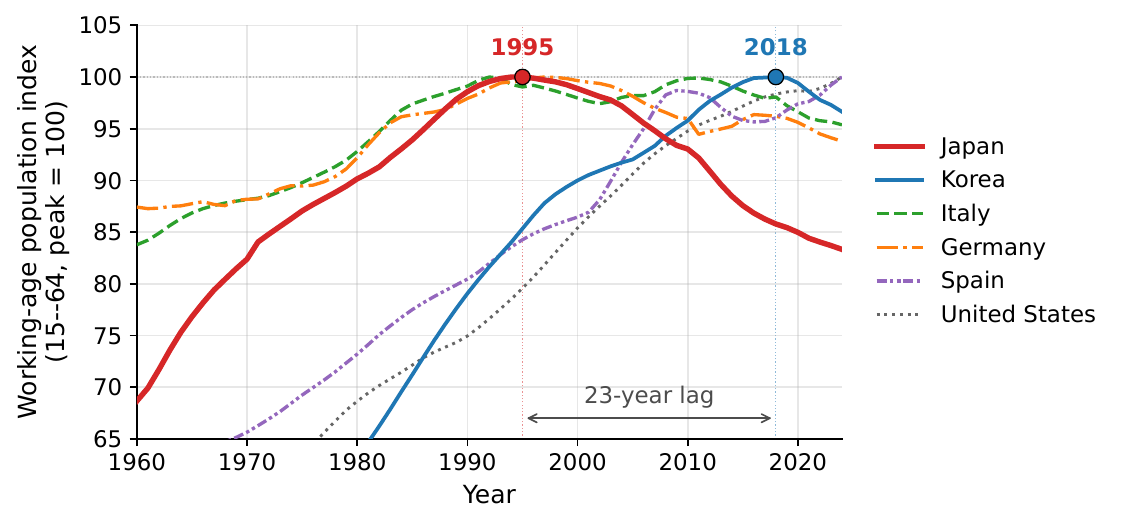}
\caption{Working-age (15--64) population, indexed to each country's peak.}
\label{fig:oecd_workage_trajectory}
\par\smallskip
\begin{minipage}{0.85\textwidth}
\footnotesize \emph{Notes:} Japan peaked in 1995 and has declined by approximately 17\% in the three decades since; Korea peaked in 2018, twenty-three years behind. Italy (1992) and Germany (1997) both peaked in the 1990s and subsequently reached secondary local peaks --- Italy in 2010, Germany in 2016---through immigration, before resuming decline; Spain, whose fertility is as low as Italy's, has so far been held at its peak by immigration; the United States is still growing under current immigration patterns. Source: \citet{WorldBank-WDI}.
\end{minipage}
\end{figure}

Japan is two decades ahead of the rest of the OECD on this trajectory (Figure~\ref{fig:oecd_workage_trajectory}): its working-age population has contracted since the mid-1990s and its total population since around 2008, so the years from 2010 on are its first phase of sustained national decline. Through exactly this phase, inter-prefectural migration into the Tokyo metropolitan area (MA)\footnote{Throughout the paper, the Tokyo metropolitan area refers to the four prefectures of Tokyo, Saitama, Chiba, and Kanagawa.} has not slowed. The annual net inflow of 20-to-29-year-olds roughly doubled, from about $62{,}000$ in 2010 to over $110{,}000$ by 2023--25, dipping only in the two pandemic years, while Osaka, Nagoya, and Fukuoka plateaued or turned negative: net concentration is now unipolar (Section~\ref{sec:descriptive_longterm}). The inflow is overwhelmingly young, and Tokyo MA's share of the nation's 20--29-year-olds, flat through the post-bubble decades ($31.2\%$ in 2010), rose to $33.9\%$ by 2020 as national decline set in (Section~\ref{sec:descriptive}). That Tokyo keeps concentrating while the nation contracts --- and fastest among the young --- is a striking pattern that existing theory has yet to explain.

\paragraph{Hierarchical industry location and demographic decline.} The mechanism behind it, we argue, is an extensive-margin reorganization of the urban system, sharp at the \emph{city} level but blurred at the administrative-region level where urban and rural zones are pooled. It rests on the \emph{hierarchy property of industrial location} established by \citet{Mori-Nishikimi-Smith-JRS2008,Mori-Smith-JRS2011}: industry location is nested by city size, so that a smaller city's set of locally hosted industries is approximately a subset of any larger city's, and the most specialized industries are confined to the few largest cities, because hosting them requires a correspondingly large population \citep{Christaller-1933,Fujita-Krugman-Mori-EER1999,Hsu-EJ2012}. Throughout the paper, ``city'' refers to the \emph{Urban Agglomeration} (UA; maximal contiguous cluster of high-density grid cells; Appendix~\ref{app:ua_map}) --- the spatial unit on which central place theory \citep{Christaller-1933,Fujita-Krugman-Mori-EER1999,Tabuchi-Thisse-JUE2011,Hsu-EJ2012,Davis-Dingel-JIE2020} predicts the industry hierarchy and at which \citet{Mori-Akamatsu-Takayama-Osawa-2025} validate the hierarchy property; Section~\ref{sec:descriptive_concavity} verifies the nesting on Japan's establishment data. The framework we develop (Section~\ref{sec:framework}) extends this property into a single primitive: each industry has a minimum population threshold below which it cannot be locally sustained, so a city's size determines the \emph{range} of industries it can host --- from the \emph{specialized} (\emph{high-order}) industries whose hosts are rare, such as cardiovascular surgery above $10$M or TV-program production above $1$M, down through lawyers and wedding halls at $50$k to internists, bakeries, and funeral services at $30$k, where the industries shade into everyday \emph{necessities}: the \emph{ubiquitous} (\emph{low-order}) end, present in every city (Online Appendix~\ref{oa:app:characteristic}, Table~\ref{oa:tab:char_ntt}). Because necessities are sustainable at modest scale, their thresholds bunch closely together at the low end of the size axis, while specialized industries are spaced far apart at the top; the count of locally hosted industries is therefore concave in city size.

When a city's population falls, it drops below the thresholds of its own highest-order industries first; these do not disappear but \emph{consolidate upward} to a larger city that still clears their threshold, ultimately Tokyo at the apex. A declining city therefore sheds industries whole (the extensive margin) rather than shrinking each industry it has (the intensive margin), and what differs across the size distribution is the \emph{pace}: a given proportional population loss crosses \emph{many} thresholds at once in a small city but only \emph{one at a time} in a large core. This is \emph{concavity on jobs}: the hosted-industry count is concave in city size, so a given decline costs a small city more industries than a large one. And because skills do not transfer well across industries \citep{Neal-JOLE1995,Neffke-Henning-SMJ2013}, for a worker, whether her industry is present in the city is whether her job exists there at all.

Japan's $448$ Urban Agglomerations show exactly this (Section~\ref{sec:descriptive_concavity}): the 2009 hosted-industry count is sharply concave in UA size, smaller UAs lost a larger share of their industries by 2021, the exits come from the top of each UA's own set, and population followed (Figures~\ref{fig:concavity_on_jobs_UA} and~\ref{fig:ua_exit_quintile}). None of this began in 2010: small cities were losing population long before the nation was, and in every ten-year window since the 1980s smaller cities have lost more industries, and from the top of their sets. What decline has added is depth --- in small cities, the more necessity-like of the industries they host exit too, a deepening that began with their population loss in the mid-1990s and continues through the 2010s --- a change in which industries a city has at all, invisible to any measure of the intensive margin (Online Appendix~\ref{oa:app:exits_1981}).

What is left behind after an industry exits matters as much as the exit itself. Skills carry only between industries whose requirements are close, so whether a \emph{related} industry remains decides how far down the hierarchy a displaced worker who stays falls. We call industries in the same 2-digit JSIC group --- a supermarket and a convenience store, a hospital and a clinic --- related. In small cities one often does not remain: among industries that exited over 2009--2021, no related industry was left for $26\%$ of exits at UAs of $10$--$30$k, against $1\%$ above one million (Figure~\ref{fig:related_left}).

If an industry's exit decides whether a worker's job exists, it also decides what she earns, because the industry hierarchy is also a skill hierarchy --- the equilibrium logic of the skilled-sorting literature \citep{Behrens-etal-JPE2014,Davis-Dingel-JIE2020,Diamond-AER2016} --- in which higher-order industries pay systematically higher wages. Japanese wage data bear this out: wages differ by industry far more than by region, and fall as an industry's hosting count rises, with the gap between neighboring industries widest at the specialized end of the hierarchy (Section~\ref{sec:descriptive_concavity}, Table~\ref{tab:bss_framework_primitives}). A worker whose industry has consolidated away can no longer do that work at home; staying means falling down the hierarchy to whatever is left locally, and to its wage. That drop is the incentive to follow the job to Tokyo, where it is recovered, and it is what we call the \emph{premium}: the premium is the very drop that sent her there. Its size depends on what the origin's decline has left for her at home, and that differs sharply between large and small origins (Figure~\ref{fig:wage_premium_concept}, Section~\ref{sec:framework}).

At a declining large core, thresholds are widely spaced, so however deep the decline, the number of industries lost barely changes; what matters is \emph{which} industry goes, and the larger the core, the more specialized that industry. A related industry remains one step below, so a stayer falls one step --- but near the top of the hierarchy the wage gap between one industry and the next is large (Section~\ref{sec:descriptive_concavity}), so that step costs more the larger the core. The premium is therefore largest at the large-origin corner, the largest declining cores.

At a declining small origin, the industries at risk are necessities whose thresholds lie close together, so both pace and size matter: as population falls, the faster the decline and the smaller the origin, the more thresholds are crossed at once (concavity on jobs), and what decides how far a stayer falls is whether a related industry is left locally to step down to. In a small city it often is not (Figure~\ref{fig:related_left}): mostly because the industry stood alone in its group to begin with --- small cities host many of their industries without a same-group neighbor --- and partly because the neighbors went too. The stayer then falls all the way down the hierarchy, to its floor --- a job that does not use her skill, or no job. Slow decline leaves the related industries and the step shallow. The premium is therefore largest at the small-origin corner, the smallest, fastest-declining origins.

Seen from both ends, this is one mechanism. Under population decline, the regions lose their industries and shrink, the smallest cities first and larger ones after them, while industries and the workers who follow them consolidate into larger cities and, above all, into Tokyo. Japan shrinks and concentrates at the same time, for the same reason.

The mechanism is the population--industry linkage on which central place theory \citep{Christaller-1933,Fujita-Krugman-Mori-EER1999,Hsu-EJ2012,Davis-Dingel-JIE2020} builds the urban hierarchy: population size determines the set of locally hosted industries, so population variation is an \emph{extensive-margin} phenomenon. The economics of agglomeration has centered instead on the \emph{intensive} margin --- the city-size wage premia \citep{Combes-Duranton-Gobillon-JUE2008,Andersson-Burgess-Lane-JUE2007,de-la-Roca-Puga-REStud2017,Dauth-etal-JEEA2022} read off cross-sectional differences across stable city sizes; these forces do not switch off under decline, but nothing in them ties a worker's gain to what her origin has lost. Quantitative spatial models \citep{Allen-Arkolakis-QJE2014,Diamond-AER2016,Hsieh-Moretti-AEJM2019,Kleinman-Liu-Redding-ECTA2023} hold the national population stationary and the dynamic spatial literature \citep{Rossi-Hansberg-Wright-REStud2007,Desmet-Rossi-Hansberg-AER2014,Desmet-Nagy-Rossi-Hansberg-JPE2018,Caliendo-Dvorkin-Parro-ECTA2019,Duranton-Puga-HB2004,Glaeser-Gottlieb-JEL2009,Duranton-Puga-ECTA2023} centers on growth; the closest, \citet{Giannone-etal-DP2024}, models Japanese depopulation in a life-cycle spatial equilibrium, but there too regional population is an intensive-margin outcome of amenity and productivity differences. Building the extensive margin into these models is the natural next step; this paper supplies the empirical pattern such a model must reproduce.

\paragraph{Empirical strategy.} \emph{(i) The wage premium as a differential.} We measure the premium by the worker's real, housing-inclusive wage growth. Even so, Tokyo's real wage is underestimated: the amenities available only in Tokyo never enter the data, and the deflator prices a common national basket, so the measured cost of living misses what Tokyo's higher prices buy. What the framework pins down is therefore not the premium's \emph{level} but how it \emph{changes} across origins --- how it rises toward the corner of each size class --- and we read the premium as this differential.

\smallskip\noindent \emph{(ii) Tokyo as the terminal destination.} Consolidation climbs the hierarchy in stages, and a worker may reach Tokyo directly, after intermediate moves, or stop short. Rather than identify each step, we use two properties of the \emph{terminal} rent at Tokyo. It is path-independent: the origin-to-Tokyo wage gap depends only on the origin--apex size differential, not on the route. And Tokyo is the one destination the mechanism singles out: whatever industry a worker has lost, Tokyo has it, and for the specialized industries a large core loses, \emph{only} Tokyo has it. This is a prediction, not a convenience. If decline simply pushed workers out, to be absorbed by whichever major city they reached, moves to Osaka, Nagoya, or Fukuoka would carry the same premium as moves to Tokyo, and adding them to the treatment would sharpen the pattern. Instead, the large-origin premium collapses under the broadened treatment, the small-origin premium --- recoverable at any city larger than the origin --- partly survives, and the over-identification test, which the Tokyo-bound design passes, rejects (Section~\ref{sec:robustness}).

\smallskip\noindent \emph{(iii) Cells at the prefecture grain.} The worker panel --- the Japanese Panel Study of Employment Dynamics (JPSED) --- records origins at the prefecture level; a prefecture can mix small and large UAs, which lose from different heights of the hierarchy, and origins of the same size decline at very different rates (Table~\ref{oa:tab:pref_ua_composition}). Splitting the origins into four cells on size and decline rate \emph{jointly}, each above or below its mean over the $39$ origin regions (Section~\ref{sec:strategy_wage}), recovers this distinction to a workable degree: the size $\times$ decline interaction takes a separate slope in each (small/large) $\times$ (slow/fast) cell. Decline is measured on the young (15--29), because rural decline proceeds through their out-migration and they carry the Tokyo migration we study. The regression also needs a measure of what consolidation has taken from each origin --- the loss the premium recovers. The most direct would be a count of the industries lost. But hosting is a property of cities, and at the prefecture grain an industry counts as lost only once its \emph{last} host city drops it; the count that results is compressed, ranging only from $10$ to $38$ across the rural origins while population differs ninefold (Section~\ref{sec:strategy_mechanism}). We therefore use size and decline in its place: under the hierarchy they encode what a place stands to lose --- its size, \emph{which} industries are at stake; its decline, \emph{how many} go --- and the test is how the premium varies across origins with the two.

\smallskip\noindent \emph{(iv) Endogeneity and instrumentation.} The origin decline rate $g_o$ is endogenous: migration feeds back into measured decline, and local shocks may drive both. We instrument it with three predetermined shocks dated before the panel: a demographic Bartik --- the change in the young population that the prefecture's five-yearly birth counts for 1985--2000 (the cohorts aged 15--29 in 2010 and 2019) and national survival rates imply in the absence of migration --- the 1965--85 change in the prefecture total fertility rate, and 1980--85 capital deepening (R-JIP; \citealp{RJIP-2017}). Drawn from unrelated pieces of history, the three give independent estimates of the cell coefficients, and Section~\ref{sec:strategy_id} explains what their agreement tests.

\paragraph{Empirical findings.} Using COVID-clean wave transitions (2015/16--2018/19) of the JPSED, we confirm the predictions for rural-origin workers aged 29 or younger, who account for the bulk of inter-prefectural migration. The object is the marginal effect of origin size on the Tokyo MA wage premium, and across the size\,$\times$\,decline cells it moves as the framework predicts: the premium rises toward the corner of each size class --- at the largest declining cores, where the larger the origin, the more specialized the exiting industries, and at the smallest, fastest-declining origins, where the smaller the origin, the more industries are lost at once, often leaving no related industry to step down to (Section~\ref{sec:results_main}). Mildly declining small origins keep their related industries, and the marginal effect shows no size gradient, as predicted. This pattern survives an over-identified 2SLS on three predetermined historical shocks (Sections~\ref{sec:strategy_id} and~\ref{sec:results_causal}), and for workers aged 30--64 the small-origin gradient disappears and the large-origin one is much weaker, consistent with migration being concentrated in the 20s. What the estimates deliberately do not deliver is a decomposition: the premium is not apportioned among the channels through which decline reaches workers --- the industry exits we measure are one --- nor is population decline separated from the earlier shocks that redrew the distribution of people and industries, chiefly deindustrialization and the opening of the high-speed transport network, which reached small cities in the 1990s. The analysis is set in 2010--2020 because, those shocks having run their course, it is the first decade in which decline acts on the regions without a new shock of that kind (Section~\ref{sec:descriptive_concavity}). The claim is correspondingly narrower, and that is what makes it hard to evade: any account of the premium that does not run through what the origin has lost must also explain why it rises exactly where the framework puts it and shows no gradient where it does not.

\paragraph{Robustness and international context.} The cell-level result is robust to alternative decline windows, destination and outlier definitions (Section~\ref{sec:robustness}), and to the weighting choice (Online Appendix~\ref{oa:app:weighting}). Of the cross-country comparators, only Korea has a labor market contracting deeply enough for a worker-level test, and there the decline-keyed premium reappears on the axis Korea can identify (Section~\ref{sec:external}, Online Appendix~\ref{oa:app:external}) --- a shared trajectory, not evidence that Japan is uniquely vulnerable.

\paragraph{Contribution and policy.} Our worker-level evidence establishes a decline-activated channel invisible to the static cross-section. The two types of declining origin differ in what they lose: a large core, only its specialized industries; a small origin, the industries just above the necessities as well. For Japan's regional policy debate \citep{Coulmas-2007,Matanle-Rausch-2011,Masuda-2014,OECD-Japan-2016,Hoshi-2026,Mori-2026,Mori-Murakami-DP2025}, the implication is this: decline removes industries from small cities first, only in the smallest does it reach the industries daily life depends on, and small cities are many --- too many to hold all of them above their thresholds while the population keeps falling. Below a threshold an employment subsidy does not bring an industry back; the binding constraint is market size, not the cost of labor. Short of restoring fertility, what policy can do is adapt: keep the industries daily life requires by gathering a shrinking population in places that can still hold them. The questions are then which places to keep as such cores and how to design them so that moving into them works (Section~\ref{sec:discussion_policy}).

The remainder of the paper is organized as follows. Section~\ref{sec:descriptive} documents Japan's decline and Tokyo's concentration, and the hierarchy and wage facts, Section~\ref{sec:framework} the framework, Section~\ref{sec:data} the data, Section~\ref{sec:strategy} the empirical strategy, Section~\ref{sec:results} the results, Section~\ref{sec:robustness} robustness, Section~\ref{sec:external} the cross-country evidence, and Section~\ref{sec:conclusion} the conclusion.
\section{Stylized Facts}\label{sec:descriptive}

\subsection{Working-age decline and Tokyo concentration}\label{sec:descriptive_longterm}

Japan's working-age population has been shrinking since the mid-1990s, and the decline has hit the 20--29 cohort first: that pool has nearly halved, from $\sim 18$M in the early 1970s to $\sim 11$M in 2025 (Online Appendix~\ref{oa:app:descriptive}). Even so, population has kept concentrating into the Tokyo MA throughout the postwar period, carried almost entirely by workers in their twenties. In 2025 Tokyo MA's net in-migration (Basic Resident Registry) was $+111$k at ages $20$--$29$ and near zero or negative at every age from $30$ up --- Tokyo loses population at the older ages --- and the same asymmetry has held throughout 2010--2025, with the 20s inflow roughly doubling from $+62$k in 2010. The other three centers --- the Osaka and Nagoya MAs and Fukuoka prefecture --- have drawn no such inflow at any age: the Osaka and Nagoya MAs lose population on net in their twenties and above, and Fukuoka prefecture's surplus is a few thousand a year, an order of magnitude below Tokyo MA's (Online Appendix~\ref{oa:app:descriptive}).

\begin{figure}[ht!]
\centering
\includegraphics[width=0.88\textwidth]{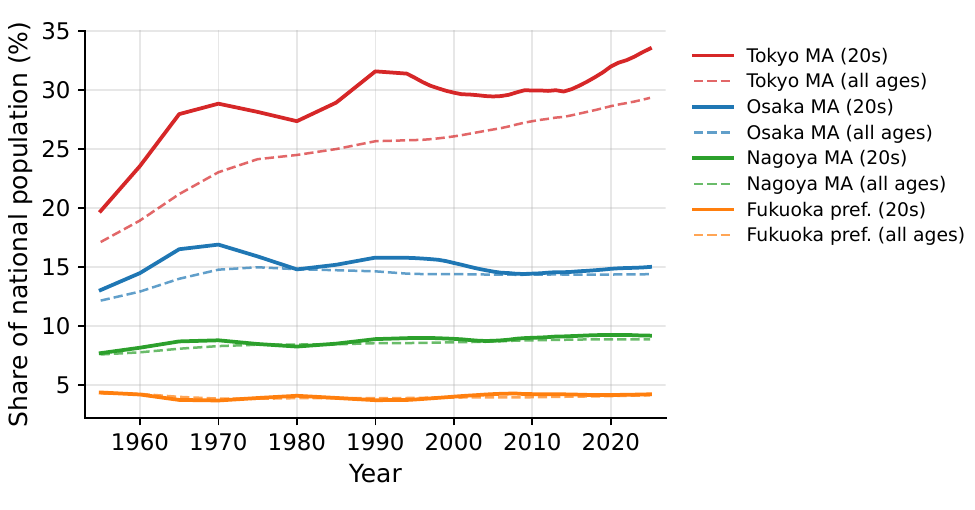}
\caption{Share of national population in the three major metropolitan areas and Fukuoka prefecture, 1955--2025.}
\label{fig:share_4mas_long}
\par\smallskip
\begin{minipage}{0.88\textwidth}
\footnotesize \emph{Notes:} The three major metropolitan areas are the Tokyo, Osaka, and Nagoya MAs; Fukuoka is at the prefecture level. Solid lines: ages 20--29; dashed lines: all ages. Sources: Population Census (Statistics Bureau, Ministry of Internal Affairs and Communications; every ten years, 1955--1990; total population including foreign residents) and Basic Resident Registry (annual, 1994--2025; Japanese residents only). The 1990 Census and 1994 JBRR endpoints are connected by a straight line spanning the source switch. The two sources differ in whether foreign residents are counted, but this creates only a small break at the switch: each line is a region's share of the national total, so foreign residents enter both the region (numerator) and the nation (denominator) and mostly cancel. Tokyo MA's 20s share has consistently exceeded its all-ages share by 4--5 percentage points and the gap has widened over time.
\end{minipage}
\end{figure}

The share of Japan's 20--29 population residing in the Tokyo MA has risen from $\sim 20\%$ in 1955 to $\sim 34\%$ in 2025 (Figure~\ref{fig:share_4mas_long}).\footnote{The share dips in the 1970s and around 2000 as the baby-boom generation and its echo pass out of the 20--29 band; these are age-composition effects, not pauses in the concentration.} Its \emph{acceleration} comes in two episodes that differ in how broadly the gains were shared. In the 1960s rural workers entered the expanding manufacturing and service sectors of all three major metropolitan areas, so Osaka and Nagoya expanded their shares alongside Tokyo (Online Appendix~\ref{oa:app:netflow_destinations}). The post-2010 episode --- the focus of this paper --- is Tokyo-specific: only Tokyo MA's young share resumes its climb, while Osaka, Nagoya and Fukuoka have been flat since the 1980s.

By single year of age, new graduates (age $22$) account for about $30\%$ of Tokyo MA's net 20s inflow, and the post-graduate ages $23$--$29$ for about half --- the component that has grown fastest since 2010 (Online Appendix~\ref{oa:app:descriptive:age22}).

\subsection{Hierarchy of industries and concavity on jobs}\label{sec:descriptive_concavity}

\paragraph{The hosted-industry sets are hierarchically nested.} The mechanism this paper proposes for the young migration to Tokyo that continues through the decline era operates at the level of the city, and rests on the \emph{hierarchy property of industrial location}: a smaller city's hosted-industry set forms a near-subset of any larger city's, with the most specialized industries confined to the few largest cities. Hosting the specialized end of the range requires a large population, so the cities that do are few: the urban system is thin at the top.\footnote{Of Japan's $448$ UAs, only $10$ exceed one million residents.} Under this hierarchy, as a city's population declines its hosted set changes systematically with size, and with it whether jobs in a given industry exist locally at all --- an \emph{extensive-margin reorganization of the urban system}. ``City'' throughout is the \emph{Urban Agglomeration} (UA; a maximal contiguous cluster of high-density grid cells), the unit on which central place theory predicts the industry hierarchy; a population map of the 2010 system is in Appendix~\ref{app:ua_map}. The pattern originates in \citet{Christaller-1933}, was first formalized by \citet{Fujita-Krugman-Mori-EER1999}, with microfoundations in \citet{Tabuchi-Thisse-JUE2011}, \citet{Hsu-EJ2012} and \citet{Davis-Dingel-JIE2020}; empirically it is established for Japan by \citet{Mori-Nishikimi-Smith-JRS2008,Mori-Smith-JRS2011}, for the United States by \citet{Hsu-EJ2012}, for both by \citet{Mori-Akamatsu-Takayama-Osawa-2025}, and for restaurants by \citet{Schiff-JoEG2015}. We verify the nesting directly with the Hierarchy-Property (HP) test of \citet{Mori-Akamatsu-Takayama-Osawa-2025} (procedure and prefecture-grain version in Appendix~\ref{app:hierarchy}), on all $587$ 3-digit JSIC industries of the 2009 Economic Census across the $448$ UAs, fixed 2010 delineation.\footnote{The establishment-level census data used throughout the paper are the questionnaire information of the Economic Census for Business Frame (2009) and the Economic Census for Business Activity (2021), both conducted jointly by the Ministry of Internal Affairs and Communications and the Ministry of Economy, Trade and Industry, and provided by the Ministry of Internal Affairs and Communications under Article 33 of the Statistics Act. The wage records of the Basic Survey on Wage Structure are provided by the Ministry of Health, Labour and Welfare under the same provision.}\footnote{\label{fn:industry_counts}Three industry counts appear in the paper. The 2009 Economic Census classifies establishments into $587$ 3-digit industries, and the hierarchy test uses all of them, since it needs a single year. Comparing 2009 with 2021 requires reconciling two revisions of the JSIC, which leaves $494$ unified 3-digit industries common to both censuses --- each hosted somewhere in Japan in both years (Online Appendix~\ref{oa:app:data}); every exit and entry statistic uses these. Of the $494$, $490$ have wage observations in the Basic Survey on Wage Structure ($483$ for workers aged 20--29), and these are the industries behind the wage hierarchy of Table~\ref{tab:bss_framework_primitives}. Appendix~\ref{app:hierarchy} runs the hierarchy test at the prefecture grain on the published prefecture tabulation of the 2009 census, which reports $518$ 3-digit industries.} For each industry $i$, hierarchy predicts that the set of UAs hosting $i$ is a subset of the set of UAs hosting any more-ubiquitous industry $j$ (one with more hosting UAs than $i$). How well this subset relation holds in the data is tested by $\widehat{\mathrm{HP}}_i \in [0, 1]$: the share of coordination actually realized in the data, relative to its theoretical maximum --- the number of times more-ubiquitous industries would co-locate with $i$ in $i$'s host UAs under perfect hierarchy ($\widehat{\mathrm{HP}}_i = 1$ means perfect nesting). We compare $\widehat{\mathrm{HP}}_i$ against a null that reshuffles each industry's UA assignment at random, preserving its hosting count ($1{,}000$ replications).

\begin{figure}[!ht]
\centering
\includegraphics[width=0.88\textwidth]{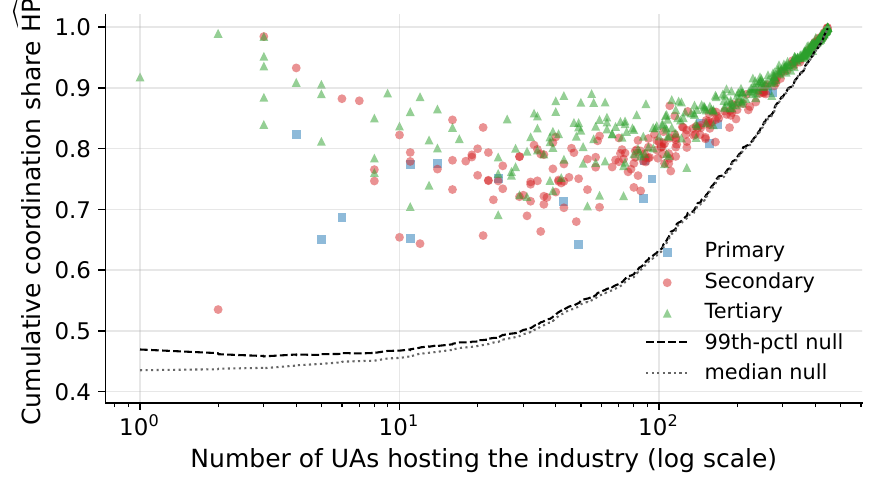}
\caption{Hierarchy-Property test at the UA grain.}
\label{fig:hp_test_ua}
\par\smallskip
\begin{minipage}{0.88\textwidth}
\footnotesize \emph{Notes:} 2009 Economic Census establishments on the 448 UAs of the fixed 2010 delineation, 587 3-digit JSIC industries. Each point's cumulative-coordination share $\widehat{\mathrm{HP}}_i$ against its hosting count (log scale), colored by broad sector (primary / secondary / tertiary). Dashed and dotted lines: 99th-percentile and median of a hosting-count-preserving random-permutation null ($1{,}000$ replications). Actual shares lie above the null across the entire range; all 587 industries reject at $p<0.01$ (mean actual $\widehat{\mathrm{HP}}=0.870$ vs.\ null mean $0.730$). Prefecture-grain version in Appendix~\ref{app:hierarchy} (Figure~\ref{fig:hp_test}).
\end{minipage}
\end{figure}

Figure~\ref{fig:hp_test_ua} plots $\widehat{\mathrm{HP}}_i$ against the number of UAs hosting industry $i$ (specialized industries on the left), colored by broad sector. Every one of the $587$ industries exceeds its null at $p < 0.01$; the mean is $0.870$ against a null mean of $0.730$. The property holds within each sector --- mean $\widehat{\mathrm{HP}}_i$ of $0.764$ for the $17$ primary industries, $0.826$ for the $210$ secondary, $0.901$ for the $360$ tertiary, all far above the null. Primary's lower mean reflects its location-specific nature: agriculture, forestry and fishery follow arable land, forest and coastline rather than the city-size hierarchy, yet remain far above the null. The nesting is thus a genuine ordering of \emph{which} cities host \emph{which} industries, not a by-product of the size--count correlation.

\begin{figure}[tp]
\centering
\begin{subfigure}{0.49\textwidth}
\centering
\caption{Hosted industries, 2009}
\label{fig:concavity_on_jobs_UA_2010}
\includegraphics[width=\textwidth]{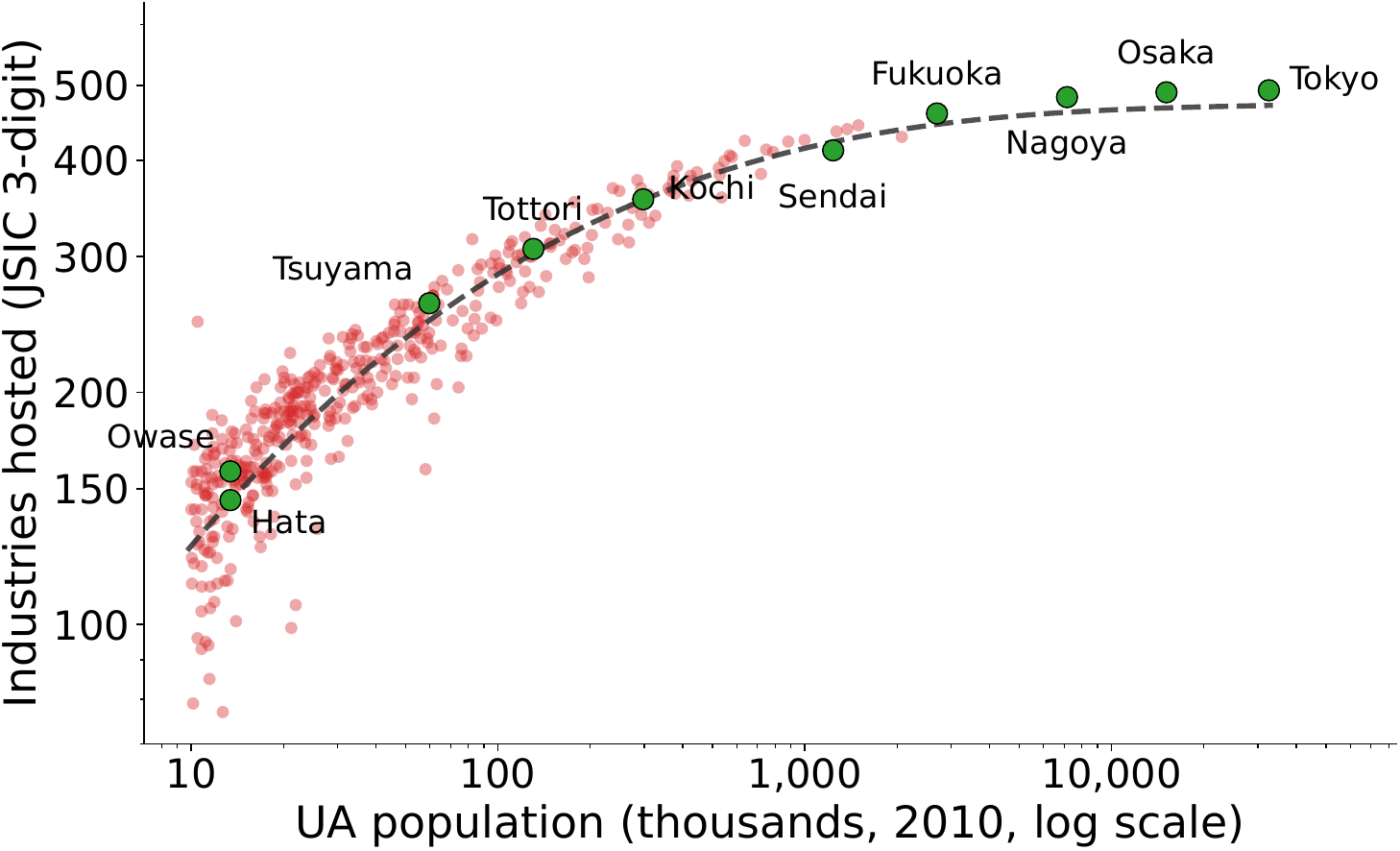}
\end{subfigure}\hfill
\begin{subfigure}{0.49\textwidth}
\centering
\caption{Industry-exit rate, 2009 $\to$ 2021}
\label{fig:concavity_exit_rate}
\includegraphics[width=\textwidth]{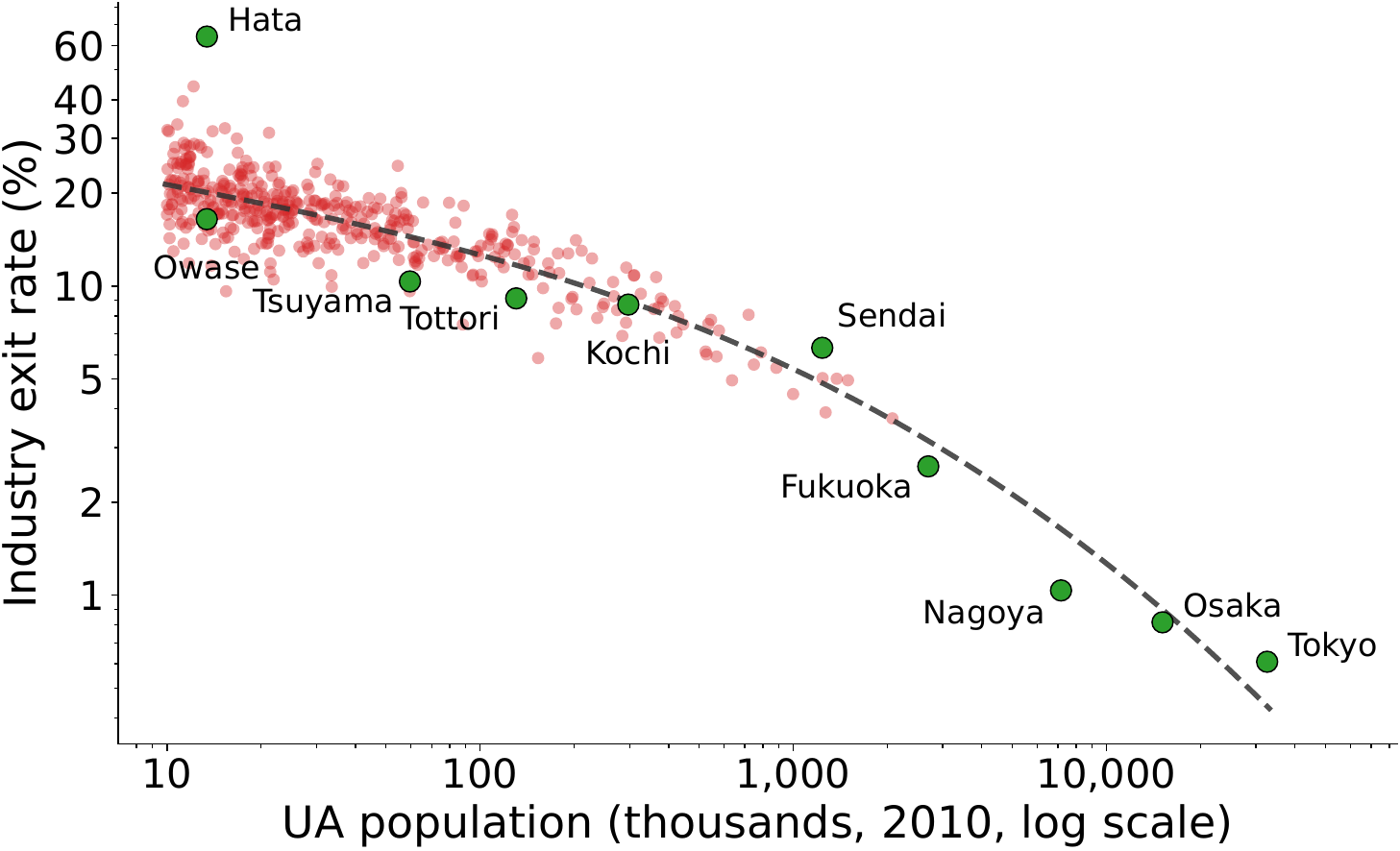}
\end{subfigure}
\par\vspace{0.6em}
\begin{subfigure}{0.49\textwidth}
\centering
\caption{Population growth, 2010 $\to$ 2020}
\label{fig:ua_pop_growth_2010_2020}
\includegraphics[width=\textwidth]{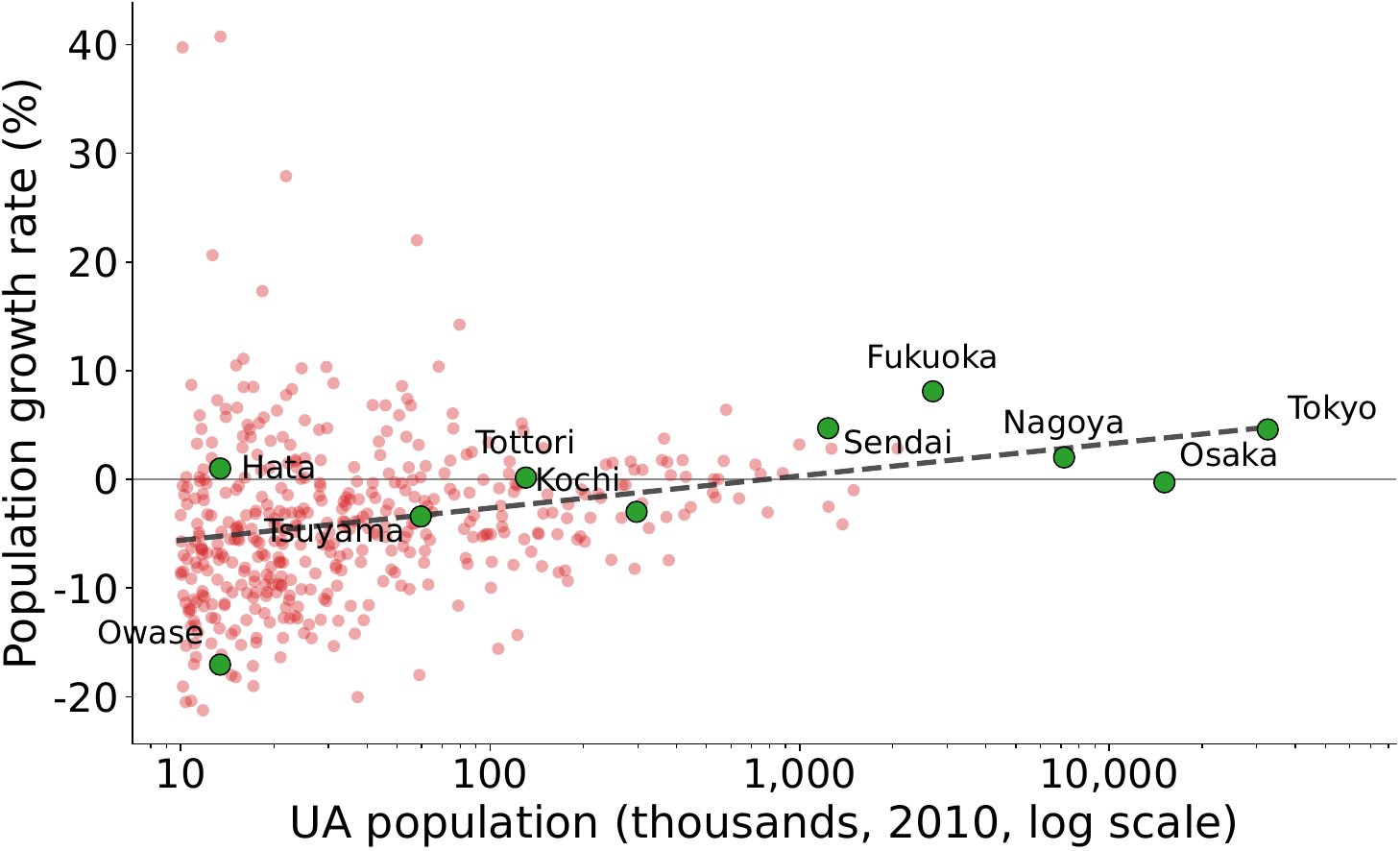}
\end{subfigure}\hfill
\begin{subfigure}{0.49\textwidth}
\centering
\caption{Employment change by margin, 2009 $\to$ 2021}
\label{fig:ua_emp_margin}
\includegraphics[width=\textwidth]{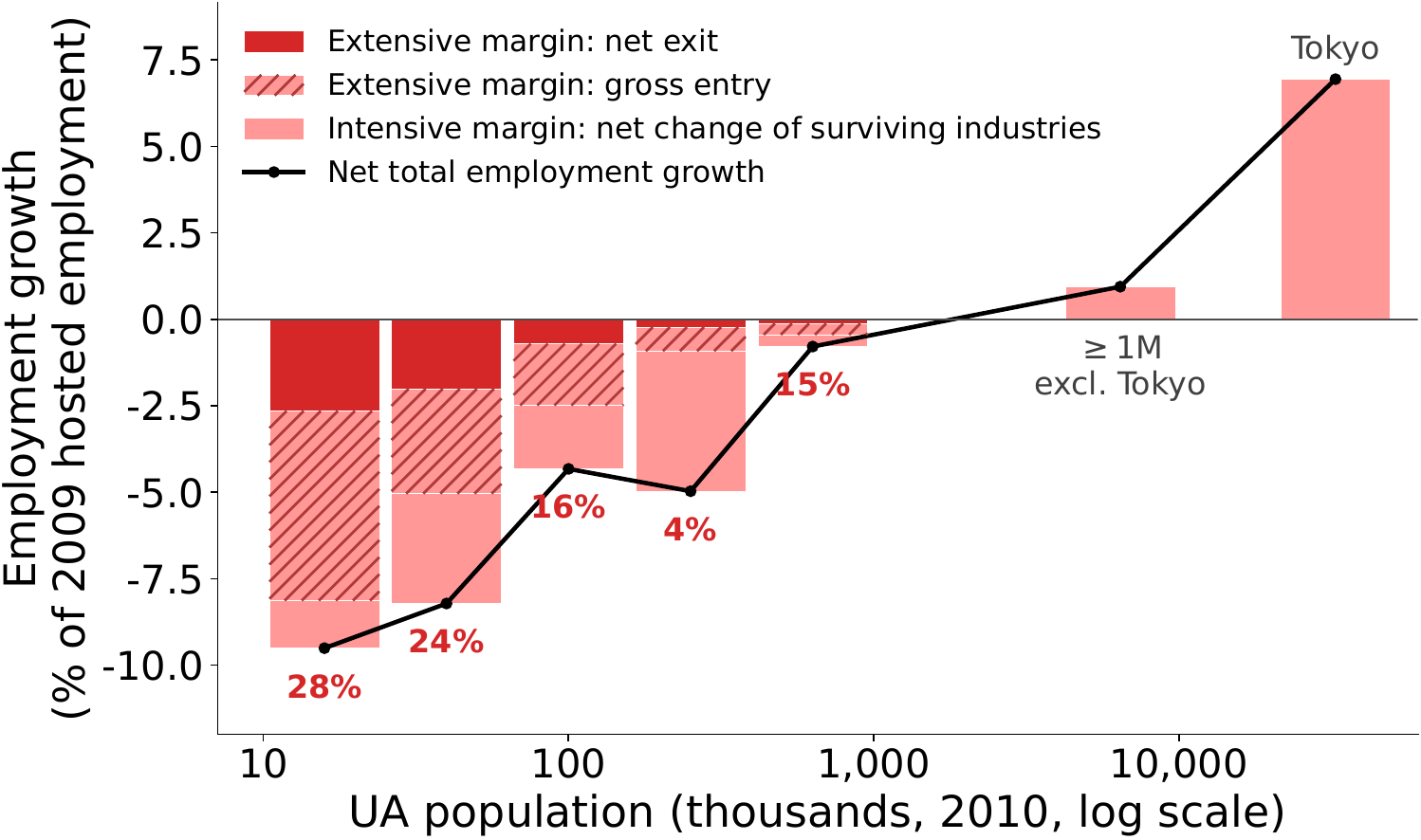}
\end{subfigure}
\caption{Concavity on jobs, population, and employment at the Urban Agglomeration grain.}
\label{fig:concavity_on_jobs_UA}
\par\smallskip
\begin{minipage}{0.97\textwidth}
\footnotesize \emph{Notes:} All panels: each point is one of the $448$ Urban Agglomerations, on fixed 2010 UA polygons, against 2010 UA population (log scale). (a)~The count of 3-digit JSIC industries hosted in 2009; the dashed curve is a shape-constrained cubic fit (concave and monotone in the log-log coordinates); labeled points are representative cities. (b)~The share of each UA's 2009-hosted industries no longer hosted in 2021; the dashed curve is the same shape-constrained cubic fit, here in the exit rate against log population. (c)~UA population growth 2010--2020, with a log-linear fit. (d)~The change of each UA's total employment 2009--2021. The full red bar (hatched band included) is the 2009 employment of the industries the UA lost (the extensive margin's gross exit); the hatched band is the 2021 employment of the newly entering industries (gross entry), drawn inside the exit bar as the part it offsets, so the unhatched red that remains is the net exit; pink is the net change of the surviving industries (the intensive margin); the black line is net total employment growth. Red percentages: the net exit's share of the net decline. Bars are employment-weighted within log-size bins, with the $\geq 1$M bin split into Tokyo and the other $\geq 1$M UAs. The 39-region aggregation of the same data is in Figure~\ref{fig:concavity_cells} (Section~\ref{sec:strategy_wage}). Source: Economic Census, 2009 and 2021 (establishments, harmonized universe and classification; Online Appendix~\ref{oa:app:data}) and Population Census grid data (Statistics Bureau, Ministry of Internal Affairs and Communications).
\end{minipage}
\end{figure}

\paragraph{Concavity on jobs.} The hierarchy implies that each industry has a population threshold below which it cannot be locally sustained. Figure~\ref{fig:concavity_on_jobs_UA} traces the chain of predictions that this primitive generates on Japan's $448$ UAs. Panel~(a) plots UA population against the count of locally hosted industries, and the relationship is sharply concave: a linear fit adds $65$ industries per log-unit of population among the $438$ UAs below one million, but only $23$ among the ten UAs above one million --- a slope $2.8$ times steeper at small UAs. The shape reflects the distribution of these thresholds across industries: high-order specialized industries have widely spaced thresholds, while ubiquitous low-order industries have thresholds tightly clustered at the bottom of the population range (Online Appendix~\ref{oa:app:characteristic} gives the industry content of this threshold distribution, class by class). The framework's first prediction follows: under a given proportional contraction, smaller UAs cross many of these densely clustered low-end thresholds simultaneously while larger UAs cross only a few widely spaced high-end ones, so small UAs should lose a disproportionately larger share of their hosted industries. Panel~(b) confirms the prediction --- tracking each UA's 2009-hosted industries to 2021 on fixed 2010 polygons, smaller UAs do lose a measurably larger share ($21\%$ at the smallest UAs versus near zero at the top; band slopes $-3.78$ versus $-1.57$ percentage points per log-unit of population below and above $1$M). Within the extensive margin, the analysis focuses on the \emph{exit} side, for two reasons. For a worker, an industry is a job: human capital is largely specific to the industry or occupation \citep{Neal-JOLE1995,Kambourov-Manovskii-IER2009,Gathmann-Schonberg-JOLE2010}, so each industry a city loses is a kind of job that a worker who stayed can no longer hold, and an entering industry opens jobs of a different kind --- it does not give her back her own (Section~\ref{sec:framework} builds the framework on this distinction). And industries also enter UAs over the same window, but among the industries newly available to enter each UA, those that actually enter show no direct relation to population change (Section~\ref{sec:strategy_mechanism}). Combining panel~(b) with this last fact yields a further prediction on the residential side: because the industries that small UAs lose under decline are the low-order ones that anchor everyday life there --- general merchandise stores, hospitals, kindergartens, specialty retail; the most ubiquitous of all, food retail and clinics, hold on at the UA grain and fail only in settlements below it (Online Appendix~\ref{oa:app:exits_1981}) --- small UAs should themselves contract in population most severely. Panel~(c) confirms this final step: the smaller the UA, the faster it shrinks --- small UAs lost $10$--$20\%$ of their residents over 2010--2020 while the largest metros (Tokyo, Fukuoka, Sendai) grew (log-linear gradient $+1.29$ pp per log-population, zero growth at $\approx 730{,}000$, $74\%$ of UAs declining).

\paragraph{The employment consequences run through the extensive margin.} Panel~(d) of Figure~\ref{fig:concavity_on_jobs_UA} translates the industry counts into employment. The census counts workers by industry at the two end points, 2009 and 2021, and does not trace individuals, so splitting the change into margins requires a convention; we adopt the simplest: the workers of exiting industries leave with them, and entering industries are staffed from elsewhere. The full red bar is then the extensive margin's \emph{gross exit} (the 2009 employment of the industries the UA lost); the hatched band inside it is the \emph{gross entry} (the 2021 employment of the industries it gained), which offsets that much of the exit; the unhatched red is the \emph{net exit}; and pink is the intensive margin, the net change of the surviving industries.\footnote{With $\Delta T$ the net change in total employment, $X$ the exiting industries' 2009 employment, $E$ the entering industries' 2021 employment and $\Delta S$ the surviving industries' change, $\Delta T = \Delta S + E - X$ exactly.} Aging accounts for only a limited part of the extensive loss: advancing each UA's 2010 age structure ten years with national survival rates, about a quarter of the exited industries' 2009 workforce would have left work by 2021 in any case --- about a fifth of the net change in the two smallest size bins, under $8\%$ above $100$k.

Two features matter for what follows; both describe the employment the exiting and entering industries held, not the number of workers who moved. First, the extensive margin deepens as size falls, and on balance it removes employment. At the smallest UAs the industries that exited held $7.7\%$ of 2009 employment, while the industries that entered hold $5.1\%$ of it in 2021: the entries replaced about two-thirds of the jobs the exits took out, a ratio that holds in every size band below one million. The smaller the city, the larger the share of its employment loss that takes the form of whole industries disappearing. This was not always so. In the 1986--96 window the entering industries carried about twice the employment of the exiting ones; the balance turned negative in the 1996--2006 window, a decade before the national decline began (Online Appendix~\ref{oa:app:exits_1981}). Second, total employment grows only at Tokyo ($+7\%$): the apex, where the wage analysis sends its movers, is the one place that gains while every band below one million loses.

\begin{figure}[!ht]
\centering
\includegraphics[width=0.72\textwidth]{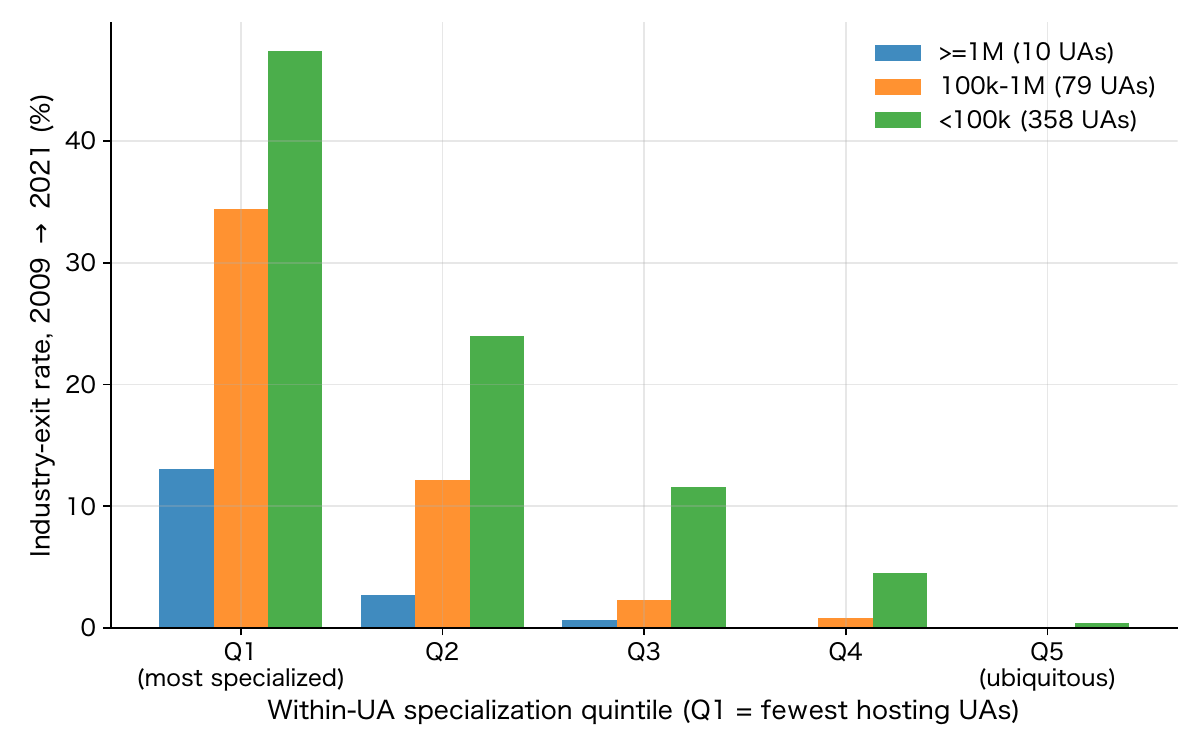}
\caption{Industry-exit rate by within-UA specialization quintile and size band, 2009--2021.}
\label{fig:ua_exit_quintile}
\par\smallskip
\begin{minipage}{0.85\textwidth}
\footnotesize \emph{Notes:} Quintiles are defined with Q1 = most specialized nationally, split by UA size band. Within each UA, its 2009-hosted 3-digit JSIC industries are ranked by their national hosting count --- how many of the $448$ UAs host the industry, a measure of how rare it is nationally --- and split into five equal groups (quintiles), Q1 being the most specialized (fewest hosting UAs). One UA is excluded because it hosted fewer than $25$ industries in 2009, too few to split into fifths; the figure therefore covers $447$ UAs. Exits concentrate at the Q1 (most specialized) end, and this concentration --- the rise in exit rate from Q5 to Q1 --- is steeper at small UAs. Source: Economic Census, 2009 and 2021; 2010 UA polygons fixed.
\end{minipage}
\end{figure}

\paragraph{What declining cities lose is their own specialized margin.} Figure~\ref{fig:concavity_on_jobs_UA}(b) shows that smaller origins lose a larger \emph{share} of their hosted industries; Figure~\ref{fig:ua_exit_quintile} shows \emph{which} industries they lose. Within each UA (2010 polygons fixed) we rank its 2009-hosted 3-digit industries by national hosting count into specialization quintiles (Q1 = most specialized, fewest hosting UAs) and plot the 2009--2021 industry-exit rate by quintile. The exits come overwhelmingly from each UA's locally specialized industries: pooled across the $448$ UAs, those in the most-specialized fifth of a UA's hosted industries (Q1) exit at $42.3\%$ versus only $0.3\%$ for the most ubiquitous fifth (Q5), and the bulk of UA-level exits come from Q1 and Q2 combined. The gradient holds in every size band but is far steeper at small UAs: the locally specialized industries exit at $47.4\%$ at UAs below $100$k versus $13.1\%$ at UAs of one million or more. This is the dynamic content of concavity on jobs: a declining city sheds its locally rarest industries first, and in a small city those industries sit lower in the national hierarchy, nearer the necessities, and it sheds a larger fraction of them. The figure shows the same exits as Figure~\ref{fig:concavity_on_jobs_UA}(b), with the order in which a city loses its industries added. It is the exit side of the extensive margin, the side demographic decline drives (Section~\ref{sec:strategy_mechanism}); the entry side is in Online Appendix~\ref{oa:app:extmargin_hazards}.

\paragraph{Threshold crossings, not small numbers.} Figure~\ref{fig:ua_exit_quintile} shows the exits coming from the top of each city's own hosted set in every size band. Two more facts bear out that these exits are threshold crossings, and both hold in every ten-year window since 1981 (Online Appendix~\ref{oa:app:exits_1981}). First, the exits are not the arithmetic of small establishment counts. One might object that small cities lose more industries simply because they host each with fewer establishments, so that a single closure is enough for an exit. So consider only the industries a city hosts with a single establishment, where one closure means one exit in every city alike. Even then the exits follow the threshold: the more specialized the industry, the more likely its single establishment closes. Over 2009--2021, in cities below $30{,}000$, $67\%$ of the single-establishment industries in the city's most specialized fifth exited against $18\%$ in its most ubiquitous fifth (fifths defined over all its hosted industries, as in Figure~\ref{fig:ua_exit_quintile}); in cities above one million, $65\%$ against $11\%$ --- the same ordering that Figure~\ref{fig:ua_exit_quintile} shows for all industries. What decides an exit is whether the city's market still clears the industry's threshold. That is a fact about presence, not about size: it shows only on the extensive margin, and no count of establishments or share of employment would reveal it. Second, the exits are permanent: $71$--$80\%$ of the industries a city loses in a window are still absent at the next census.

\paragraph{The pattern predates the decline; what decline changed is its depth.} None of this began in 2010. In every ten-year window since the 1980s, smaller cities have lost industries at higher rates, from the top of their sets, and the losses have been permanent; and since the mid-1990s the industries small and mid-sized cities gain no longer make up for those they lose, in count or in jobs (Online Appendix~\ref{oa:app:exits_1981}). What decline has changed is depth. The most specialized fifth of a small city's industries has exited at much the same rate since the 1980s; as small cities began losing population in the mid-1990s, the second and third fifths --- which used to stay --- began to exit too, and the smaller the city, the deeper the loss reaches. The 2010s continue at that depth, now with the nation declining as well. Like the properties above, this is a change on the extensive margin --- in which industries a city has, not in how large any of them is --- and no measure of the intensive margin would register it.

The 1990s rise in small cities' exit rates cannot be traced to a single cause: it coincides with deindustrialization, the completion of the high-speed transport network and the 1997 financial crisis, and the effects of successive shocks mix across windows. By the 2010s those shocks had largely run their course --- the trunk network in place for decades, manufacturing's retreat from the regions long done --- and the national decline had begun; the analysis that follows is set in 2010--2020 because it is the first decade in which decline acts on the regions without a new shock of that kind --- none that redrew the location of industries nationwide; the regional shock of the 2011 earthquake is taken up in Section~\ref{sec:strategy_id}.

\paragraph{What a city loses, its workers lose: related industries.} The exits above are counted in industries; for the worker they are counted in jobs. Skills do not carry well across industries: displaced workers who find work in the same industry keep the wage returns to their experience and tenure, those who switch industries lose them \citep{Neal-JOLE1995}, and when workers do switch, they move between industries whose skill requirements are close \citep{Neffke-Henning-SMJ2013,Kambourov-Manovskii-IER2009,Gathmann-Schonberg-JOLE2010}. Whether an industry is present in a city is therefore whether that job exists there for the worker. Once the industry is gone, so is the job, for everyone who stays; what remains in the city decides only how far down the hierarchy a stayer falls. We call industries in the same 2-digit JSIC group --- a supermarket (581) and a convenience store (589) under food retail (58), a hospital (831) and a clinic (832) under medical services (83) --- \emph{related}: in a related industry, part of the worker's skill and experience still counts; in an unrelated one, little of it does.

\begin{figure}[!ht]
\centering
\includegraphics[width=0.66\textwidth]{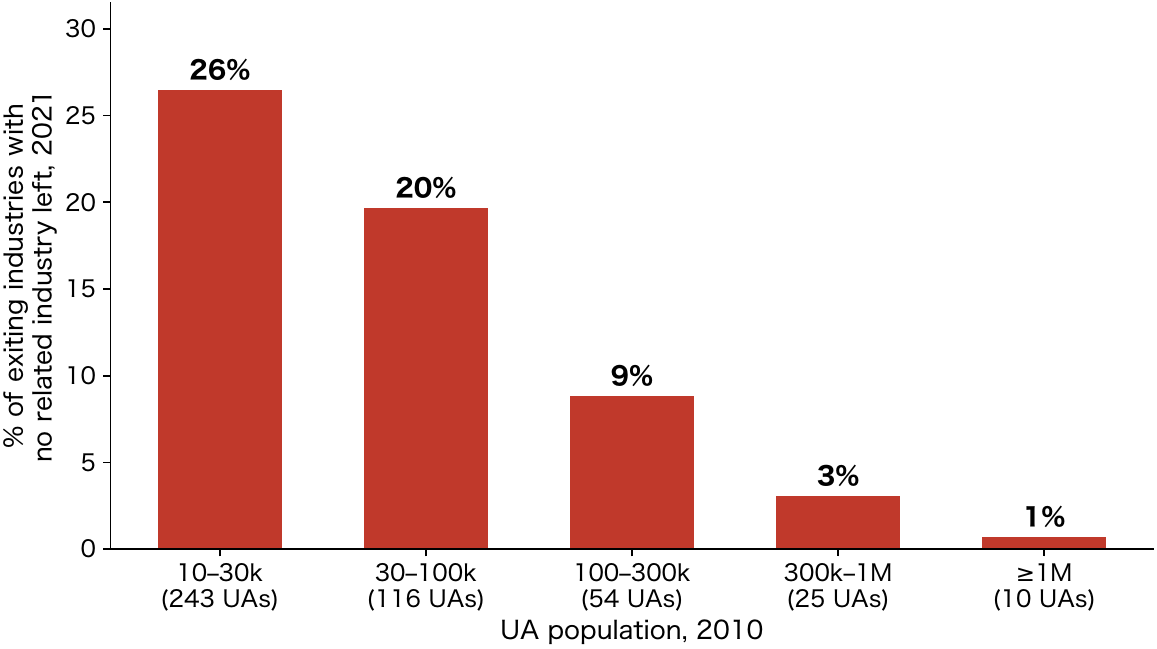}
\caption{Share of exiting industries with no related industry left, by UA size, 2009--2021.}
\label{fig:related_left}
\par\smallskip
\begin{minipage}{0.85\textwidth}
\footnotesize \emph{Notes:} Unit: an exit, i.e.\ a 3-digit JSIC industry hosted by a UA in 2009 with no establishment left in 2021 ($448$ UAs on fixed 2010 polygons $\times$ the $494$ 3-digit industries common to the 2009 and 2021 censuses; footnote~\ref{fn:industry_counts}). Vertical axis: the share of these exits after which no related industry remained in the UA. Related means the same 2-digit group (e.g.\ 58 food retail: supermarket 581, convenience store 589; 83 medical: hospital 831, clinic 832); an industry counts as remaining if it was present in the UA in 2021, whether it survived from 2009 or entered after. Bands by 2010 UA population; the number of UAs in each band is in parentheses. Online Appendix~\ref{oa:app:related_io} repeats the figure with related industries defined by input-output trade. Source: Economic Census, 2009 and 2021.
\end{minipage}
\end{figure}

In small cities there is often no related industry left. Figure~\ref{fig:related_left} takes every exit over 2009--2021 across the $448$ UAs and asks whether any industry in the same 2-digit group was still present in 2021 (survivors or new entrants): for $26\%$ of exits at UAs of $10$--$30$k none was, falling to $20\%$ at $30$--$100$k, $9\%$ at $100$--$300$k, $3\%$ at $300$k--$1$M, and $1\%$ above one million. Defining related industries by input-output trade instead of by classification --- two industries related when the sectors they belong to buy from or sell to each other substantially --- gives the same gradient (Online Appendix~\ref{oa:app:related_io}). The reason is the threshold distribution above, which leaves nothing related in two ways. A small city hosts many of its specialized industries alone, with no related one beside them: of the most specialized fifth of the industries a UA of $10$--$30$k hosted in 2009, one in four had no same-group neighbor in that UA, against one in fifty above one million.\footnote{Take each UA's most specialized fifth of 2009-hosted industries (the Q1 of Figure~\ref{fig:ua_exit_quintile}). An industry in it is \emph{alone} if the UA hosted no other industry of the same 2-digit group in 2009. The share of alone industries, pooled over the UAs of a size band, is $25\%$ at $10$--$30$k and $2\%$ above one million.} And where related ones are present, they sit in the dense necessity band, with thresholds so close that they cross them together. This is the asymmetry the framework builds on: the job itself is gone in both; in a large city a stayer falls one step down the hierarchy, to a related industry, while in a small city there is often nothing to step down to (Section~\ref{sec:framework}).

\begin{table}[!ht]
\centering
\caption{BSS wages: industry versus region, and the wage--hosting-count gradient.}
\label{tab:bss_framework_primitives}
{\small\begin{tabular}{lcc}
\toprule
 & All ages & Ages 20--29 \\
\midrule
\emph{(1) Share of the variance of individual $\log w^{r}$ explained by industry and region ($R^2$):} & & \\
\quad Industry fixed effects (JSIC 3-digit) & 0.196 & 0.188 \\
\quad Region fixed effects ($39$ regions) & 0.042 & 0.047 \\
\quad Industry and region fixed effects & 0.227 & 0.229 \\
\quad Industry $\times$ region fixed effects & 0.267 & 0.298 \\
\midrule
\emph{(2) OLS across industries: adjusted industry $\log w^{r}$ on $\log(\text{hosting count})$:} & & \\
\quad $\hat\beta_{\log\text{hc}}$: $\log w$ per log hosting count (hierarchy predicts $<0$) & \textbf{-0.0330$^{***}$} & -0.0198$^{***}$ \\
\quad \quad (HC1 SE) & (0.0057) & (0.0042) \\
\quad JSIC 3-digit industries & 490 & 483 \\
\bottomrule
\end{tabular}}
\par\smallskip
\begin{minipage}{0.95\textwidth}
\footnotesize \emph{Notes:} \emph{Wage}: BSS real contracted wage per scheduled hour (excluding overtime and bonuses; prefecture-by-year deflator, housing-inclusive), general workers aged $15$--$70$, BSS 2015--2019 pooled (one JSIC revision); the sample of Figure~\ref{fig:jsic_hosting_vs_bss_wage}. (1) $R^2$ of OLS regressions of the individual log real hourly wage on industry dummies (JSIC 3-digit, $490$), on region dummies ($39$ regions: the Tokyo, Osaka, and Nagoya MAs aggregated, prefectures elsewhere), on both, and on industry $\times$ region cells; the remainder is variation among workers in the same industry and region. (2) OLS across the $490$ JSIC 3-digit industries of the industry's adjusted log real hourly wage --- the industry fixed effect from a regression of individual log wages on gender-by-age, education, tenure and its square, employment form, and year fixed effects, evaluated at the sample means (Figure~\ref{fig:jsic_hosting_vs_bss_wage}) --- on the log of the number of the $448$ UAs hosting it in the 2021 Economic Census; no prefecture fixed effects, since where an industry locates is part of what the gradient measures; HC1 standard errors. The second column repeats both panels for workers aged $20$--$29$ ($483$ industries in (2)). The individual-level version of this regression, and the variants of (1), are in Online Appendix~\ref{oa:app:data}.
\end{minipage}
\end{table}

\paragraph{The hierarchy is also a wage hierarchy.} For a worker, losing her industry is losing her job; of what that costs her, the part we can measure is the wage. Two facts about wages follow. Wage variation is organized by industry rather than by place; and within the industry dimension, pay rises with order: specialized high-order industries pay systematically more than ubiquitous low-order ones. In the Basic Survey on Wage Structure (BSS; Ministry of Health, Labour and Welfare; 2015--2019, pooled; general workers, real hourly wage), industry fixed effects (JSIC 3-digit) explain $20\%$ of the variance of individual log wages, region fixed effects ($39$ regions: the three MAs aggregated, prefectures elsewhere) $4\%$, and the two together $23\%$ --- holding the industry fixed, the region adds $3$ percentage points; holding the region fixed, the industry adds $19$ (Table~\ref{tab:bss_framework_primitives}, row (1)). The remaining three quarters is variation among workers in the same industry and region. For individual workers, \emph{which industry one works in} therefore matters more for the wage than where one works within it.

Within the between-industry dimension, the wage falls systematically as an industry's hosting count rises --- specialized industries (hosted in few UAs) pay more than ubiquitous industries (hosted in many) --- consistent with skilled workers sorting into the more skill-intensive sectors that survive only in the largest cities \citep{Davis-Dingel-JIE2020} and with wages rising with skill \citep{Behrens-etal-JPE2014,Diamond-AER2016}. We quantify the gradient across industries: regressing each industry's adjusted log real hourly wage --- the industry fixed effect from a regression of BSS individual wages on gender-by-age, education, tenure and its square, employment form, and year fixed effects --- on the log hosting count gives a slope of $-0.033$ (SE $0.006$) across the $490$ industries with BSS wage data (footnote~\ref{fn:industry_counts}; Table~\ref{tab:bss_framework_primitives}, row (2)), and $-0.020$ (SE $0.004$) for workers aged $20$--$29$. Figure~\ref{fig:jsic_hosting_vs_bss_wage} plots this industry by industry: each dot is a 3-digit industry, the horizontal axis the number of UAs hosting it (log scale), the vertical axis its adjusted log hourly wage. The gradient is clear, and the highest-paying industries sit at the specialized end. Because the wage is linear in the \emph{log} of the hosting count, one step along the hierarchy is worth more at the specialized end: the same wage step accompanies a doubling of the hosting count anywhere on the ladder, so the step is as large between industries hosted in $10$ and $20$ UAs as between those hosted in $200$ and $400$ --- and since only $25$ industries sit between $10$ and $20$ hosts against $119$ between $200$ and $400$, the step from one industry to its neighbor is larger at the specialized end. The realized gaps between neighboring industries show the same pattern: ordering industries by hosting count, the wage gap between rank-adjacent industries averages $0.20$ log points among those hosted in fewer than $30$ UAs and $0.09$ among those hosted in $400$ or more (within-group standard deviations $0.16$ and $0.09$). The wage gaps between industries are thus widest at the top of the hierarchy and narrowest among the necessities. How this wage hierarchy combines with the concavity-on-jobs mechanism is taken up in Section~\ref{sec:framework}; the underlying BSS sample construction is in Online Appendix~\ref{oa:app:data}.

\FloatBarrier
\section{Conceptual Framework}\label{sec:framework}

\begin{figure}[!ht]
\centering
\begin{minipage}{0.95\textwidth}\centering
{\footnotesize (a) Large declining regional core}\\[1mm]
\includegraphics[width=\textwidth]{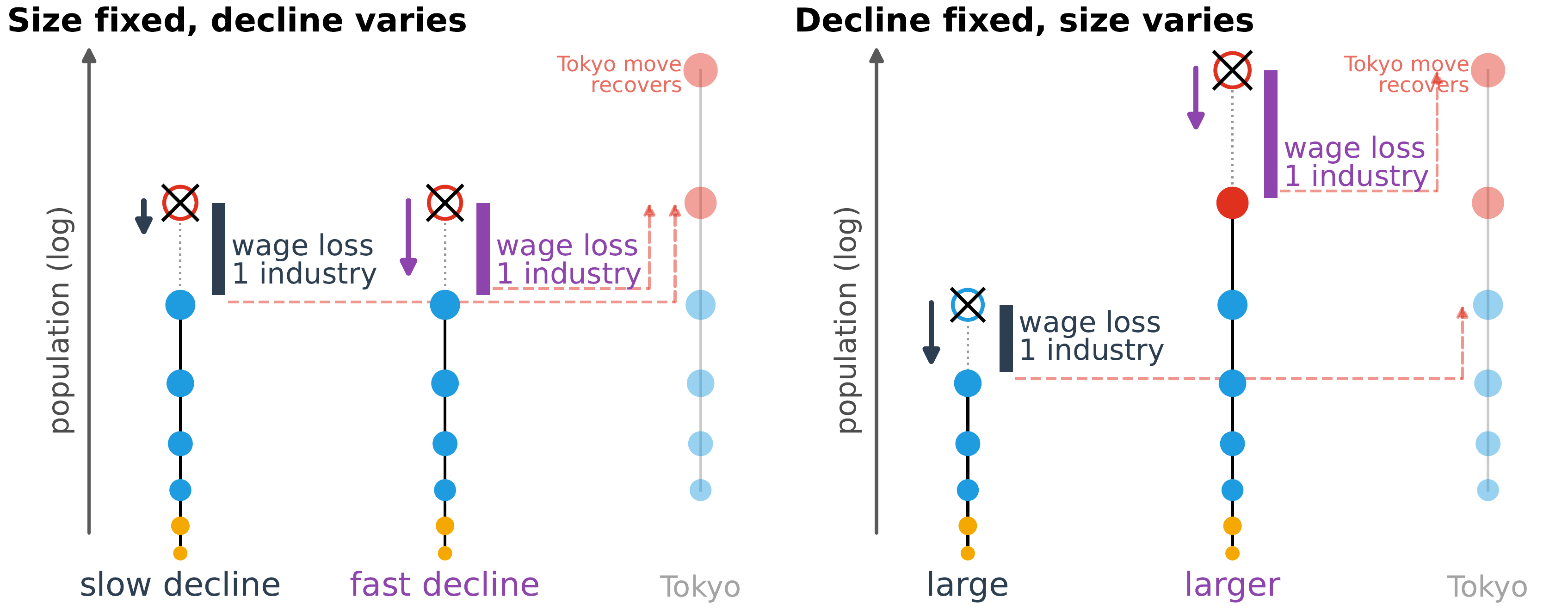}\\[3mm]
{\footnotesize (b) Small declining origin}\\[1mm]
\includegraphics[width=\textwidth]{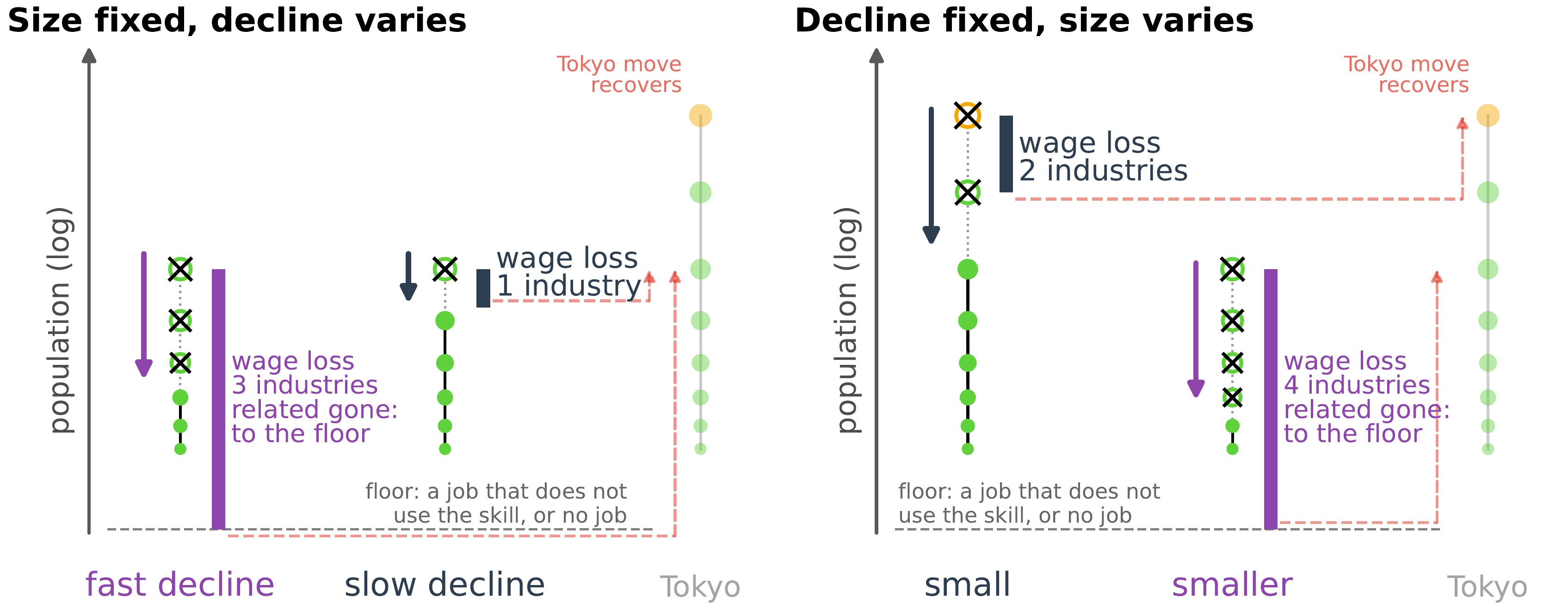}
\end{minipage}
\caption{Conceptual illustration of the wage premium a Tokyo move recovers: large and small origins.}
\label{fig:wage_premium_concept}
\end{figure}

\paragraph{Extensive margin: concavity on jobs and upward consolidation.} The framework predicts that the migration wage premium loads on the extensive margin --- whether an industry is locally present at all --- rather than on the intensive margin --- how large the industries that remain are. Combined with the wage hierarchy of Section~\ref{sec:descriptive_concavity}, this pins down what consolidation costs a worker, illustrated in Figure~\ref{fig:wage_premium_concept}. The logic is the same in every city. When an industry consolidates away, its jobs go with it: a worker attached to it can no longer do that work at home, in a large city and a small one alike. If she stays, she takes whatever job her origin still hosts, and its wage --- one step down the hierarchy if a related industry remains, all the way to the floor if none does. If she follows the job up the hierarchy --- to a larger city, and ultimately to Tokyo, which hosts every industry\footnote{In many-region economic-geography models the number of cities an economy can stably support is set by its overall size: as the population grows, more cities emerge; conversely, once the population becomes small enough, only a single city can be supported in the end \citep{Fujita-Krugman-Mori-EER1999,Mori-Akamatsu-Takayama-Osawa-2025}.} --- she gets her industry and its wage back. The wage she recovers by moving to Tokyo and returning to her industry --- the \emph{premium} --- is therefore the drop a stayer would take. How large that drop is depends on what consolidation has removed from the origin, and that differs sharply between small and large origins; the framework's predictions are therefore stated separately for the two size sides, and read, within each side, as how the premium varies with origin size at a given pace of decline (Section~\ref{sec:strategy_wage}).

Figure~\ref{fig:wage_premium_concept} draws the two cases. In each panel the vertical axis is population (log), and a column of dots is the set of industries one city hosts, each dot placed at the population threshold at which that industry can locate --- far apart near the top (specialized industries), packed close together near the bottom (necessities). The downward arrow is the origin's population decline; only the industries it crosses exit. The bar at an exiting industry is the stayer's wage drop, down to the nearest industry still hosted; the faint column at the right is Tokyo, and the dashed arrow the move that recovers the drop. The left pair in each panel shows two cities of the same size declining at different paces; the right pair, two cities declining at the same pace but of different sizes.

At a \emph{large declining regional core}, near the top of the hierarchy, thresholds are widely spaced, so however deep the decline runs, the \emph{number} of industries lost barely changes: a few cross their thresholds, one at a time, and how fast the core declines says little about what it loses. What separates large origins is instead \emph{which} industry goes, and that is read by size: the larger the core, the more specialized the industries it stands to lose (administrative headquarters, high finance such as trust banking and futures trading), from the high-order end of the hierarchy where the wage gaps between industries are largest. A related industry remains one step below, so a worker who stays falls only one step down the hierarchy --- but that step is taller the larger the core. The premium a Tokyo move can carry therefore rises with origin size, largest at the large-origin corner.

At a \emph{small declining origin}, the industries it stands to lose are low-order ones, whose thresholds sit close together, so here both the pace of decline and the origin's size matter. The industries a small origin loses are not at the floor, however: a UA of 10--30k still hosts about $160$ of the $494$ industries, and Tottori, the smallest origin prefecture, $361$ across its UAs, so what a small origin loses first is the top of its own set, well above its floor. The faster the decline, the more thresholds are crossed at once; and the smaller the origin, the more tightly its remaining necessities are packed in population terms (concavity on jobs), so the same proportional decline removes more of them. What matters for the stayer is whether a related industry is left locally to step down to. The smaller the UA, the more often none is (Figure~\ref{fig:related_left}): a small city hosts many of its specialized industries with no neighbor from the same group beside them, and because a city loses its industries from the top down, a deeper decline takes the second and third fifths as well (Section~\ref{sec:descriptive_concavity}), where such neighbors sit. A worker who stays after a fast decline therefore often has no related industry to step down to and falls all the way down the hierarchy, to its floor: a job that does not use her skill, or no job. Under slow decline fewer thresholds are crossed and the related industries remain; and because at the low-order end the wage gap between neighboring industries is small (Section~\ref{sec:descriptive_concavity}), the stayer's step down is shallow, and about equally shallow whatever the origin's size, so a Tokyo move has little to recover: under slow decline, being smaller does not raise the premium. The premium therefore grows toward the small-origin corner: the smaller \emph{and} faster-declining the origin, the larger the premium a Tokyo move can carry.

Because the exiting industries are necessities, the framework further predicts that their joint exit takes with it the residents' \emph{daily-life infrastructure}: losing these, residents leave, pushing the origin below still more thresholds, so the smallest origins shrink at an accelerating pace and eventually disappear. Smallness and fast decline thus come as a package, and the corner is where both are extreme.

\paragraph{What the framework does not rule out.} Tokyo's thick labor market may carry a premium of its own --- the sorting and matching rents of the urban-wage-premium literature \citep{Combes-Duranton-Gobillon-JUE2008,Andersson-Burgess-Lane-JUE2007,de-la-Roca-Puga-REStud2017,Dauth-etal-JEEA2022} --- and a shrinking population may change those rents as well. The framework does not deny this. Such rents raise the wage of every mover to Tokyo by much the same amount, whatever the origin, so they enter the \emph{level} of the premium, which the analysis leaves uninterpreted. What the framework predicts, and what the analysis tests, is the \emph{difference} in the premium across origins of different size and decline. Thick-market rents cannot produce that difference, because they do not depend on what the origin has lost. Section~\ref{sec:strategy_mechanism} first checks, on the census, that the young-population decline rate used in the wage regression moves the extensive margin and not the intensive one; Section~\ref{sec:strategy_wage} then sets out how the predictions are taken to the worker data.

\section{Data and Sample}\label{sec:data}

Four data sources support the analysis: (i) the Basic Resident Registry for long-run prefecture-to-prefecture migration flows used in the descriptive evidence (Section~\ref{sec:descriptive}); (ii) the JPSED individual panel for the worker-level wage outcome and Tokyo-MA migration treatment; (iii) the Population Census for prefecture-level size and decline as the explanatory variables; and (iv) three predetermined cross-prefecture shocks (demographic Bartik, fertility, capital deepening) used as instruments in the identification strategy (Section~\ref{sec:strategy_id}).

\paragraph{(i) Long-run migration flows (JBRR).} The Basic Resident Registry (\emph{J\=umin Kihon Daich\=o}, JBRR; \citealp{JBRR}) supplies the bilateral OD migration flows used in the long-term descriptive analysis (Section~\ref{sec:descriptive}).

\paragraph{(ii) Outcome and migration treatment (JPSED).} The Recruit Works \emph{Japanese Panel Study of Employment Dynamics} \citep[JPSED;][]{Recruit-JPSED} is an annual individual-level panel covering 2016--2025 (ten waves; SSJDA distribution). For each worker--year, we extract residence prefecture, gender, age, four-level education (high-school, 2-year college, 4-year university, graduate), industry at the 2-digit JSIC code, occupation at the 1-digit JSOC code, and annual main-employment income; we additionally use the cross-sectional and longitudinal pair weights provided by JPSED, which are post-stratified by the Recruit Works Institute to match the demographic distribution of the Statistics Bureau's Labour Force Survey (\emph{R\=odoryoku Ch\=osa}), by gender $\times$ age $\times$ employment status; the pair weight additionally adjusts for inter-wave attrition. Statistics throughout the paper are weighted by these JPSED weights so that the sample represents the underlying working population (full source-code mapping in Online Appendix~\ref{oa:app:data}). The wage outcome is real log wage growth between consecutive waves,
$\Delta\log w_{i,t} \equiv \log w^r_{i,t} - \log w^r_{i,t-1}$, where $w^r$ is real main-employment income --- nominal income deflated by a prefecture-by-year price index. The Tokyo-move \emph{premium} is the difference in this growth between workers who move to the Tokyo MA and those who stay (Section~\ref{sec:strategy_wage}). We deflate because the migration premium is a welfare object --- the worker's gain in real consumption --- rather than a nominal pay change: Tokyo's cost of living exceeds the origin's, so a nominal premium overstates the welfare-relevant gain. The deflator combines the Statistics Bureau's national all-items Consumer Price Index, which removes economy-wide inflation, with its cross-prefecture Regional Difference Index, which prices the origin-to-destination cost-of-living gap. The published regional index gives housing little weight, because owner-occupiers' imputed rent is left out; our baseline therefore adds the index's housing component, so that Tokyo's higher housing costs count in the deflator (details in Online Appendix~\ref{oa:app:cpi_deflation}); a variant without this addition serves as a robustness check \citep{CPI-Japan}. The analysis using fully nominal wage growth is in Online Appendix~\ref{oa:app:cpi_deflation}.

\paragraph{Analytical sample.} The wage-growth sample consists of consecutive-wave pairs from rural-origin workers (35 rural prefectures, i.e.\ origin outside the Tokyo, Osaka, and Nagoya MAs and Fukuoka prefecture), with both wages valid and the top/bottom $1\%$ of $\Delta\log w$ trimmed. The move treatment is migration into the Tokyo MA, and the reference group is rural workers whose destination remains rural. The baseline restricts the move sample to Tokyo-bound migrants; alternative destinations (and alternative origin definitions) are explored as robustness in Section~\ref{sec:robustness}. Workers who at $t-1$ were in school and working part-time are dropped: their $t-1$ wage is a student side-job wage, not the labor-market wage from which the change should be measured. The sample is likewise restricted to \emph{post-first-job} movers: the specification needs a labor-market wage before the move, which new graduates do not have. Retrospective tables are in Online Appendix~\ref{oa:app:retro}.

\paragraph{Pre-COVID baseline.} We restrict the baseline panel to the four pre-COVID transitions --- wage changes over the calendar-year pairs $(t-1, t)$ with $t \in \{2016, 2017, 2018, 2019\}$, that is, 2015/16 through 2018/19.\footnote{Throughout the paper, $t$ denotes the calendar year a wage refers to. JPSED wave labels run one year ahead of the calendar year they report: the wave fielded in January of year $t+1$ covers calendar year $t$, so the calendar transition $(t-1, t)$ is measured from the consecutive wave pair labeled $(t, t+1)$, and the four baseline transitions draw on the waves labeled 2016--2020.} This drops the COVID-contaminated transitions and the single post-COVID transition currently available, because the post-pandemic years are not a comparable steady state: Tokyo's domestic net in-migration at all ages collapsed during the pandemic and has not returned to its pre-COVID level (about $119{,}000$ in 2024 against a pre-pandemic peak of $146{,}000$ in 2019; \citealp{JBRR}), and the 20--29 inflow fell by a fifth in 2020--21 before recovering. The pre-COVID baseline yields $47{,}536$ person-wave pairs ($358$ Tokyo-MA migrants; $4{,}617$ pairs and $95$ migrants among the $\leq 29$ sub-sample, workers aged 29 or younger; the youngest in the sample are 16). Table~\ref{tab:descriptive} reports descriptive statistics. Movers' raw mean wage growth is negative, and movers are younger and more educated than stayers; both features are taken up in Section~\ref{sec:strategy_wage}. Full sample-construction details are in Online Appendix~\ref{oa:app:data}.

\paragraph{(iii) Origin size and decline (Population Census).} The framework's two origin coordinates, size $S_o$ and decline $g_o$ (Section~\ref{sec:framework}), are measured as follows. The size is the origin's population of all ages in 2010 (Population Census; \citealp{POP-CENSUS-2020}): it fixes the origin's position in the hierarchy at the outset --- which industries it hosts, and so what it stands to lose as decline proceeds. The decline is the signed percentage change in the origin's young (15--29) resident population over 2010--2019, $g_o \equiv (S^{\mathrm{y}}_{o,2019}/S^{\mathrm{y}}_{o,2010} - 1) \cdot 100$, with the 2010 base from the Census and the 2019 endpoint from the annual counts of the Basic Resident Registry \citep{JBRR}; negative values mean contraction. This window is the decline the baseline wage transitions (2015/16--2018/19) respond to: it opens at the onset of the national contraction and closes with the last transition, long enough for decline to accumulate into industry exits.

We measure decline on the young cohort because that is where an origin's decline originates. Inter-prefectural migration is concentrated in the twenties (Section~\ref{sec:descriptive_longterm}), and when the young leave, the births they would have had leave with them: the natural decrease that follows is itself a product of their departure. Their decline therefore comes first; the decline of the origin as a whole, and of its market, follows. The all-ages headcount hides this for now, because the elderly population is still growing: over 2010--2020 the rural 15--29 population fell $13\%$ while the $65+$ population rose $19\%$, leaving the population of all ages down only $5\%$.

\paragraph{(iv) Identification instruments.} For the identification strategy in Section~\ref{sec:strategy_id}, we additionally draw on prefecture-level birth records spanning 1965--2004 \citep{Vital-Stats-Japan}, national age-specific survival rates \citep{NIPSSR-2023}, and the Regional-Level Japan Industrial Productivity Database for prefecture-level real capital stock and employment, from which the 1980--85 growth of capital per worker is computed \citep{RJIP-2017}; all three are predetermined relative to the JPSED wage panel.

\begin{table}[htbp]
\centering
\caption{Descriptive statistics of the analytical sample.}
\label{tab:descriptive}
{\small\begin{tabular}{lrr}
\toprule
 & Mean/Count & SD \\
\midrule
\multicolumn{3}{l}{\emph{Sample size}} \\
N person-wave pairs & 47,536 & \\
\quad of which $\leq 29$ & 4,617 (9.7\%) & \\
Unique persons & 21,472 & \\
\midrule
\multicolumn{3}{l}{\emph{Treatment}} \\
Movers to Tokyo MA & 358 (0.75\%) & \\
\quad of which $\leq 29$ & 95 (2.06\%) & \\
\midrule
\multicolumn{3}{l}{\emph{Demographics}} \\
Age (years) & 46.6 & 12.7 \\
Female & 40.3\% & --- \\
University-or-higher & 32.7\% & --- \\
\midrule
\multicolumn{3}{l}{\emph{Outcome}} \\
Wage at $t-1$ (10,000 yen) & 343 & 256 \\
$\Delta\log w$ & 0.0083 & 0.4832 \\
\midrule
\multicolumn{3}{l}{\emph{Outcome and composition by mover status (raw group means)}} \\
$\Delta\log w$: movers & -0.089 & 0.562 \\
$\Delta\log w$: movers, $\leq 29$ & -0.021 & 0.649 \\
$\Delta\log w$: stayers & +0.0091 & 0.4825 \\
Age: movers / stayers & 38.8 / 46.7 & \\
University-or-higher: movers / stayers & 53.1\% / 32.5\% & \\
\midrule
\multicolumn{3}{l}{\emph{Origin drivers (35 rural prefectures)}} \\
Young-population decline $g_o$, 2010--2019 (\%) & -10.2 & 3.9 \\
\quad range & -16.4 to -1.3 & \\
Population $S_o$, 2010 (million): median (range) & 1.38 (0.58--5.50) & \\
\midrule
\multicolumn{3}{l}{\emph{Person-wave pairs by origin cell (Section~\ref{sec:strategy_wage})}} \\
\quad Cell I (large $\times$ slow) & 10,797 (22.7\%) & movers 90 \\
\quad Cell II (large $\times$ fast) & 15,726 (33.1\%) & movers 148 \\
\quad Cell III (small $\times$ slow) & 7,565 (15.9\%) & movers 43 \\
\quad Cell IV (small $\times$ fast) & 13,448 (28.3\%) & movers 77 \\
\midrule
\multicolumn{3}{l}{\emph{Coverage}: pre-COVID (4 transitions)} \\
\bottomrule
\end{tabular}}
\par\smallskip
\begin{minipage}{0.9\textwidth}
\footnotesize \emph{Notes:} Rural origins, JPSED pre-COVID wave transitions (2015/16--2018/19); unweighted statistics (estimation uses the JPSED longitudinal weights). $g_o$ and $S_o$ as defined in the text. Cells: origins split by demeaned size and demeaned decline (over the $39$ origin regions) into (large/small) $\times$ (slow/fast), the classification of Section~\ref{sec:strategy_wage}. Mover--stayer gaps are raw group means of the housing-inclusive real wage growth, with no adjustment.
\end{minipage}
\end{table}

\section{Empirical Strategy}\label{sec:strategy}

The strategy has two steps. Section~\ref{sec:strategy_mechanism} checks, on the census alone, that the decline rate entering the wage regression does remove industries from an origin, and not merely shrink them. Section~\ref{sec:strategy_wage} then specifies the wage regression that tests the predictions of Section~\ref{sec:framework}.

\paragraph{Grain: cities inside prefecture data.} Hosting thresholds are properties of UAs, so the exits the framework predicts happen city by city. The worker data (JPSED), however, record a worker's residence only by prefecture, so from here on the analysis is at the prefecture level. The two steps handle this gap differently. The first step uses no worker data, so it can observe exits at the UA level and pool them to the prefecture without losing them: an industry counts as lost from a prefecture only when every UA in it that hosted the industry in 2009 has lost it by 2021. The second step, the wage regression, has no such option: each origin's UAs are collapsed into one size and one decline. On the destination side this costs nothing, since the Tokyo MA is effectively the single Tokyo UA; the cost falls on the origin side, and what it costs, cell by cell, is documented in Figure~\ref{fig:concavity_cells} and Online Appendix~\ref{oa:app:cell_aggregation}.

\begin{figure}[ht!]
\centering
\includegraphics[width=0.7\textwidth]{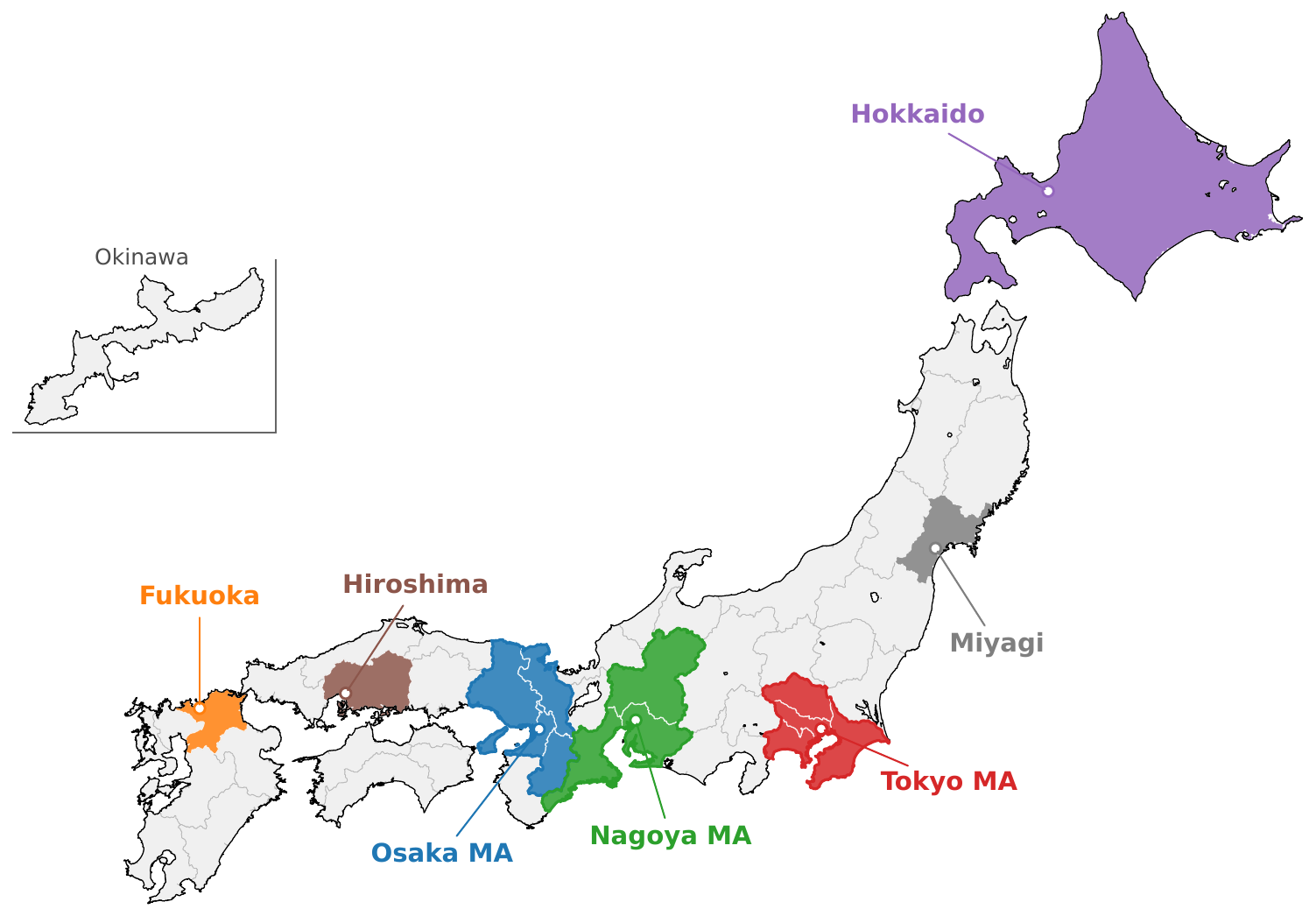}
\caption{Map of the seven metropolitan centers used in the paper.}
\label{fig:metros_reference}
\par\smallskip
\begin{minipage}{0.85\textwidth}
\footnotesize \emph{Notes:} Tokyo, Osaka, and Nagoya MAs are multi-prefecture aggregates; Fukuoka, Hiroshima, Miyagi (Sendai), and Hokkaid\=o (Sapporo) are reported at the prefecture level. The destination in the main wage regression is the Tokyo MA alone; the three major MAs (Tokyo, Osaka, Nagoya) and Fukuoka prefecture define the metropolitan centers excluded from the rural origin set. White circles mark the central ward of each center's principal city --- the representative point of the prefecture's largest metropolitan area.
\end{minipage}
\end{figure}

The destination in the main wage regression is the Tokyo MA alone; the robustness checks below then vary the destination set and the rural-origin set (Fukuoka as a rural origin, Miyagi/Sendai as an MA). Figure~\ref{fig:metros_reference} maps the seven metropolitan centers the paper refers to; Appendix~\ref{app:pref_map} gives a reference map of all prefectures with the three major MAs outlined.

\subsection{Which margin does decline move? Exit, contraction, and entry}\label{sec:strategy_mechanism}

\paragraph{Three margins of the response.} Section~\ref{sec:descriptive_concavity} showed how jobs disappear through the extensive margin: a declining city crosses its industries' hosting thresholds, and the industries exit with their jobs. But exit is not the only thing that happens to a city's industries while its population falls. Surviving industries can shrink, and new industries can enter. Before reading the wage premium through the exit channel, we check which of these demographic decline actually moves. The unit is the origin prefecture: the index $o$ denotes one of the $35$ \emph{rural prefectures} --- those outside the Tokyo, Osaka, and Nagoya MAs and Fukuoka ($34$ once Okinawa, which lies outside the UA delineation, is set aside). For each we measure three things --- the industries it lost, the industries that newly entered, and the contraction of the industries that survived --- and regress each on the decline rate and size. All three are built from the establishment records of the 2009 and 2021 Economic Censuses on the harmonized classification (Online Appendix~\ref{oa:app:data}), aggregated to the fixed 2010 UA polygons and then to the prefecture holding each UA (Online Appendix~\ref{oa:app:indloss_ua}); the same margins at the UA grain are in Online Appendix~\ref{oa:app:lastcopy}.

How should the UA-level exits of Section~\ref{sec:descriptive_concavity} be carried to prefecture origins? An industry counts as lost from an origin when no UA of the origin hosts it any more: at least one of its UAs hosted the industry in 2009 and none does in 2021. The exit-side extensive measure is thus the exit rate --- the share of the industries the origin hosted in 2009 that it had lost by 2021, the same object as Figure~\ref{fig:concavity_on_jobs_UA}(b) at the UA grain:
\begin{equation}
\mathrm{ExitRate}_o \;=\; 100 \times \frac{\#\bigl\{\, j \,:\, j \text{ hosted in 2009 by at least one of } o\text{'s UAs and by none in 2021} \,\bigr\}}{\#\bigl\{\, j \,:\, j \text{ hosted in 2009 by at least one of } o\text{'s UAs} \,\bigr\}},
\label{eq:exits}
\end{equation}
which ranges from $2.2$ to $10.2\%$ over the rural origins (median $6.0\%$; $10$ to $38$ of the $494$ unified 3-digit industries).

The entry-side extensive measure is the \emph{entry hazard}: of the industries the origin did not host in 2009, the share that one of its UAs had come to host by 2021,
\begin{equation}
\mathrm{EntryHz}_o \;=\; 100 \times \frac{\#\bigl\{\, j \,:\, j \text{ hosted in 2009 by none of } o\text{'s UAs and by at least one in 2021} \,\bigr\}}{494 - \#\bigl\{\, j \,:\, j \text{ hosted in 2009} \,\bigr\}},
\label{eq:entryhz}
\end{equation}
entries per $100$ industries \emph{at risk} of entering. The at-risk denominator is essential. A small origin hosts few industries to begin with, so nearly the whole classification is available to enter; Tokyo's portfolio is close to saturation ($493$ of the $494$ unified industries), so almost nothing is. A raw entry count therefore varies with the origin's position in the hierarchy mechanically, before any economics; the hazard removes this and asks how likely an \emph{available} industry is to arrive.

The intensive measure is the survivor contraction
\begin{equation}
\mathrm{Contr}_o \;=\; -100 \sum_{j \in \mathrm{Surv}_o} \omega_{oj}\, \Delta\log e_{oj},
\label{eq:contr}
\end{equation}
the employment-weighted contraction of the industries hosted in \emph{both} years, where $e_{oj}$ is the employment of industry $j$ in prefecture $o$'s UAs and $\omega_{oj}$ its 2009 employment share; the minus sign makes a shrinking survivor set a positive number, so that, like the exit rate, a larger value means a larger loss (every rural prefecture contracts, range $6.6$--$21.6$ log points, median $14.6$).

All three are testable in one line:
\begin{equation}
\mathrm{Margin}_o = \alpha + \beta_g\, g_o + \beta_S \log S_o + u_o,
\qquad \mathrm{Margin} \in \{\mathrm{ExitRate}, \mathrm{Contr}, \mathrm{EntryHz}\},
\label{eq:margin_fs}
\end{equation}
estimated on the $34$ rural origin prefectures with UA-based measures (Okinawa hosts no UA in the 2010 delineation), with heteroskedasticity-robust (HC1) standard errors.

\paragraph{Results: young decline moves the extensive margin.} Table~\ref{tab:margin_firststage} reports the estimates. The exit rate responds to young-population decline --- $0.21$ percentage points higher per percentage point of decline, about $0.65$ more industries lost, conditional on size --- and to size as the concavity mechanism implies ($-3.6$ percentage points per log point of population). Survivor contraction responds to neither. The young who migrate away are no longer the customers of the industries that survive locally; those industries serve the region's remaining all-age population, so their contraction is tied to all-age decline rather than to the young measure that drives migration --- as Online Appendix~\ref{oa:app:lastcopy} confirms. The entry hazard completes the asymmetry: it is decline-neutral and loads on size alone ($+10$ per log point). The entering industries are indeed of a kind with no direct link to the origin's demography --- guest houses riding the inbound-tourism boom, nursing and social care riding population aging, electricity generation riding the solar feed-in tariff, non-store retail riding e-commerce. New industries spread from the top of the urban hierarchy downward, indifferent to the origin's decline. Online Appendix~\ref{oa:app:extmargin_hazards} examines this entry margin in detail. Of the three margins, then, young-population decline moves only exit, and most strongly at small origins, as Figure~\ref{fig:concavity_on_jobs_UA}(b) showed at the UA grain.

\begin{table}[!htbp]
\centering
\caption{Margin-response regressions: which margin does demographic decline move?}
\label{tab:margin_firststage}
\sbox{0}{% GENERATED by run_indloss_nocell.py mechanism_first_stage() -- do not edit.
\begin{tabular}{lccc}
\toprule
 & Exit rate (\%, extensive) & $\mathrm{Contr}_o$ (intensive) & Entry hazard (\%) \\
\midrule
Young-population decline $g_o$ & $-0.21^{***}$ & $-0.12$ & $-0.01$ \\
 & $(0.06)$ & $(0.15)$ & $(0.23)$ \\[2pt]
Log origin size $\log S_o$ & $-3.60^{***}$ & $-1.49$ & $+10.04^{***}$ \\
 & $(0.37)$ & $(1.19)$ & $(1.38)$ \\
\midrule
$R^2$ & $0.70$ & $0.05$ & $0.51$ \\
$N$ (rural prefectures) & $34$ & $34$ & $34$ \\
\bottomrule
\end{tabular}
}
\usebox{0}
\par\smallskip
\begin{minipage}{\wd0}
\footnotesize \emph{Notes:} OLS on the $34$ rural prefectures; robust standard errors in parentheses. All three measures are built from the establishment records of the Economic Census (2009 and 2021) on the harmonized universe and classification (Online Appendix~\ref{oa:app:data}), aggregated to the fixed 2010 UA polygons; $g_o$ is the 2010--2019 young (15--29) decline rate (2010 Population Census base, 2019 Basic Resident Registry endpoint), signed as the growth rate (negative under decline), and $S_o$ the 2010 origin population. The exit rate (equation~\eqref{eq:exits}) and $\mathrm{Contr}_o$ are oriented so that a larger value means more decline (a larger share of the 2009-hosted industries lost, deeper contraction); the entry hazard is entries per $100$ industries not hosted in 2009. Of the three margins, only the exit rate responds to young-population decline. $^{***}$, $^{**}$, $^{*}$: significance at the 1\%, 5\%, 10\% level.
\end{minipage}
\end{table}

\paragraph{Why the exit rate is not the treatment.} If demographic decline moves the exit rate, why not carry it into the analysis that follows as the treatment? Because the measure is compressed: across the $34$ prefecture origins the rate ranges only from $2$ to $10\%$ ($10$ to $38$ industries), while the same origins differ in population by a factor of $9.4$ and in decline rate from $-1.3$ to $-16.4\%$. The analysis below therefore works directly with the two continuous drivers that govern the hierarchy, size and young-population decline (the specification follows in Section~\ref{sec:strategy_wage}), with what the exit measure does identify documented in Online Appendix~\ref{oa:app:indloss_ua}.

One caution remains. Showing that young decline moves only the exit margin does not make a premium keyed to $g_o$ free of other origin-side pressures that move with decline and that we cannot strip out --- a thinning local market, fading amenities; an exit measure would carry them too, since the exits are outcomes of the same decline. What separates the extensive reading from such pressures is the \emph{shape} the premium should take across origins: at small origins absent under slow decline and present under fast, at large origins following size rather than decline. Such pressures are the intensive-margin forces of the agglomeration literature, and the models that microfound them --- sharing, matching, and learning in a market whose size varies continuously \citep{Duranton-Puga-HB2004} --- have no thresholds in them. A premium they drive would rise smoothly with decline at every origin, neither switching on only under fast decline at small origins nor following size at large ones. That shape requires thresholds, crossed in bunches at small origins and one at a time at large ones. Whether the premium follows the threshold shape is what the regression of Section~\ref{sec:strategy_wage} is built to read.

\subsection{The wage regression}\label{sec:strategy_wage}

The task now is to test, in a wage regression, what Section~\ref{sec:framework} predicts: that the premium a Tokyo move carries differs across origins with what consolidation has taken from them --- which industries, and how many at a given pace of decline --- and so between large and small origins. As Section~\ref{sec:strategy_mechanism} concluded, the regression works with the two drivers of the hierarchy, size and decline, measured at the prefecture level as the worker data require. The framework's predictions are stated on exactly these two coordinates: an origin's size decides \emph{which} industries it loses, and its pace of decline decides whether their thresholds are crossed within the window at all (Figure~\ref{fig:wage_premium_concept}). We therefore sort the rural origins into four groups by size and decline rate, as follows, and ask whether the premium differs across them as the framework predicts.

\paragraph{Cell construction.} The framework's predictions (Section~\ref{sec:framework}) are operationalized as four cells, splitting each origin prefecture's demeaned log size and demeaned young-population (15--29) decline rate at zero,
\begin{equation}
\log\tilde S_o = \log S_{o,2010} - \overline{\log S}_{39},
\qquad
\tilde g_o = g_{o,2010,2019} - \overline{g}_{39},
\label{eq:demean}
\end{equation}
where $\overline{\log S}_{39} = \frac{1}{39}\sum_{r=1}^{39} \log S_{r,2010}$ and $\overline{g}_{39} = \frac{1}{39}\sum_{r=1}^{39} g_{r,2010,2019}$ are the means over the $39$ regions: the three metros aggregated, Fukuoka, and the $35$ rural prefectures. The sign of $(\log\tilde S_o, \tilde g_o)$ assigns each origin to one of four cells:
\begin{center}
\renewcommand{\arraystretch}{1.3}
\begin{tabular}{r|cc}
\toprule
                              & Slow decline ($\tilde g_o > 0$) & Fast decline ($\tilde g_o < 0$) \\
\midrule
Large ($\log\tilde S_o > 0$)  & Cell I    & Cell II \\
Small ($\log\tilde S_o < 0$)  & Cell III  & Cell IV \\
\bottomrule
\end{tabular}
\end{center}
The size split at $\overline{\log S}_{39}$ corresponds to a total population of $\approx 1.81$ million. The decline center is $\overline{g}_{39} = -10.1\%$; since all rural prefectures are in absolute decline, the ``slow-decline'' side means less-declining-than-the-39-region-mean. Descriptive statistics of the sample by cell --- person-wave pairs and Tokyo-MA movers --- are in Table~\ref{tab:descriptive}.

\paragraph{Cells at the prefecture-origin grain.} Each cell is a set of prefecture origins, and what it tracks is decided by the UAs those origins contain (Table~\ref{oa:tab:pref_ua_composition} lists each origin's composition; Online Appendix~\ref{oa:app:pref_ua_composition}). A large origin (cells I/II) mixes one large UA with many small ones. Its prefecture size reflects the large core --- Sapporo in Hokkaid\=o ($2.07$ million in 2010, $38\%$ of the prefecture's population), Sendai in Miyagi ($1.24$ million, $53\%$) --- because population adds up: a prefecture with a million-plus core cannot be small. The converse fails: five of the eleven large-side prefectures --- Ibaraki, Tochigi, Fukushima, Niigata, Nagano --- have no UA above $530$k and are large only as a sum of many small UAs and the population outside any UA (Table~\ref{oa:tab:pref_ua_composition}). Prefecture size therefore does what the framework needs for the six large-side origins that have a core: it ranks them by the size of that core, and what their cells capture is the core's loss of specialized industries. For the five without a core it ranks nothing of the kind. Four of these sit near the mean decline rate, which is why the large-side gradient is read away from that point, at the depths where the cores are (Section~\ref{sec:results_main}, finding (ii)). A decline \emph{rate} does not add up; it averages. The prefecture's $g$ is the change in the prefecture's whole young population --- the core, the small UAs, and the areas outside any UA together --- which amounts to a population-weighted average of their declines, and the core's is the mildest of these: Sendai's young population fell $4.8\%$ over 2015--2020 (the earliest window the 1\,km mesh allows) while Miyagi's fell $15.5\%$ over 2010--2019. The prefecture's $g$ therefore does not measure the core's decline --- and the core's own $g$ was in any case uninformative (Section~\ref{sec:framework}: a few one-off exits at widely spaced thresholds). Whether a large origin reads as cell I or cell II is therefore decided by its small UAs, not by its core, and we read the two large-origin cells jointly (Section~\ref{sec:results_main}, finding (ii); Online Appendix~\ref{oa:app:cell_aggregation}).

A small origin hosts only small UAs --- no Sapporo or Sendai --- and here the decline dimension is usable. At the UA level $g$ is informative at a small UA (Section~\ref{sec:framework}: dense thresholds, so the depth of decline sets how many industries go), and with only small UAs inside, the prefecture's $g$ is their own decline. Cells III and IV differ not in the size of their top UA, which is similar, but in $g$: where the origin's largest UA still retains its young, the aggregated decline is mild and the origin lands in cell III; where even the largest UA is losing its young, in cell IV. Cell III thus tracks consolidation to Tokyo out of a small core that still holds its population --- the core being the last host within the origin, since smaller UAs drop an industry before it does; cell IV, consolidation to Tokyo from small fast-declining UAs across the board.

\begin{figure}[htbp]
\centering
\begin{subfigure}{0.72\textwidth}
\centering
\caption{Hosted-industry count}
\label{fig:concavity_static}
\includegraphics[width=\textwidth]{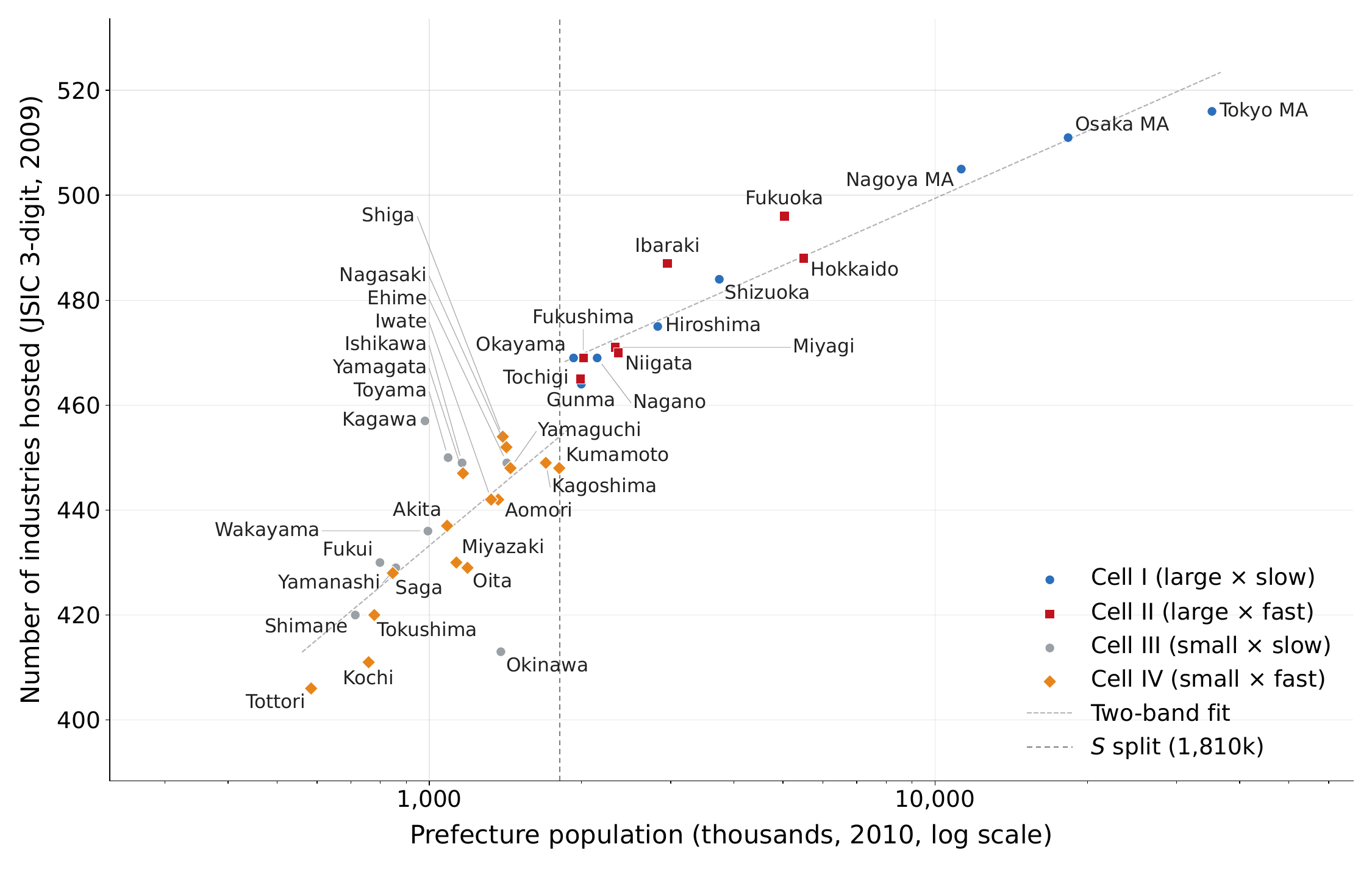}
\end{subfigure}
\par\vspace{0.3em}
\begin{subfigure}{0.72\textwidth}
\centering
\caption{Industry-exit rate}
\label{fig:concavity_exit}
\includegraphics[width=\textwidth]{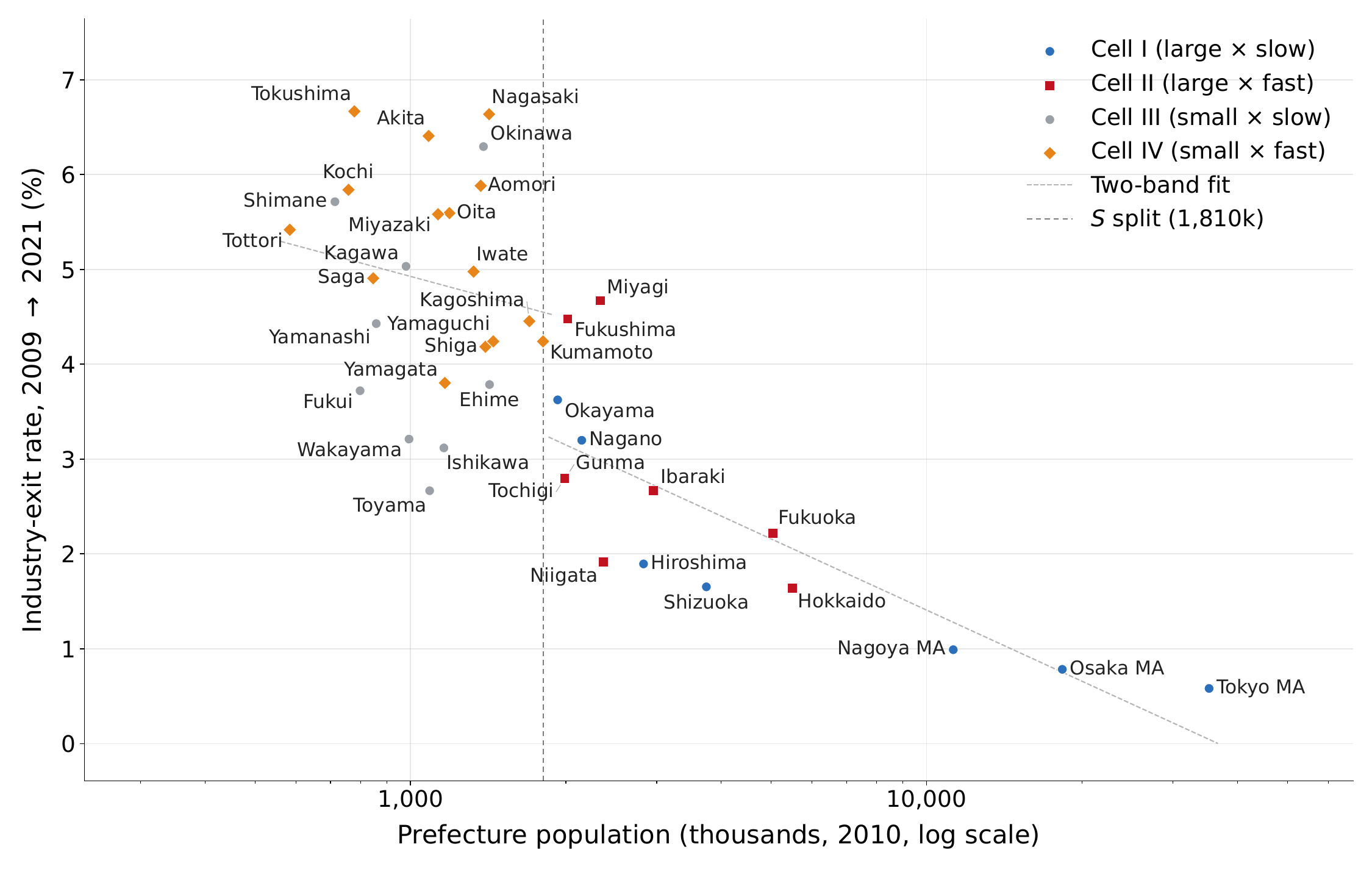}
\end{subfigure}
\caption{Concavity on jobs at the prefecture-origin grain.}
\label{fig:concavity_cells}
\par\smallskip
\begin{minipage}{0.85\textwidth}
\footnotesize \emph{Notes:} Each point is one of the $39$ origin regions (the three metros aggregated, Fukuoka, and the $35$ rural prefectures), colored by its four-way framework cell --- the same cells the wage regression estimates. The vertical dashed line is the $S$ split at $1{,}810$k, separating large origins (cells I/II) from small ones (cells III/IV). (a)~The count of 3-digit JSIC industries each origin region hosts in the 2009 Economic Census, against its 2010 total population (log $x$); the count rises with size but concavely --- the two-band fit split at the $S$ split gives $+35.0$ industries per log-unit of population among small origins versus $+18.5$ among large ones, a $1.9\times$ steeper slope at the small end. (b)~The share of each region's 2009-hosted industries no longer hosted by 2021, against its 2010 total population (log $x$); the exits concentrate among the fast-declining small origins (cell IV), which shed the largest fraction, while the largest MAs (Tokyo/Osaka/Nagoya, cell I) shed essentially none. The sharper UA-grain version is Figure~\ref{fig:concavity_on_jobs_UA} (Section~\ref{sec:descriptive_concavity}).
\end{minipage}
\end{figure}

Figure~\ref{fig:concavity_cells} colors each of the $39$ regions by its cell, and shows that the descriptive primitive of Section~\ref{sec:descriptive_concavity} --- documented at the UA grain --- survives at this coarser grain: the hosted-industry count is concave in size (Panel~(a), slope $+35.0$ industries per log-unit of population below the $S$ split versus $+18.5$ above --- a $1.9\times$ steeper gradient at smaller origins), and the $2009$--$2021$ industry exit rate concentrates at the small fast-declining (cell IV) origins (Panel~(b)), while the large origins take lower exit rates than the small ones.

\begin{figure}[tbp]
\centering
\includegraphics[width=0.98\textwidth]{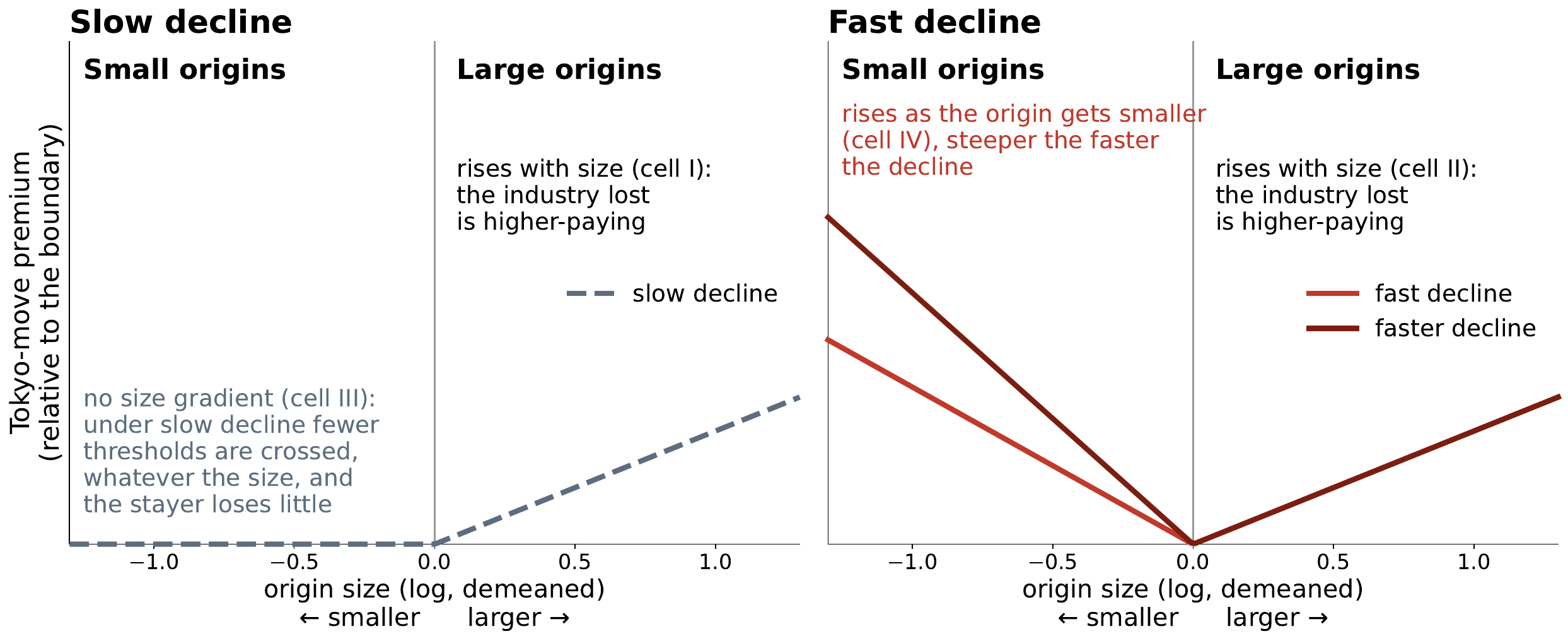}
\caption{Predicted shape of the Tokyo-move premium across origin size and decline.}
\label{fig:prediction_schematic}
\par\smallskip
\begin{minipage}{0.95\textwidth}
\footnotesize \emph{Notes:} Schematic. Horizontal axis: origin size (log, demeaned), with the small/large boundary at zero; vertical axis: the Tokyo-move premium relative to its value at the boundary, so the slope of each line is $\mathrm{ME}(S)$. Left panel, slow decline; right panel, fast decline (two lines, the darker the faster). Right halves: cell I (left panel) and cell II (right panel); left halves: cell III (left panel) and cell IV (right panel).
\end{minipage}
\end{figure}

\paragraph{Prediction.} Figure~\ref{fig:prediction_schematic} summarizes what Section~\ref{sec:framework} predicts at the prefecture-origin grain, as the premium's dependence on origin size at slow and at fast decline. The horizontal axis is origin size (log, demeaned), zero marking the small/large boundary; the vertical axis is the Tokyo-move premium relative to its value at the boundary, so that the slope of each line is how the premium changes with size. On the large side (the right halves) the premium rises with size whatever the pace of decline: a larger core loses a taller step. Decline carries no signal there, so cells I and II are read jointly, as one rising line. On the small side the two panels differ. Under slow decline (left panel, cell III) fewer thresholds are crossed and the related industries remain. The stayer steps down to one of them and loses little, so what a Tokyo move from such an origin recovers is not a loss the framework measures, and no size pattern is predicted for it: the line has no gradient. Under fast decline (right panel, cell IV) the smaller the origin, the more thresholds it crosses and the more likely that no related industry is left, so the premium rises toward the smallest origins; and because a faster decline crosses more thresholds at any small-origin size --- most of all at the smallest, where they are packed tightest --- the faster the decline, the larger the premium and the steeper the rise. The regression below asks whether this shape appears in the data.

\paragraph{Specification.} We estimate the following baseline specification on the rural sample, where $i$ indexes individual workers, $t$ indexes calendar years (annual; the JPSED wave labeled $t+1$ reports calendar year $t$, Section~\ref{sec:data}), and $M_{i,t} \equiv \mathbf{1}\{\text{Move}_{i,t}\}$ indicates that worker $i$ migrated into the Tokyo MA between years $t-1$ and $t$.\footnote{The omitted reference group (\emph{stayer}) is rural-to-rural: rural workers in the wage sample whose destination is also rural (not in any of the four MAs), $98.80\%$ of them staying in their own origin prefecture --- not moving at all --- and only about $1\%$ moving to a different rural prefecture ($97.04\%$ stay-in-place in the $\leq 29$ cohort). The treatment is a move into the Tokyo MA, and the comparison is with staying, because the question is why Tokyo keeps concentrating while the nation shrinks: the premium of interest is what a worker gains by leaving a shrinking origin for the one destination that still gains the young (Section~\ref{sec:descriptive}). Moves to the other metros are examined in Section~\ref{sec:robustness}.} $S^{+} \equiv \max\{\log\tilde S_{o(i)}, 0\}$ and $S^{-} \equiv \min\{\log\tilde S_{o(i)}, 0\}$ are the positive and negative parts of the (continuous) demeaned origin log size of equation~\eqref{eq:demean}, and $g^{\pm}$ analogously for the demeaned origin decline (the origin index $o(i)$ on $S^{\pm}, g^{\pm}$ is suppressed for readability):
\begin{align}\label{eq:main_spec}
\Delta\log w_{i,t} &= \beta_0 + \beta_1 M_{i,t} \notag\\
&\quad + \beta_3^{+} M_{i,t}\, S^{+} + \beta_3^{-} M_{i,t}\, S^{-} \notag\\
&\quad + \beta_4^{+} M_{i,t}\, g^{+} + \beta_4^{-} M_{i,t}\, g^{-} \notag\\
&\quad + \beta_{Sg}^{++} M_{i,t}\, S^{+} g^{+}
       + \beta_{Sg}^{+-} M_{i,t}\, S^{+} g^{-} \notag\\
&\quad + \beta_{Sg}^{-+} M_{i,t}\, S^{-} g^{+}
       + \beta_{Sg}^{--} M_{i,t}\, S^{-} g^{-} \notag\\
&\quad + \mathbf{X}_{i,t}'\boldsymbol{\gamma} + \alpha_{o(i)} + \alpha_t + \varepsilon_{i,t},
\end{align}
where $\Delta\log w_{i,t} = \log w^{r}_{i,t} - \log w^{r}_{i,t-1}$ is real (housing-inclusive) log wage growth between consecutive calendar years (deflation as in Section~\ref{sec:data}); $\mathbf{X}_{i,t}$ collects demographic controls (age, age squared, a dummy for age $\leq 29$, university-or-higher dummy, gender); and $\alpha_{o(i)}$ and $\alpha_t$ are origin-prefecture and year fixed effects.

Stayers' premium is zero by definition, so the origin fixed effects $\alpha_{o(i)}$ absorb all differences across prefectures in stayers' wage growth, and the worker controls and year effects absorb the rest. Against that baseline, the coefficient on Move is the movers' excess wage growth over stayers --- the premium --- and the Move-interacted coefficients tell how far that premium differs with the origin's size and decline: they are the slopes of the lines in Figure~\ref{fig:prediction_schematic}. Because those slopes should differ between the large and small sides and between slow and fast decline, size and decline enter split at their means, with a separate coefficient in each of the four cells.

\paragraph{Read the gradient, not the level.} The outcome is real, housing-inclusive wage growth, because it is real income --- not nominal pay --- that tells whether a move has left a worker better off. Its \emph{level}, however, is not read. The deflator prices a common national basket --- goods available everywhere --- so it counts Tokyo's higher prices without crediting what they buy, the amenities and variety found only there; Tokyo's cost of living is overstated, and the level of the premium is biased downward. This bias is common to movers from every origin, so it cancels when origins are compared. What we read is therefore the gradient: the marginal effect of origin size on the premium, $\mathrm{ME}(S) = \partial(\text{premium})/\partial\log\tilde S$, which in the piecewise specification is $\widehat{\mathrm{ME}}(S)_{\text{cell}} = \hat\beta_3^{\pm} + \hat\beta_{Sg}^{\pm\pm}\,\tilde g$, taken separately on the two sides of the size split. In these terms the predictions of Figure~\ref{fig:prediction_schematic} read: on the large side the premium rises with size among the large cores, in both the slow- and the fast-decline cell, $\mathrm{ME}(S) > 0$, jointly across cells I and II; in cell IV it rises as the origin gets smaller, $\mathrm{ME}(S) < 0$, the more steeply the faster the decline; in cell III, where decline is too mild to cross many thresholds, no significant gradient. Section~\ref{sec:results_main} reports the estimates, with the WLS table in Appendix~\ref{app:marginal_wls} and the complete set of marginal-effect analyses in Online Appendix~\ref{oa:app:marginal_effects}.

\paragraph{Weighting.} Estimation uses the JPSED longitudinal pair weights (Section~\ref{sec:data}), so that the estimates describe the population behind Japan's wage-growth panel; the few pairs without a panel weight take the wave-$t$ cross-sectional weight. Unweighted estimates, the diagnostic \citet{Solon-Haider-Wooldridge-2015} recommend, leave the cell estimates qualitatively unchanged (Online Appendix~\ref{oa:app:weighting}; the weighting choice is discussed in Online Appendix~\ref{oa:app:data}).

\paragraph{Inference.} All specifications use cluster-robust standard errors at the origin prefecture, since our key treatments (log origin size and decline rate) are constant within prefecture. With only $35$ origin clusters ($34$ in the 2SLS, $43$ in the robustness check that widens the origin set), conventional cluster-robust sandwich estimators are downward-biased, so we report \emph{score-based wild cluster bootstrap} standard errors with $B = 999$ Rademacher replications \citep{Cameron-Gelbach-Miller-2008,Davidson-MacKinnon-2010}, the standard small-cluster procedure. The same wild bootstrap is used for the marginal-effect SEs.

\subsection{Identification}\label{sec:strategy_id}

The cells read the premium against measured decline, and measured decline is not exogenous. Three standard concerns stand in the way. \emph{Selection on unobservables}: movers self-select, so the Move main effect carries no causal interpretation. What the cells read is a shape across origins, and selection contaminates it only if the intensity of selection varies across origins in that same shape. On the small side the shape is a size gradient that switches on with decline. Selection that varies with size alone shifts $\mathrm{ME}(S)$ by the same amount at every depth of decline and cannot produce the switch; only selection varying with size and decline \emph{jointly} could, and that is what the instruments below address. On the large side the shape is a size gradient at any pace of decline, which selection on size alone could also produce, so a separate check is needed: the destination test of Section~\ref{sec:robustness}. Selection on size alone would raise the premium of moves to Osaka or Nagoya just as much as that of moves to Tokyo; the mechanism instead ties the large-origin premium to Tokyo alone, the one destination hosting the specialized industries a large core loses --- and the large-origin gradient does vanish once the destination is broadened. \emph{Omitted local shocks}: a decline-correlated shock can move both wages and migration (the 2011 earthquake did exactly this to Miyagi; Online Appendix Table~\ref{oa:tab:ghat_flip}). \emph{Reverse causality}: the Tokyo premium itself induces out-migration, and the accumulated outflow deepens the measured $g_o$. All three contaminate the observed decline variation with components not driven by demographic fundamentals.

The design meets them in sequence: most forms are removed by the specification itself, before any instrument enters; what survives is then handled by an over-identified 2SLS that instruments $g_o$ with three predetermined historical shocks --- a demographic Bartik built from past births, the 1965--85 fertility change, and 1980--85 capital deepening, constructed below --- isolating the component of decline set decades before the panel.

\paragraph{What the specification removes.} Start with what the specification removes by construction: each device eliminates one source of contamination and hands the endogeneity that remains, precisely delimited, to the next device. First, the outcome is the same worker's wage change. Anything that moves wage \emph{levels} --- ability, education, the origin's wage level --- enters both years and differences out; cross-prefecture level differences never enter the outcome at all. Second, the origin fixed effects. Their task is to implement, inside the regression, the fact that a stayer's premium is zero by definition: they set each origin's mean stayer wage growth as that origin's zero line, so that the Move coefficients measure heights above the origin's own baseline --- the premium. The fixed effects also remove a bias. Origins differ in how fast stayers' wages grow --- through industrial composition, human capital, the legacy of public capital, or shocks hitting the prefecture during the estimation window --- and these differences correlate with size and decline. Without the fixed effects, the cell coefficients would pick up these differences in stayer growth and misread them as differences in the premium. With them, any growth shift common to an origin's workers is absorbed into the baseline, while the cross-origin variation in the mover--stayer gap, which is what the cell coefficients read, is untouched. Third, a growth differential common to all movers --- say, Tokyo wages jumping in the panel years --- loads on the Move main effect, which we do not interpret; differentials varying with origin size alone, or with decline alone, are carried by the Move\,$\times$\,size and Move\,$\times$\,decline terms. What remains --- the only question the instruments are asked to answer --- is a force outside the mechanism that shifts the mover--stayer gap jointly in size and decline.

\paragraph{What is left for the instruments.} That residual takes two forms. Mover composition may shift with \emph{current} conditions --- today's premium, a contemporaneous shock. The instruments cut this off: dated decades before the panel, they cannot respond to who moves today. Or a \emph{persistent} trait correlated with decades-old demography may shift the mover--stayer growth gap --- for instance, hometown networks: origins that have sent migrants for decades have thicker webs of predecessors in Tokyo, which raise the wage growth of movers alone, and whose thickness is itself correlated with old demography. This would violate exclusion, and no test rules it out directly. We meet it with the over-identification built into the design: the 2SLS uses three instruments, each drawn from a different piece of history, and their construction follows below. Each of the three yields its own estimate of the cell coefficients, and Hansen $J$ asks whether the three estimates agree. A trait contaminating only one instrument shifts only that instrument's estimate, splits the three apart, and the test rejects. A non-rejection therefore admits only two readings: either all three instruments are valid, or all three estimates are shifted away from the true coefficient by exactly the same amount. The latter is a demanding coincidence --- an instrument's bias scales with how strongly the trait correlates with that instrument's own history, so equal shifts require the trait to correlate with birth cohorts, with the 1965--85 fertility change, and with 1980--85 capital deepening in exactly the proportions that equalize the three biases; under any other correlations the biases differ, the three estimates split, and the test rejects. In the data the three do agree: Hansen $J$ does not reject ($p = 0.15$), and WLS and 2SLS match in sign in every cell (Section~\ref{sec:results_causal}). With 34 clusters and 12 restrictions the test has little power, so we read the non-rejection as consistency among the three histories rather than as proof; the guarantees that hold at any instrument strength are the Anderson--Rubin sets of Section~\ref{sec:results_causal}.

\paragraph{Instruments for $g_o$.} The first instrument is a demographic Bartik: a mechanical prediction of how an origin's young population would have changed had there been no migration at all. We take each prefecture's five-yearly birth counts for 1985--2000 --- the cohorts aged 15--29 in 2010 and in 2019 --- \citep{Vital-Stats-Japan}, age the cohorts forward with national survival rates \citep{NIPSSR-2023}, and count how many would be aged 15--29 in 2010 and in 2019:
\begin{align*}
\widehat{S}_{o, T}^{\,\mathrm{young}} &= \sum_{b \in \{1985, 1990, 1995, 2000\}:\, T-29 \leq b \leq T-15}\!\!\!\!\!\!\!\!\!\!\!\!\!\!\! 5 \cdot B_{o, b} \cdot \sigma^{\mathrm{nat}}_{T-b}, \\
\widehat{g}_o^{\,\mathrm{Bartik}} &= 100 \cdot \log\bigl(\widehat{S}_{o, 2019}^{\,\mathrm{young}} / \widehat{S}_{o, 2010}^{\,\mathrm{young}}\bigr),
\end{align*}
where $B_{o, b}$ is prefecture $o$'s births in central year $b \in \{1985, 1990, 1995, 2000\}$ (scaled by $5$ for its 5-year band) and $\sigma^{\mathrm{nat}}_a$ is the national survival rate from birth to age $a$. The prediction moves only with past births and nationwide survival --- never with where anyone later moved or with what happened to wages --- so it is predetermined relative to the panel; the construction follows \citet{Maestas-Mullen-Powell-AEJM2023}.\footnote{\citet{Maestas-Mullen-Powell-AEJM2023} predict each U.S.\ state's future age structure by aging its starting age distribution forward at national mortality rates, and use the predicted $60+$ share as an instrument for population aging. We apply the same logic to the prefecture-level 15--29 change.}

To stress-test the exclusion restriction, start from what the Bartik's variation actually is. The instrument predicts each origin's young decline from its 1985--2000 birth counts, and fertility stayed high over that period precisely in the rural, agricultural prefectures --- so the Bartik, in effect, sorts origins by how rural they are. If rural character persists and depresses today's wages \emph{directly} (a thin industry mix, low productivity), rather than through the decline-driven consolidation the framework targets, then the instrument is correlated with the wage residual even though it is predetermined. We therefore add two further predetermined shocks whose variation comes from different pieces of history, and ask whether all three instruments deliver the same Move\,$\times$\,Decline effect; their agreement is what defeats the concern. A confounder such as rurality could produce the result through one instrument, but to produce the \emph{same} effect through all three it would have to be correlated with all three histories in the same way; if it is not, the three estimates diverge and the over-identification test rejects. The two additional shocks are constructed in Appendix~\ref{app:iv_multi}. The first is the change in the prefecture-level total fertility rate from 1965 to 1985 \citep{Vital-Stats-Japan}; the 1965 rate is unavailable for Okinawa, so the 2SLS sample is the $34$ other rural prefectures. The second is the rate of capital deepening over 1980--85 \citep{RJIP-2017}: 1980s public investment allocated politically rather than by productivity, which, as the Solow decomposition in \citet{Mori-2026} documents, did not raise productivity afterwards --- prefectures that received more of it grew less, and are losing their young faster today. The three pick up economically distinct cross-prefecture variation in $g_o$: the demographic Bartik captures cohort-size mechanics, the 1965--85 TFR change isolates the per-woman fertility transition, over a window that ends where the Bartik's birth counts (1985--2000) begin, and the 1980--85 capital-deepening shock captures spatial-economic structure inherited from the political-investment cycle (the full argument for their independence is in Online Appendix~\ref{oa:app:bartik}).

The 4-way piecewise refinement is identified via an 18-instrument 2SLS: each of the three shocks is split the same way as the six $g$-containing terms of equation~\eqref{eq:main_spec} --- the two linear pieces and the four cell terms --- giving $3 \times 6 = 18$ instruments, one per shock for each endogenous term. An origin's side of the decline split is its observed one, the same as its cell; the instruments predict how deep its decline is on that side. The 2SLS thus instruments the depth of decline within a cell, not the cell assignment. Observed and predetermined decline fall on opposite sides of the mean in six of the 34 origins, and the cells survive their removal (Online Appendix~\ref{oa:app:cell_assignment}). First-stage diagnostics, Hansen $J$ over-identification tests, and cell-level coefficient estimates are reported in Section~\ref{sec:results_causal} and Online Appendix~\ref{oa:app:bartik}, together with Anderson--Rubin confidence sets that remain valid at any first-stage strength and a sensitivity that controls for the prefecture's 1985 output level (log real value added from R-JIP, \citealp{RJIP-2017}, interacted with Move), to rule out that the premium reflects a prefecture's initial poverty rather than its decline.

\paragraph{Why not an event design.} A reader used to natural experiments may want a more exogenous source of variation than demographic history: some event that changed the population of certain cities and not of otherwise similar ones, so that the two groups can be compared. Two reasons speak against it. First, in Japan the events that move a region's population --- a large employer's arrival or exit, a base, a university, an inflow of foreign workers --- are shocks to local labor demand. Such an event moves local wages directly at the same time as it moves the population; with a wage outcome, it violates the exclusion restriction from the start. Second, the object of interest is not the response to a one-off shock but the cumulative process by which long-run demographic change pushes cities below industry thresholds; a one-time change in the population level identifies the short-run response to that shock --- a different object from decline-driven exit. In short, slow, cumulative decline is itself the treatment this paper studies, and the predetermined demographic variation isolates exactly that.

\section{Main Results}\label{sec:results}

\paragraph{Exits by cell: the same ordering as at the UA grain.} Before turning to the wage regression, Table~\ref{tab:cell_exit_tier} repeats the exit-by-quintile reading of Figure~\ref{fig:ua_exit_quintile} at the prefecture-origin grain: each origin's 2009-hosted industries (an industry is hosted by the origin if any of its UAs hosts it, the rule of Section~\ref{sec:strategy_mechanism}) are ranked into quintiles by national hosting count, and the table reports, for each cell, the share of the industries in each quintile that had been lost by 2021, pooling the industries of the cell's prefectures. Because of the nesting, a large origin's Q1 reaches the genuinely rare industries while a small origin's Q1 is still relatively ubiquitous nationally. As at the UA grain, exits come from the top: in every cell $76$--$88\%$ of them are from Q1 and $97$--$100\%$ from Q1--Q2, while Q4--Q5 essentially never lose an industry. And they deepen toward cell IV: the Q1 exit rate rises monotonically from $16\%$ at cell I to $29\%$ at cell IV, and at small origins the erosion reaches Q2 as well ($6$--$8\%$ against $2$--$3\%$ at large origins) --- the concavity-on-jobs gradient at the prefecture grain. The wage regression below prices these threshold crossings.

\begin{table}[htbp]
\centering
\caption{Industry-exit rate by origin cell and specialization quintile, $2009$--$2021$.}
\label{tab:cell_exit_tier}
\small
\begin{tabular}{lccccc}
\toprule
 & \multicolumn{5}{c}{Within-prefecture specialization quintile} \\
\cmidrule(lr){2-6}
Origin cell & Q1 (spec.) & Q2 & Q3 & Q4 & Q5 (ubiq.) \\
\midrule
Cell I (large,slow) & 16.0\% & 2.0\% & 0.2\% & 0.0\% & 0.0\% \\
Cell II (large,fast) & 17.8\% & 3.4\% & 0.0\% & 0.0\% & 0.0\% \\
Cell III (small,slow) & 24.2\% & 6.0\% & 0.3\% & 0.0\% & 0.0\% \\
Cell IV (small,fast) & 29.0\% & 8.0\% & 0.9\% & 0.1\% & 0.0\% \\
\bottomrule
\end{tabular}
\par\smallskip
\begin{minipage}{0.92\textwidth}
\footnotesize \emph{Notes:} Cells as defined in Section~\ref{sec:strategy_wage} (5, 6, 9, and 15 prefectures in cells I--IV). Each number in the table is one exit rate, computed over all the prefecture--industry pairs of that cell and quintile together, not an average of prefecture-level rates. Industry hosting and exits from the Economic Census (2009 and 2021) at the 3-digit JSIC level, on the union of each prefecture's fixed 2010 UA polygons (each UA assigned to the prefecture holding most of its area): an industry counts as hosted in a year if any of the prefecture's UAs hosts it, and as exited if hosted in 2009 and by none in 2021. Each prefecture's 2009-hosted industries are ranked by \emph{national} hosting count --- the number of the $448$ fixed 2010-polygon UAs hosting the industry in 2009 --- and split into within-prefecture quintiles Q1--Q5, from Q1 (the prefecture's rarest, lowest-count industries) to Q5 (its most ubiquitous). Because each prefecture ranks only the industries it hosts, the quintile cutoffs differ across prefectures.
\end{minipage}
\end{table}

\subsection{Baseline regression}\label{sec:results_main}

All specifications below estimate equation~\eqref{eq:main_spec} on the rural sample defined in Section~\ref{sec:data} ($47{,}536$ person-wave pairs, $358$ Tokyo-MA migrants; $4{,}617$ pairs and $95$ migrants in the $\leq 29$ sub-sample). Origin decline is measured over the long-run pre-COVID window 2010--2019 (alternative 2010--2025 window in Section~\ref{sec:robustness}). Table~\ref{tab:main} reports five specifications, building up from a parsimonious baseline (Spec 1) to the full triple-interaction specification (Spec 4) on the all-age rural sample, and analogously for the $\leq 29$ sub-sample (Spec 5).

\begin{table}[htbp]
\centering
\caption{Main regression: log wage growth on Tokyo MA migration.}
\label{tab:main}
\resizebox{\textwidth}{!}{%
\begin{tabular}{lccccc}
\toprule
 & (1) & (2) & (3) & (4) & (5) \\
 & Full & Full & Full & Full & $\leq 29$ \\
 & rural & rural & rural & rural & only \\
\midrule
$\text{Moved} \to \text{MA}$ & -0.138$^{***}$ & -0.136$^{***}$ & -0.151$^{***}$ & +0.074 & +0.325 \\
 & (0.034) & (0.033) & (0.032) & (0.126) & (0.328) \\[3pt]
$\text{Move} \times \text{Size}$ &  &  & +0.108$^{*}$ &  &  \\
 &  &  & (0.057) &  &  \\[3pt]
$\text{Move} \times \text{Decline}$ &  & +0.003 & +0.006 &  &  \\
 &  & (0.012) & (0.010) &  &  \\[3pt]
$\text{Move} \times \text{Size}^{+}$ (large) &  &  &  & -0.554$^{*}$ & -1.094 \\
 &  &  &  & (0.304) & (0.755) \\[3pt]
$\text{Move} \times \text{Size}^{-}$ (small) &  &  &  & +0.476$^{**}$ & +0.854 \\
 &  &  &  & (0.197) & (0.538) \\[3pt]
$\text{Move} \times \text{Decline}^{+}$ (slow decl) &  &  &  & -0.063$^{*}$ & -0.130 \\
 &  &  &  & (0.035) & (0.092) \\[3pt]
$\text{Move} \times \text{Decline}^{-}$ (fast decl) &  &  &  & +0.082$^{***}$ & +0.180$^{**}$ \\
 &  &  &  & (0.030) & (0.083) \\[3pt]
$\text{Move} \times \text{Size}^{+} \times \text{Decl}^{+}$ (cell I: large $\times$ slow decl) &  &  &  & +0.216$^{*}$ & +0.615$^{**}$ \\
 &  &  &  & (0.114) & (0.265) \\[3pt]
$\text{Move} \times \text{Size}^{+} \times \text{Decl}^{-}$ (cell II: large $\times$ fast decl) &  &  &  & -0.170$^{***}$ & -0.336$^{**}$ \\
 &  &  &  & (0.059) & (0.156) \\[3pt]
$\text{Move} \times \text{Size}^{-} \times \text{Decl}^{+}$ (cell III: small $\times$ slow decl) &  &  &  & -0.146 & -0.180 \\
 &  &  &  & (0.103) & (0.181) \\[3pt]
$\text{Move} \times \text{Size}^{-} \times \text{Decl}^{-}$ (cell IV: small $\times$ fast decl) &  &  &  & +0.144$^{**}$ & +0.503$^{***}$ \\
 &  &  &  & (0.064) & (0.155) \\[3pt]
\midrule
Year FE & Yes & Yes & Yes & Yes & Yes \\
Demographic ctrls & Yes & Yes & Yes & Yes & Yes \\
JPSED weight & Yes & Yes & Yes & Yes & Yes \\
N & 47,536 & 47,536 & 47,536 & 47,536 & 4,617 \\
\bottomrule
\end{tabular}
}
\par\smallskip
\begin{minipage}{\textwidth}
\footnotesize \textit{Notes:} Wage-growth sample: rural-origin COVID-clean wave transitions (2015/16--2018/19); the origin decline rate $g_o$ is measured over 2010--2019. Standard errors in parentheses are a score-based wild cluster bootstrap ($B = 999$, Rademacher) clustered at the origin prefecture ($35$ clusters), which corrects the downward bias of the usual clustered errors when clusters are few. \emph{Reading the 4-way piecewise rows (Specs 4--5).} The framework's content is read through the marginal effect of origin size, $\mathrm{ME}(S)$, which the cell interactions determine at each depth of decline (Appendix~\ref{app:marginal_wls}). On the small-origin side $\mathrm{ME}(S)$ is negative and steepens with decline in cell IV and shows no gradient in cell III, as predicted. On the large-origin side it is positive at the depths where the large cores sit, with cells I and II read jointly (finding (ii)). The Move main effect is the premium at the knot --- the point of average size and decline, $\log\tilde S_o = \tilde g_o = 0$ --- and the linear $\text{Move} \times \text{Size}^{\pm}$ and $\text{Move} \times \text{Decline}^{\pm}$ rows are its size and decline gradients there. None of these is read: the knot is not where the gradient is evaluated (Section~\ref{sec:results_main}, finding (ii)). $^{***}$, $^{**}$, $^{*}$: significance at the $1\%$, $5\%$, $10\%$ levels.
\end{minipage}
\end{table}

\begin{figure}[!ht]
\centering
\includegraphics[width=0.9\textwidth]{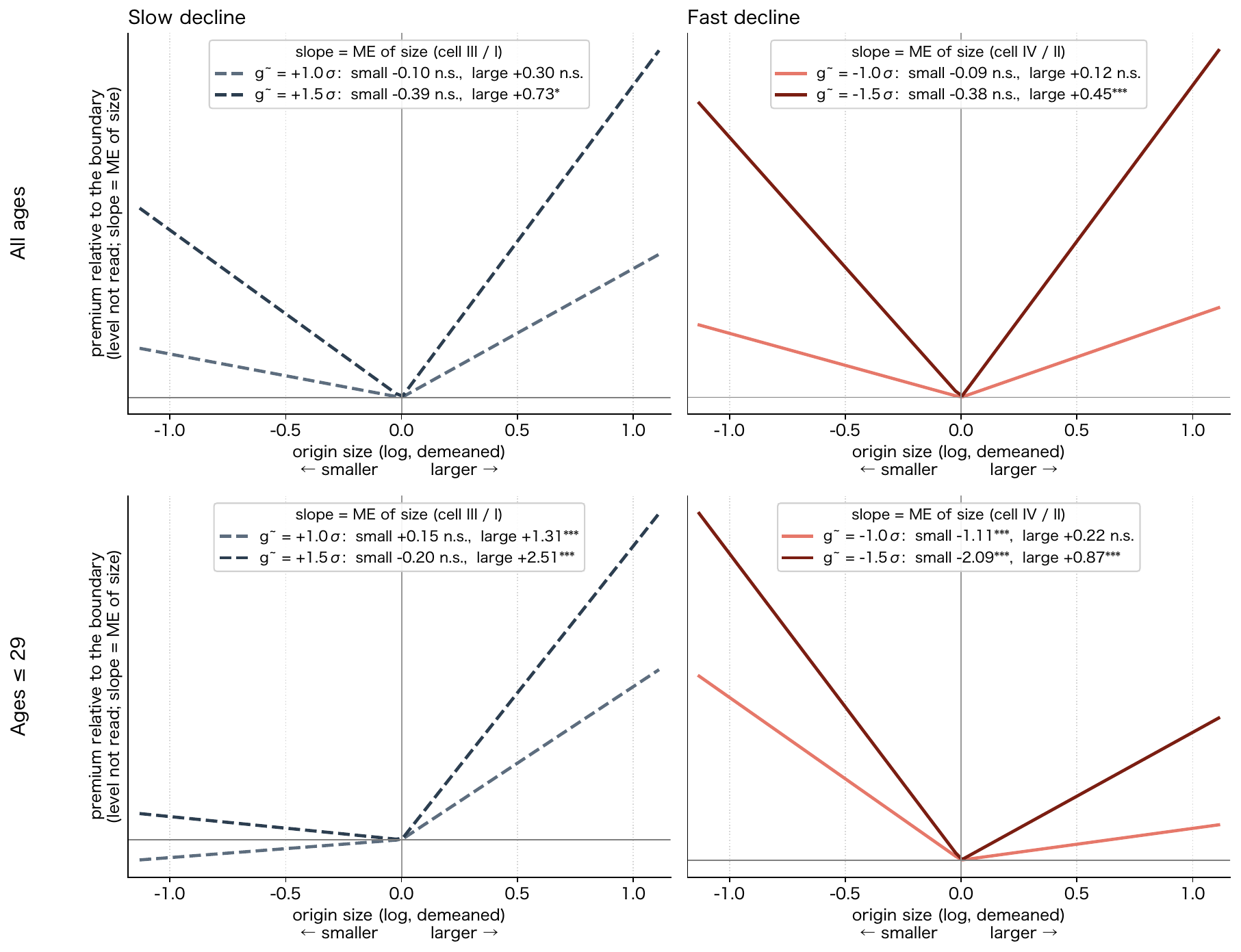}
\caption{Estimated Tokyo-move premium across origin size, at slow and fast decline: all ages and ages $\leq 29$.}
\label{fig:estimate_slow_fast}
\par\smallskip
\begin{minipage}{0.95\textwidth}
\footnotesize \emph{Notes:} Same frame as Figure~\ref{fig:prediction_schematic}. Lines are the premium implied by the piecewise WLS of Table~\ref{tab:main} (Spec 4, all ages, top row; Spec 5, ages $\leq 29$, bottom row), evaluated along demeaned origin size at fixed decline --- slow decline at $\tilde g = +1\sigma_g$ and $+1.5\sigma_g$ (left, dashed), fast decline at $\tilde g = -1\sigma_g$ and $-1.5\sigma_g$ (right, solid; darker is faster) --- and drawn relative to the boundary $\log\tilde S = 0$ at the same $\tilde g$. The slope of each line is the marginal effect of origin size, $\widehat{\mathrm{ME}}(S)$, reported in the legend for the small and the large side with significance from the same wild cluster bootstrap as Appendix~\ref{app:marginal_wls} (n.s.\ = not significant). $\sigma_g$ is the within-sample standard deviation of $\tilde g$, as in Appendix~\ref{app:marginal_wls} ($3.95$ all ages, $3.90$ ages $\leq 29$); the appendix table reports the standard errors. Levels are not interpreted (Section~\ref{sec:strategy_wage}); only the slopes are read.
\end{minipage}
\end{figure}

Specs 1--3 of Table~\ref{tab:main} estimate what a pooled design would --- one premium, one global slope --- and none of it is read, for the reasons of Section~\ref{sec:strategy_wage}: the Move main effect ($-0.151^{***}$ in Spec 3) is a level, biased downward by the real-wage measure; and a single slope averages over cells of opposite sign, which is why the global decline slope is null and the global size slope ($+0.108^{*}$) is only the average of the opposite-signed cell slopes read below. The same holds for the Move main effects and the linear size and decline terms of Specs 4--5 (e.g.\ Move\,$\times$\,Size$^{+} = -0.554^{*}$ against Move\,$\times$\,Size$^{-} = +0.476^{**}$). The framework's content is in the four cell interactions of Specs 4--5, read as cell-specific slopes --- the marginal effect of size, $\mathrm{ME}(S)$ --- and drawn in Figure~\ref{fig:estimate_slow_fast} in the frame of Figure~\ref{fig:prediction_schematic}.

Two findings stand out. \emph{(i) The premium is concentrated in young workers.} The cell-level coefficients are two to four times larger in cells I, II and IV in the $\leq 29$ sub-sample than in the all-age sample (cell III, where no gradient is predicted, is insignificant in both) --- column (5) against column (4) of Table~\ref{tab:main}: the hierarchy-consolidation mechanism operates most strongly among young workers, for whom migration determines career formation. Education does not split it further: university graduates earn no larger a premium than other movers, and the cell coefficients are unchanged when the margin is allowed for (adding Move\,$\times$\,Univ to Spec 5 gives an insignificant $+0.10$; Online Appendix~\ref{oa:app:move_univ}).

\emph{(ii) The cells carry the predicted shape, and the two origin sides are read differently.} Figure~\ref{fig:estimate_slow_fast} draws the marginal effects for both samples: the horizontal axis is origin size as in Figure~\ref{fig:prediction_schematic}; each line is the premium implied by the WLS estimates at a given pace of decline, slow in the left-hand panels and fast in the right-hand ones; the upper row is all ages (Spec 4), the lower ages $\leq 29$ (Spec 5), where the premium concentrates and which we read in the text; and the slope of a line is $\mathrm{ME}(S)$ --- a slope per log unit of origin size, read for sign and ordering rather than as a quantity. On the \emph{large-origin} side the two cells are read jointly (Section~\ref{sec:strategy_wage}),\footnote{Ignoring the $g$ that splits cells I and II --- merging the two cells into a single size term --- would misspecify the model. The evidence is in Table~\ref{tab:main} itself: the two cells' slopes are sharply different ($+0.615^{**}$ against $-0.336^{**}$). That is, the prefecture-aggregated $g$, contaminated by the mixing of large and small UAs, still explains the slopes, so a specification that drops it fails. The $g$ itself is not --- and cannot be --- interpreted, nor is the fitted gradient at the knot $\tilde g = 0$.} as a single rising line, and $\widehat{\mathrm{ME}}(S)$ is read away from the knot $\tilde g = 0$ (Table~\ref{tab:main}'s Size$^{\pm}$ rows are this knot). The reason is where the large cores sit (Section~\ref{sec:strategy_wage}): most of them --- Hiroshima, Takasaki, Hamamatsu, Okayama --- on the slow side at $\tilde g = +0.4$ to $+0.9\sigma_g$ ($\sigma_g$, the standard deviation of $\tilde g$ across the $39$ origin regions, is $3.95$ percentage points), and Sapporo and Sendai on the fast side at $-1.4\sigma_g$ --- around $\pm 1\sigma_g$, the depths at which we evaluate ($\pm 1$ and $\pm 1.5\sigma_g$). There $\widehat{\mathrm{ME}}(S)$ is positive --- $+1.31^{***}$ at $\tilde g = +1.0\sigma_g$ (cell I), $+0.87^{***}$ at $\tilde g = -1.5\sigma_g$ (cell II; Appendix~\ref{app:marginal_wls}). In both cells, then, the larger the core, the larger the premium: the right halves of Figure~\ref{fig:estimate_slow_fast} rise in both the slow- and the fast-decline panels, as predicted. On the \emph{small-origin} side the III-versus-IV contrast is sharp. Here too $\widehat{\mathrm{ME}}(S)$ is read away from the knot, for a different reason: $\tilde g = 0$ is a mild decline, on the side on which too few thresholds have been crossed for the gradient to have switched on, so cell IV is read at $-1\sigma_g$ and $-1.5\sigma_g$. In cell IV $\widehat{\mathrm{ME}}(S)$ is negative at every evaluation point and deepens with decline --- $-1.11^{***}$ at $\tilde g = -1.0\sigma_g$, $-2.09^{***}$ at $-1.5\sigma_g$: the smaller and the faster-declining the origin, the larger the premium. In cell III the marginal effects are null at every point: no significant gradient within the cell. In the left halves of Figure~\ref{fig:estimate_slow_fast} the fast-decline lines rise toward smaller origins, steeper the faster, and the slow-decline lines show no gradient --- the shape of Figure~\ref{fig:prediction_schematic}; the all-age row shows the same shape, with the small-side slopes not significant and the large-side ones significant only at the deeper point.

Taken together, the pattern across origins is the one the extensive margin predicts: on the large side the premium follows size across the large cores, at slow and fast decline alike --- the one step a stayer falls is taller the larger the core; on the small side it switches on with decline --- the related industries go only under fast decline, and only then does the stayer fall to the floor. Cell III is the telling part. The framework predicts no gradient there: under slow decline a small origin loses fewer industries and its stayers step down to a neighboring one --- a step of much the same length at every size, so the premium does not vary with size. None appears: the cell III slope is not significant at any evaluation point. A null cannot by itself show that the slope is zero, and this one is imprecise (s.e.\ $0.50$ at $+1\sigma_g$); what the estimates do establish is the contrast the framework predicts on the small side, that the slope differs from cell IV's ($p = 0.03$).\footnote{The difference between the two cells' marginal effects of size, evaluated at $\pm\sigma_g$, and its standard error are computed from the same wild-bootstrap draws as Table~\ref{tab:main} (Spec 5): $-1.26$ (s.e.\ $0.57$).}

The WLS cell $\times$ decline-evaluation table is in Appendix~\ref{app:marginal_wls} (Table~\ref{tab:marginal_wls}); the full version with the unweighted columns is in Online Appendix~\ref{oa:app:marginal_effects}, Table~\ref{oa:tab:marginal}. At ages 30--64 the small-origin gradient disappears and the large-origin one is much weaker, consistent with the premium's concentration in young workers. Robustness of the four cells across alternative decline windows, Tokyo definitions, and outlier treatments is reported in Section~\ref{sec:robustness}; the triple-IV 2SLS is in Section~\ref{sec:results_causal}.

\subsection{Instrumenting decline: three predetermined histories}\label{sec:results_causal}

The WLS estimates above are subject to the migration-driven endogeneity of $g_o$ (Section~\ref{sec:strategy_id}). We instrument $g_o$ with the three predetermined shocks introduced in Section~\ref{sec:strategy_id} --- the demographic Bartik, the 1965--85 TFR change, and the 1980--85 capital deepening --- each split by size side and decline sign as the regressors are, so that every endogenous interaction has its own instrument; the over-identifying restrictions are tested with Hansen $J$. Table~\ref{tab:iv_summary} reports the headline WLS, 2SLS, and Hansen $J$ statistics for the linear (Spec 3) and 4-way piecewise (Spec 5) coefficients.

\begin{table}[htbp]
\centering
\caption{WLS vs.\ over-identified 2SLS: three predetermined shocks.}
\label{tab:iv_summary}
{\small
\begin{tabular}{lccc}
\toprule
                                                             & WLS              & 2SLS (3 IV)      & Hansen $J$ ($p$-value) \\
\midrule
\multicolumn{4}{l}{\emph{Linear Spec 3 (Move\,$\times$\,Decline, full rural):}}                                                  \\
$\beta_4$                                                    & $+0.006$         & $-0.012$        & $0.59$ (df $= 2$)      \\
                                                             & $(0.010)$        & $(0.009)$        &                        \\[3pt]
\midrule
\multicolumn{4}{l}{\emph{4-way piecewise (Spec 5, $\leq 29$ sub-sample, 2010--2019 long-run window):}} \\
Cell I ($\beta_{Sg}^{++}$, large $\times$ slow decline)      & $+0.615^{**}$   & $+0.910^{***}$    & \multirow{8}{*}{$0.15$ (df $= 12$)} \\
                                                             & $(0.265)$        & $(0.197)$        &                        \\[3pt]
Cell II ($\beta_{Sg}^{+-}$, large $\times$ fast decline) & $-0.336^{**}$  & $-0.461^{***}$    &                        \\
                                                             & $(0.156)$        & $(0.131)$        &                        \\[3pt]
Cell III ($\beta_{Sg}^{-+}$, small $\times$ slow decline)    & $-0.180$         & $-0.321^{*}$      &                        \\
                                                             & $(0.181)$        & $(0.168)$        &                        \\[3pt]
Cell IV ($\beta_{Sg}^{--}$, small $\times$ fast decline) & $+0.503^{***}$  & $+0.622^{***}$    &                        \\
                                                             & $(0.155)$        & $(0.171)$        &                        \\
\bottomrule
\end{tabular}}
\par\smallskip
\begin{minipage}{0.95\textwidth}
\footnotesize \emph{Notes:} The table compares the WLS estimates (the linear Spec 3 and the 4-way piecewise Spec 5 of Table~\ref{tab:main}) with their 2SLS counterparts --- the Spec 3 row on the full rural sample, the piecewise rows on the $\leq 29$ sub-sample --- under JPSED longitudinal weights. Standard errors in parentheses (score-based wild cluster bootstrap, $B = 999$, Rademacher, at the origin prefecture), for both WLS and 2SLS. The 2SLS instruments the endogenous decline interactions with three predetermined cross-prefecture shocks --- a demographic Bartik, the 1965--85 change in the total fertility rate, and the 1980--85 rate of capital deepening (Section~\ref{sec:strategy_id}) --- so that agreement between WLS and 2SLS tests whether the results survive removing the endogenous part of decline. The 1965 TFR is unavailable for Okinawa (before reversion), so the 2SLS sample has $34$ rural prefectures. Full first-stage diagnostics are in Online Appendix~\ref{oa:app:bartik}. $^{***}$, $^{**}$, $^{*}$: significance at the $1\%$, $5\%$, $10\%$ levels.
\end{minipage}
\end{table}

Under 2SLS the piecewise pattern survives and sharpens (Table~\ref{tab:iv_summary}): every cell keeps its WLS sign, the estimates move \emph{away} from their WLS values rather than attenuating toward them, and the over-identification test fails to reject ($J$ $p = 0.15$ across $12$ over-identifying restrictions). As argued in Section~\ref{sec:strategy_id}, a persistent regional trait biasing all three instruments by exactly the same amount is an implausible coincidence; that three instruments of distinct historical origin point to the same estimates is therefore consistent with the exclusion restriction, and with it the causal reading of the cell coefficients. Because every coefficient moves in the same direction, the implied 2SLS $\mathrm{ME}(S)$ also keeps its WLS sign in every cell; we do not report these as estimates, however, for the reason taken up below --- the strength of the first stage. The same triple-IV 2SLS on the all-age rural sample delivers the same pattern ($J$ $p = 0.24$); the many-instrument diagnostics and the first stage of each shock alone are in Online Appendix~\ref{oa:app:bartik}.
\paragraph{Weak-instrument robustness.} The 2SLS has six regressors that contain $g$ --- the two linear pieces and the four cell terms --- and each has its own first stage. The first-stage $F$ statistics differ sharply: $198$ in cell II, $18$ in cell I, and $15$ in cell IV, but $7$ in cell III and $3$ for the slow-side linear term Move\,$\times$\,$g^{+}$ (Online Appendix Table~\ref{oa:tab:iv_firststage_4way}). Read against the Stock--Yogo criteria, which are derived for a single endogenous regressor and serve here as a guide --- about $9$ for the point estimates to carry little weak-instrument bias, about $22$ for conventional $t$-intervals to have correct coverage --- the three cells the paper reads clear the first bar, and their estimates indeed move away from WLS rather than toward it; cell III and the slow-side linear term do not. Only cell II clears the second bar, so the $t$-intervals on the other cells may be too narrow. We therefore complement them with Anderson--Rubin confidence sets (AR sets), which are valid whatever the first-stage strength (Online Appendix~\ref{oa:app:bartik}). They establish two things. First, the shocks move the premium: in the linear specification estimated on the $\leq 29$ sample, the AR set for the decline slope excludes zero. Because the shocks can reach wages only through the decline they predict, this is a decline-driven response, whatever the first-stage strength. Second, a model with a single slope on each size side is rejected --- its AR set is empty --- so the four-cell split is needed. What the AR sets cannot deliver is a confidence interval for each cell's coefficient. An AR set for one cell can only be obtained from a joint test of all six endogenous coefficients, and with $18$ instruments against $34$ clusters that test is uninformative. There is thus no informative weak-instrument-robust interval for each cell separately, nor for the marginal effects built from them, and the 2SLS marginal effects are not reported as estimates. What the 2SLS does add is the direction in which it moves the coefficients, taken up next.

\paragraph{The direction of the WLS-to-2SLS gap.} Instrumenting moves all four cell coefficients up in magnitude (Table~\ref{tab:iv_summary}), and the direction is informative. Measured decline contains the migration that the premium itself induces, and this pushes the WLS gradient two ways. Out-migration deepens measured $g_o$ exactly where the premium is high, which makes the observed premium look \emph{more} sensitive to decline than it is. But each migrant who arrives in Tokyo also narrows the Tokyo--origin wage gap \citep{Blanchard-Katz-1992}, and this erosion is strongest at the origins that have sent the most migrants --- which are, since measured decline includes the outflow, the origins with the deepest measured decline --- so it makes the observed premium look \emph{less} sensitive to decline. 2SLS strips the migration component out of $g_o$; that it comes out larger than WLS in every cell means the second force dominated: WLS measures the premium after arbitrage has eroded it. Cell by cell, the WLS and 2SLS confidence intervals overlap heavily, so the size of the gap is not to be read. What is to be read is the pattern: all four cells move in the same direction, the two estimators agree in sign in every cell and in significance in cells I, II and IV; cell III, where no gradient is predicted, stays insignificant under WLS and reaches only the $10\%$ level under 2SLS, within what that prediction allows.\footnote{Cell III is insignificant under WLS and significant at $10\%$ under 2SLS ($-0.32$). The instruments that identify this coefficient --- the three shocks in the $S^{-}g^{+}$ form --- have the weakest first stage of the four cells ($F = 7$), below the bias guide of about $9$ and far below the coverage guide of about $22$. With a first stage this weak the 2SLS standard error is understated, so the $10\%$ significance may be spurious; this is one of the intervals that the weak-instrument discussion above says may be too narrow.}

\section{Robustness}\label{sec:robustness}

\paragraph{Destination, origin, window, and outliers (R1--R7).} R1--R5 vary the destination and origin definitions: broadening the move treatment to all four major MAs (R1), narrowing it to Tokyo Prefecture alone (R2), reclassifying Fukuoka as a rural origin (R3) or Miyagi/Sendai as an MA (R4), and expanding the origin set to all 43 non-Tokyo-MA prefectures (R5). Across these five redefinitions the predicted pattern holds in the $\leq 29$ sample (the one departure, an insignificant sign change of cell I in the all-age sample under R2, is discussed in Online Appendix~\ref{oa:app:robustness_detail}); significance moves with the redefinition, in the directions the mechanism implies. R1 is the destination test of Section~\ref{sec:strategy_id}. When moves to Osaka, Nagoya, and Fukuoka are counted as treatment alongside Tokyo, the large-origin cells collapse toward zero, and the instruments stop agreeing with one another: the Hansen $J$ test, which the Tokyo-bound design passes ($p = 0.15$), rejects ($J = 27.8$, df $= 12$, $p = 0.006$; Online Appendix~\ref{oa:app:dest_placebo}). Extending the decline window to 2010--2025 (R6) behaves the same way --- signs intact, significance weakened --- because Tokyo-bound migration collapsed in 2020--21 and, at all ages, has not recovered (Section~\ref{sec:data}), so the post-COVID years add a period in which the migration response was suspended. The baseline trims the top and bottom $1\%$ of wage growth (Section~\ref{sec:data}); winsorizing them instead keeps the large wage changes of displaced movers in the sample and strengthens the framework cells (R7), so the trimmed baseline is the conservative choice. Nor is the deflation what drives the cells: with nominal wage growth as the outcome, the four cell coefficients are essentially unchanged (Online Appendix~\ref{oa:app:cpi_deflation}). The checks are tabulated in Online Appendix~\ref{oa:app:robustness_detail}; a further check shows that dropping the small fraction of family-driven moves leaves the cells essentially unchanged (Online Appendix~\ref{oa:app:voluntary}).

\section{International Context: Japan as Forerunner}\label{sec:external}

The mechanism activates only once working-age contraction has pushed origin industry sets across their thresholds at scale. Japan is the most advanced case among developed economies; this section places it beside four others on the same demographic trajectory --- Korea, Italy, Spain, and Germany (Figure~\ref{fig:oecd_workage_trajectory}). All four have low fertility; what separates them is immigration. Korea, like Japan, receives too little to offset its fertility deficit, and its working-age population is already contracting; Italy, Spain, and Germany have so far softened theirs through immigration: Italy's and Germany's working-age populations have fallen far less from their peaks, and Spain's has not yet fallen at all. The mechanism should therefore be operative in Korea and only beginning in the European cases. Table~\ref{tab:external_summary} summarizes two aggregate signatures by country --- the cross-region correlation between young-population decline and net out-migration, and the share of regions whose largest outflow goes to the national apex --- and bears this out: Korea shows the strongest pattern: its decline--migration correlation exceeds Japan's, and for about four in five of its regions the largest outflow goes to Seoul; Italy comes next; Spain and Germany are weaker.

\begin{table}[htbp]
\centering
\caption{Cross-country external validation summary.}
\label{tab:external_summary}
\small
\begin{tabular}{lrcrcl}
\toprule
Country & TFR & Units ($n$) & Correlation $r$ & Apex MA & Source \\
        &     &              & (young decl.\ vs.\ net mig.) & dominance & \\
\midrule
Korea   & 0.70 & 15 sido               & $+0.93^{***\,a}$\phantom{$^{)}$}             & $\sim 79\%$ & OECD RDB \\
Japan   & 1.15 & 39 (4 MA + 35)        & $+0.88^{***}\phantom{{}^{\,a)}}$            & $\sim 60\%$ & Census + JBRR \\
Italy   & 1.21 & 21 NUTS-2             & $+0.65^{**}\phantom{{}^{*\,a)}}$            & $\sim 52\%$ & ISTAT \\
Spain   & 1.13 & 19 NUTS-2             & $+0.44^{*}\phantom{{}^{**\,a)}}$            & $\sim 41\%$ & INE \\
Germany & 1.46 & 38 NUTS-2 / 16 NUTS-1 & $-0.14^{\text{n.s.}}\phantom{{}^{*\,a)}}$   & $\sim 6\%$  & Destatis \\
\bottomrule
\end{tabular}
\\[0.4em]
\begin{minipage}{0.95\textwidth}
\footnotesize \emph{Notes:} The \emph{correlation} column is, across each country's regions, the correlation between young (15--29) population decline over 2014--2024 (negative in declining regions) and the net migration rate over 2020--2024; a more positive value means faster-declining regions lose more residents to migration. ${}^{***} p<0.001$, ${}^{**} p<0.01$, ${}^{*} p<0.10$. \emph{Apex-MA dominance} is the share of origin regions whose largest outflow goes to the country's largest metropolitan area --- Seoul (Korea), Tokyo (Japan), Milan (Italy), Madrid (Spain), Berlin (Germany) --- the apex itself excluded; it measures how far internal migration converges on that single destination. $^{a}$ Sejong, a planned administrative city founded in 2012 whose young population has grown rapidly, is an outlier; excluding it, Korea $r = +0.77^{***}$ ($n = 14$). When determining each origin's dominant destination, flows smaller than $1\%$ of the origin's total outflow are ignored, so tiny flows do not decide the destination. Data sources, the regional-unit definitions (NUTS, TL, sido), and country-by-country detail are in Online Appendix~\ref{oa:app:external}.
\end{minipage}
\end{table}

Korea is also the one comparator with a worker panel that follows movers (KLIPS; \citealp{KLIPS}), so the wage regression can be replicated there (Online Appendix~\ref{oa:app:external_klips}). The four-cell design does not transfer --- Korea's origins outside the Seoul MA span only $1.15$--$3.4$ million people, so the size axis carries no identifying variation --- but the decline axis does. We estimate this decline response on the years from 2006 onward. The hierarchy between population size and industry location holds for manufacturing too, but more loosely than for tertiary industries: tertiary industries serve customers where they live, while a plant's location also answers to ports, factor costs, and siting policy. Before 2006 Korea's economy was still shifting out of manufacturing, so moves driven by that transformation would contaminate the effect of population decline; only from the mid-2000s does the tertiary sector carry the economy's weight (the timing is read from the Establishment Census; Online Appendix~\ref{oa:app:external_klips}). In this era the framework's signature appears: movers from fast-declining origins earn about $9\%$ more than movers from slow-declining ones, a gap that depends on no single origin. As in Japan, the level of the premium is not read; the difference across origins is.

Italy, Spain, and Germany weaken in step with their immigration buffers: Italy, whose young population has been falling since the early 1990s and whose south already sends its migrants to Milan, sits closest to Japan and Korea; Spain and Germany further back. The cross-country pattern is a difference in timing, not in kind: Japan and Korea are forerunners, not exceptions, and the weaker European signals mark a lag, not immunity --- nor are they evidence that Japan is uniquely vulnerable. As each of these labor markets enters contraction, the same test becomes available there (country-by-country detail in Online Appendix~\ref{oa:app:external}).

\section{Concluding remarks}\label{sec:conclusion}

\subsection{Summary}\label{sec:discussion_summary}

Japan's migration under working-age decline is dominated by young workers heading to Tokyo, a flow that has not slowed as the nation shrinks. We read it through two facts about Japan's urban system: industries are hierarchically nested across city sizes, and wage gaps between industries widen toward the top of the hierarchy. On this hierarchy, population decline is one mechanism seen from two ends. Decline removes industries from the top of each city's own set --- and, as the Economic Census shows, this exit margin is the only one that responds to young-population decline. The regions lose their industries and shrink, the smallest cities first and larger ones after them; the industries, and the workers who follow them, consolidate into larger cities and ultimately into Tokyo. The wage regression prices these threshold crossings, reading not the level of the Tokyo premium but how it changes across origins, and finds the shape the framework predicts. At large origins the premium rises with size in both the slow- and the fast-decline cell: the larger the core, the more specialized the industry it loses and the wider the wage gap to the related industry one step below, so a stayer's loss --- and the premium a Tokyo move recovers --- grows with size. At small origins the premium switches on with decline: the smaller and the faster-declining the origin, the more thresholds it crosses at once and the less likely a related industry remains, so a stayer falls further and the premium is largest at the smallest, fastest-declining origins --- and absent where decline is mild, because the related industries remain and, at the low-order end of the hierarchy where wage gaps between neighboring industries are small, the step down to them is shallow for small origins of any size. The gradients are sharpest among workers aged 29 or younger, survive an over-identified 2SLS on three predetermined historical shocks, keep their signs under every redefinition of destination, origin, window, and outliers in the young sample, and reappear in Korea, the one other economy whose labor market is already contracting. This, then, is how an economy shrinks in space: not evenly, but from its smallest cities up the hierarchy, with industries and the workers who follow them moving to ever larger cities, and the chain ending in Tokyo, the one city that hosts every industry --- which is why Tokyo's concentration continues while the nation declines.

What the paper does not measure is how much of Tokyo-bound migration runs through industry exits: the decline rate carries more than the exits we measure, and the reduced form does not apportion the premium among the channels it bundles. Quantifying that share is the task for a structural model, which we take up as the next step of this research. One feature of the driver, however, deserves emphasis. The exits this paper measures are not new. Small cities have lost industries faster than large ones in every decade since the 1980s, and from the 1990s --- as deindustrialization and the completion of the high-speed network reached them --- small cities began to lose more industries than they gained, so that the set of industries a small city holds has been shrinking since the mid-1990s (Online Appendix~\ref{oa:app:exits_1981}). The effects of these successive triggers mix across periods, and we do not separate them, because each started the same sequence of events: postwar urbanization, deindustrialization, and the high-speed transport network each drew people out of the regions on its own account, and the population they took then pushed small cities below the thresholds of their industries, which consolidated upward in turn. Those events, however, each struck once. Population decline is different: it is ongoing, it is already locked in by the age structure, and in the 2010s it acts on regions whose industrial map those shocks had already redrawn, removing from small cities the low-order industries that had survived the earlier shocks. As long as the population keeps falling, the set of industries each city can hold keeps changing, and so does the migration it sets off. The channel this paper documents is therefore not a one-off adjustment but a standing feature of the decline era: its consequences will keep accumulating, and it is one to watch closely.

\subsection{Policy implications}\label{sec:discussion_policy}

The findings inform Japan's ``Local Vitalization'' (\textit{Chih\=o S\=osei}) debate, the national program launched in 2014 after \citet{Masuda-2014} projected that, if the young kept leaving for Tokyo, nearly half of Japan's municipalities would eventually lack the population to sustain themselves \citep[see also][]{Matanle-Rausch-2011,OECD-Japan-2016}. The wage regressions are consistent with moves from fast-declining origins being pushed by the loss of local industries: the premium rises exactly where the industries workers relied on have crossed their local hosting thresholds and disappeared.

Decline proceeds in a fixed order. It removes industries from small origins first, and small origins are many --- too many to hold all of them above their thresholds while the population keeps falling; nor are large cities exempt once their own population falls --- rural Japan foreshadows urban Japan. Cell IV is the most acute concern. There the exiting industries are the necessity industries of everyday life --- general merchandise stores, hospitals, kindergartens --- so the premium movers gain is the mirror image of a thinning of daily-life infrastructure for those who stay. And those who stay are disproportionately the old and the less mobile: a declining city's housing outlasts its jobs and, as it becomes cheap, holds and draws low-income households \citep{Glaeser-Gyourko-JPE2005}, and depopulating regions age because the young leave first \citep{Giannone-etal-DP2024}. What cell IV adds is what these stayers lose --- the local presence of the necessity industries themselves: a structural risk to the livability of the smallest declining communities, not merely a labor-market reallocation. At large cores the consolidating industries are specialized and no such collapse arises; the two cases call for different responses.

The place-based policy literature makes the case for spatially targeted support along two lines. One asks where workers should be, given agglomeration spillovers: subsidizing a place largely relocates activity unless the marginal worker generates a larger externality there \citep{Kline-Moretti-ARE2014}, though once inter-regional transfers and sorting on private returns are allowed, efficiency can call for place-specific subsidies \citep{Fajgelbaum-Gaubert-QJE2020}. The other asks how to help people stuck in places that are declining, where out-migration has slowed and a new job raises employment most \citep{Austin-Glaeser-Summers-BPEA2018,Bartik-JEP2020}, with program evaluations finding effects that are weak on average and design-dependent \citep{Neumark-Simpson-HB2015}.

Both lines of this literature take the national economy's size as stable and each place's set of industries as given, so that the policy lever is a subsidy to employment in a place --- a wage subsidy, or a tax credit for hiring or investing there --- which at best shifts jobs from one place to another. Under national decline neither premise holds. The total shrinks, and what an origin loses takes the form of whole industries crossing their thresholds. Once below a threshold, a subsidy to employment does not bring the industry back, because the binding constraint is the origin's market size, not the cost of labor.

Stopping the outflow, moreover, would address only part of the shrinkage. About two thirds of the smallest UAs' 2010--2020 loss of working-age population is natural change --- the cohorts reaching 65 are larger than the cohorts reaching 15, so the working-age population falls even if no one moves\footnote{Each UA's 2010 age pyramid (municipal five-year pyramids allocated to the fixed 2010 UA polygons by mesh population shares) is aged forward ten years with national survival ratios; the counterfactual 2020 stock aged $15$--$64$ minus the actual 2010 stock is the natural component, and the actual 2020 stock minus the counterfactual is net migration. UAs below $30{,}000$ lost $15.5\%$ of their working-age population, $9.9$ points natural and $5.6$ points migration, with similar shares across the sub-$100$k bands.} --- so a policy that stopped out-migration entirely would leave most of the contraction in place. And a population shrinking through natural change alone still crosses its industries' thresholds: the exits, and the out-migration they trigger, would continue. The contraction is driven first by the age structure, which no regional policy reverses; that lies beyond this paper's reach. Short of restoring fertility, what policy can do is adapt: keep the industries daily life requires by gathering a shrinking population in places that can still hold them, arranging the smaller economy over the land so that the contraction does the least damage. The question is then no longer \emph{how much} to subsidize each place but \emph{which} places to keep above their thresholds, and which to consolidate into them.

A complementary lever works on the thresholds themselves. Remote medicine, mobile retail, and the digitalization of public services lower the population at which a necessity can be supplied locally, and such investment directly relaxes the constraint the mechanism imposes. But not every threshold can be engineered down, and daily life requires not one necessity but the whole set: a place has to be large enough to hold all of them. How large that is, is the question \citet{Mori-Ogawa-DP2025} take up, from the dual side of the problem: where the present paper asks, for each city, which industries remain locally viable as it shrinks, the companion work asks, for each necessity industry, which cities remain viable hosts. It estimates the population at which a city can still supply a full set of daily-life necessities (from obstetric care and emergency hospitals through high schools, clinics, funeral services, and dry-cleaning, and finally supermarkets and convenience stores) at about $30{,}000$ residents; of Japan's $431$ UAs on the 2020 boundaries, $198$ clear this ``sustainable-city'' threshold in 2020, but only $111$ do by 2100 and $48$ by 2150 on current projections. Together the two perspectives say what the design should look like: hold a set of regional cores above that threshold, so that each catchment area keeps a full set of necessities instead of ceding it to Tokyo by default, and consolidate the smaller settlements into those cores. Which cores to hold, and which settlements to consolidate into them, is the question Japan's ``compact-and-network'' policy poses \citep{OECD-Japan-2016}. What this paper adds is the timing: a city's position in the hierarchy pins down which industries it stands to lose next, so the coming losses are largely predictable, and the plan can be drawn before the necessities fail rather than after.

{\singlespacing
\setlength{\bibsep}{6pt plus 1pt}
\bibliographystyle{aer}
\bibliography{../references}
\par}

\clearpage
\appendix
\counterwithin{figure}{section}
\counterwithin{table}{section}
\counterwithin{equation}{section}

\begin{center}
{\LARGE\bfseries Appendix}
\end{center}
\vspace{1.5em}

\section{Reference maps}\label{app:maps}

\subsection{Prefectures}\label{app:pref_map}

\begin{figure}[!htbp]
\centering
\includegraphics[width=\textwidth]{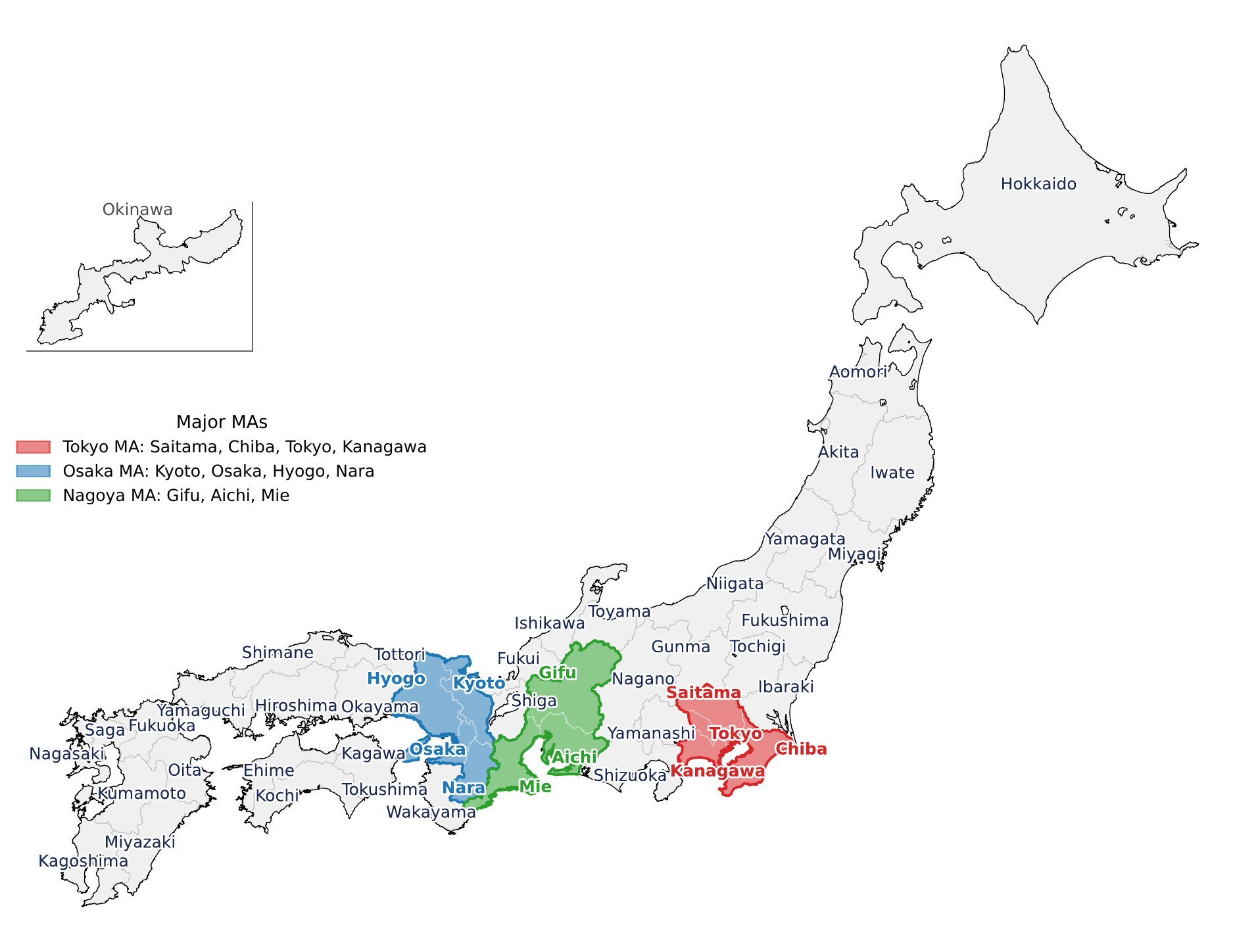}
\caption{Map of Japan's $47$ prefectures.}
\label{fig:pref_reference_map}
\par\smallskip
\begin{minipage}{0.97\textwidth}
\footnotesize \emph{Notes:} The three major MAs --- Tokyo, Osaka, and Nagoya --- are shaded and outlined, and their member prefectures are listed in the legend. Okinawa is shown as a top-left inset.
\end{minipage}
\end{figure}

Figure~\ref{fig:pref_reference_map} names all $47$ prefectures and outlines the three major MAs used throughout the paper. The Tokyo MA is Saitama, Chiba, Tokyo, and Kanagawa; the Osaka MA is Kyoto, Osaka, Hyogo, and Nara; the Nagoya MA is Gifu, Aichi, and Mie. In the main wage regression the destination is the Tokyo MA and the rural-origin set is the $35$ prefectures outside the four major MAs (the three MAs plus Fukuoka); the robustness checks (Section~\ref{sec:robustness}) reclassify Fukuoka as a rural origin and Miyagi (Sendai) as an MA. The base map matches Figure~\ref{fig:metros_reference}: mainland prefectures are clipped to the coastline, and Okinawa is drawn as an inset.

\subsection{Urban agglomerations in 2010: a population-column map}\label{app:ua_map}

\begin{figure}[!htbp]
\centering
\includegraphics[width=\textwidth]{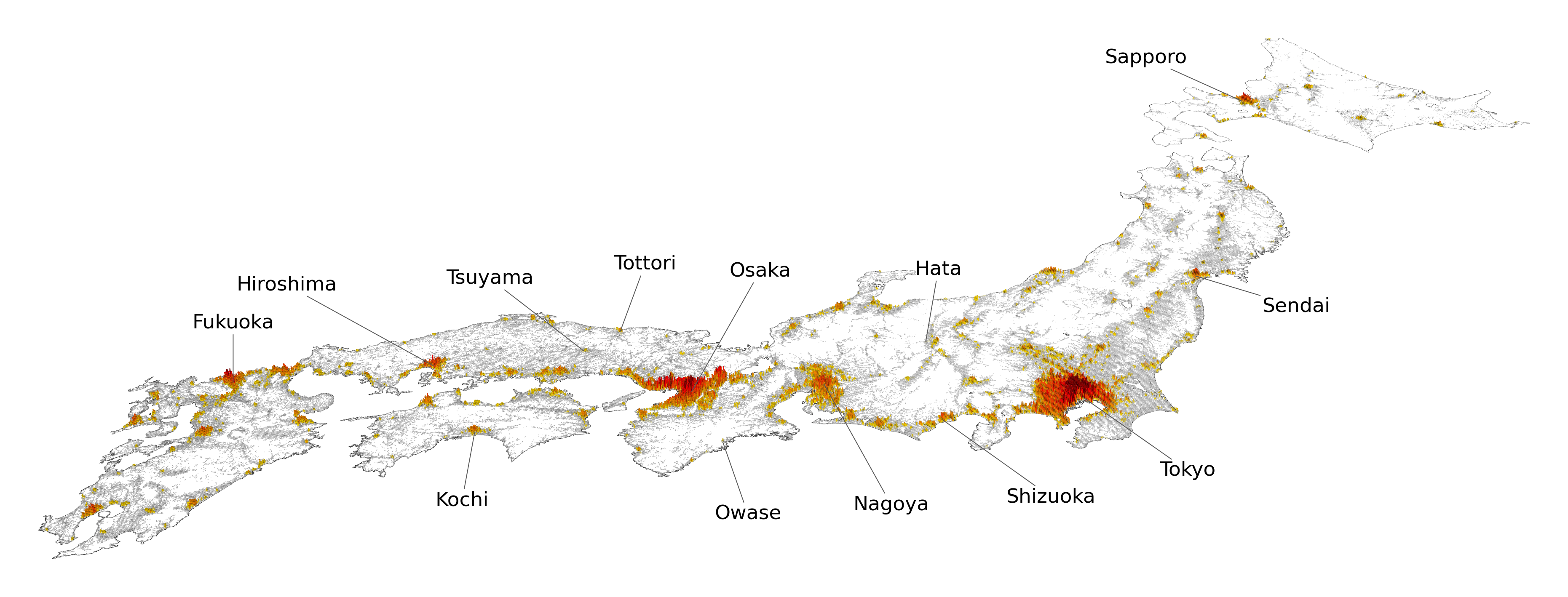}
\caption{Population-column map of Japan, 2010 Census 1\,km mesh.}
\label{fig:pop_columns_2010_ua}
\par\smallskip
\begin{minipage}{0.97\textwidth}
\footnotesize \emph{Notes:} Column height is cell population; UA cells (fixed 2010 polygons, 448 UAs) are colored on density bands from pale yellow to dark red, and inhabited non-UA cells are gray. The map shows only Japan's four main islands (Honshu, Hokkaido, Kyushu, Shikoku); smaller outlying islands are omitted. Labelled cities are the ten of Figure~\ref{fig:concavity_on_jobs_UA} plus Sapporo, Hiroshima, and Shizuoka. Source: 2010 Population Census (Statistics Bureau, Ministry of Internal Affairs and Communications) 1\,km mesh and the fixed 2010 UA delineation.
\end{minipage}
\end{figure}

Figure~\ref{fig:pop_columns_2010_ua} renders Japan's 2010 population directly from the 1\,km census mesh as a three-dimensional column map: each inhabited 1\,km cell is a column whose height is its population. An Urban Agglomeration (UA) is a maximal connected cluster of 1\,km mesh cells each with population density of at least $1{,}000$ persons per km$^2$ and with a total cluster population of at least $10{,}000$ \citep{Mori-Smith-Hsu-PNAS2020,Mori-Murakami-DP2025}; the 2010 census yields $448$ UAs. Cells belonging to a UA (the fixed 2010 polygons of Section~\ref{sec:descriptive_concavity}) are colored on the canonical density bands (dark red at the highest densities through to pale yellow); inhabited cells outside every UA are gray. Population is concentrated in a limited number of UAs of very different sizes: Tokyo is far larger than the regional cores (Osaka, Nagoya, Fukuoka, Sapporo, Sendai, Hiroshima), and those are far larger than the many small cities. The inhabited land between them, outside any UA (gray), is thinly settled.

\section{Empirical validation of the hierarchy property}\label{app:hierarchy}

The conceptual framework of Section~\ref{sec:framework} is built on the hierarchy property of industrial location established by \citet{Mori-Nishikimi-Smith-JRS2008,Mori-Smith-JRS2011}: a city of larger size hosts a strict superset of the industries hosted by any smaller city, so that as one moves up the urban size hierarchy, the set of locally viable industries grows monotonically. Both framework predictions---concavity-on-jobs at small declining origins and upward consolidation at large declining cores---rest on this primitive. This appendix tests the hierarchy property at the prefecture grain, on the 2009 Economic Census --- the same census year as the UA-grain test of the main text --- using the formal procedure of \citet{Mori-Akamatsu-Takayama-Osawa-2025}.

\paragraph{The HP test.} The test can be run on any set of locations --- UAs in the main text, prefectures here. A location \emph{hosts} industry $j$ iff at least one establishment of $j$ operates in it (the minimal-presence definition of \citealp{Mori-Akamatsu-Takayama-Osawa-2025}). For each industry $i$, sort the industries by ascending number of hosting locations, and compute the cumulative-coordination share:
\begin{equation}\label{eq:hp}
\mathrm{HP}_i = \frac{\sum_{j : |\mathrm{hosts}(j)| > |\mathrm{hosts}(i)|} |\mathrm{hosts}(i) \cap \mathrm{hosts}(j)|}{\sum_{j : |\mathrm{hosts}(j)| > |\mathrm{hosts}(i)|} |\mathrm{hosts}(i)|},
\end{equation}
The numerator counts the \emph{realized} coordination: across the locations hosting $i$, the number of (location, $j$) pairs in which a more-ubiquitous industry $j$ is co-hosted with $i$. The denominator is its theoretical maximum under perfect hierarchy---every location hosting $i$ hosting \emph{every} more-ubiquitous $j$---so $\mathrm{HP}_i \in [0,1]$ is the share of the predicted subset relations actually realized, and equals $1$ under perfect hierarchy. We construct the null distribution by reshuffling each industry's location assignment uniformly at random (preserving the number of hosting locations per industry), recomputing $\mathrm{HP}_i$ on the reshuffled assignment, and repeating $1{,}000$ times. The $p$-value is the share of null replications whose $\mathrm{HP}_i$ equals or exceeds the actual $\mathrm{HP}_i$.

\paragraph{Data and sample.} In Section~\ref{sec:descriptive_concavity} the location unit is the UA (Figure~\ref{fig:hp_test_ua}); here it is the prefecture. We use the public prefecture-level tabulation of the 2009 Economic Census for Business Frame (Table~1 of the confirmed report: private establishments by 3-digit industry and prefecture). Prefecture $o$ \emph{hosts} industry $j$ iff the tabulation reports at least one private establishment of $j$ in $o$. The sample is all $47$ prefectures $\times$ $518$ 3-digit (IND3) industries.

\begin{figure}[htbp]
\centering
\includegraphics[width=0.7\textwidth]{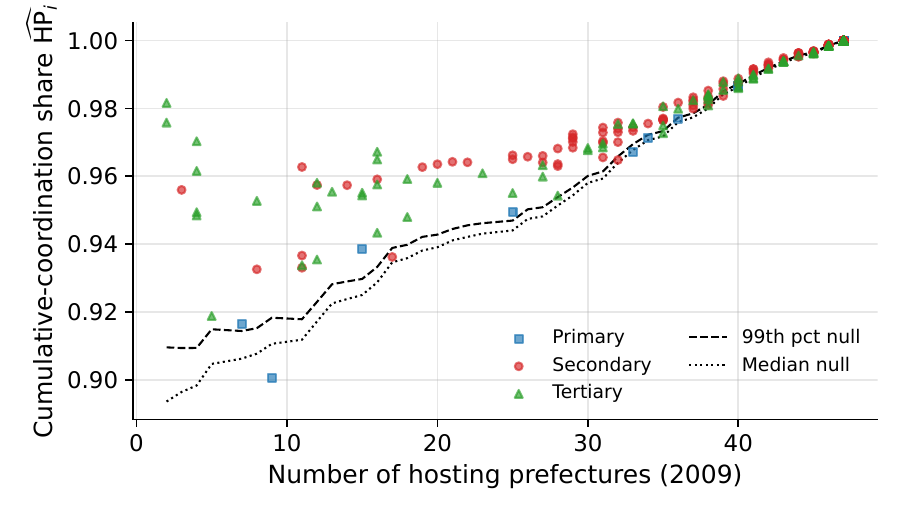}
\caption{Hierarchy-Property test at the prefecture grain.}
\label{fig:hp_test}
\par\smallskip
\begin{minipage}{0.85\textwidth}
\footnotesize \emph{Notes:} Per-industry cumulative-coordination share $\widehat{\mathrm{HP}}_i$ against the number of hosting prefectures (public prefecture tabulation of the 2009 Economic Census, private establishments, all $47$ prefectures), colored by broad sector as in the UA-grain Figure~\ref{fig:hp_test_ua}, with the 99th percentile (dashed) and the median (dotted) of the random-permutation null, which preserves each industry's hosting count, for reference. The actual shares lie at or above the 99th-percentile null line for the great majority of industries, confirming the hierarchy property at the prefecture grain.
\end{minipage}
\end{figure}

\paragraph{Results.} Run at the \emph{prefecture} grain on the 2009 Economic Census, the hierarchy property is empirically validated: $93\%$ of industries ($481$ of $518$) reject the random-permutation null at $p < 0.05$, and $90\%$ at $p < 0.01$. The mean actual cumulative-coordination share is $0.992$ versus a mean median-null of $0.988$. The absolute gap is small because prefectures host most 3-digit industries: at this grain the hosted sets are near-saturated ($320$ of the $518$ industries are present in all $47$ prefectures), leaving little room for the actual share to exceed the null. The test still rejects because the null distribution is just as narrow: under random assignment the share almost never rises as high as the actual one, so even a difference of $0.004$ is rarely matched by chance. Split by broad sector (primary, secondary, tertiary), the mean $\widehat{\mathrm{HP}}_i$ is $0.976$ for the $17$ primary industries (median null $0.975$), $0.990$ for the $207$ secondary ($0.986$), and $0.994$ for the $294$ tertiary ($0.990$)---the same level ordering (primary $<$ secondary $<$ tertiary) as at the UA grain (Figure~\ref{fig:hp_test_ua}), with the primary sector closest to its null, reflecting its location-specific geography. Figure~\ref{fig:hp_test} shows the per-industry $\widehat{\mathrm{HP}}_i$ scatter against the number of hosting prefectures. The UA-grain version of the test---the sharper, primary statement of the hierarchy property---is reported in the main text (Figure~\ref{fig:hp_test_ua}).

\begin{figure}[htbp]
\centering
\includegraphics[width=0.92\textwidth]{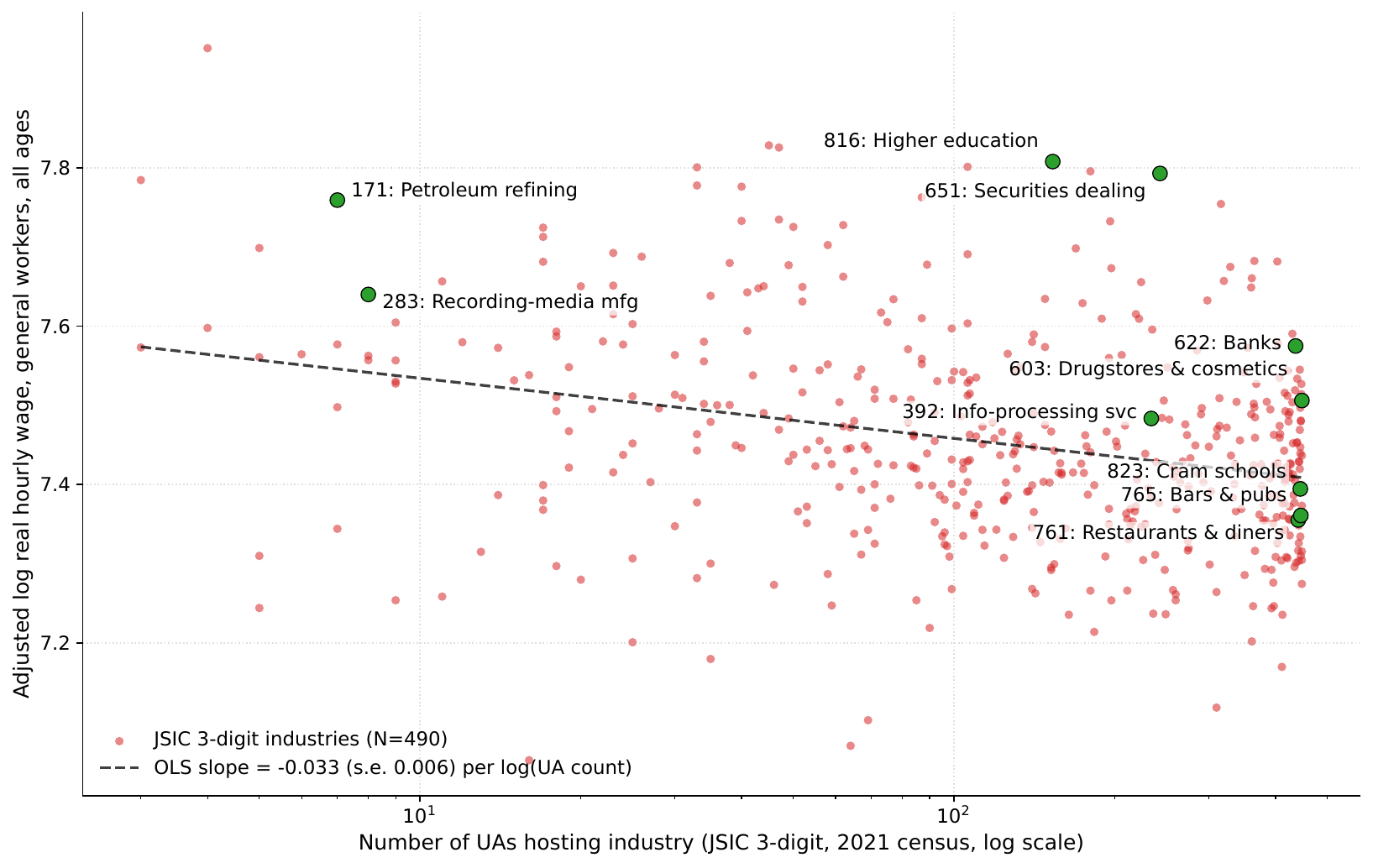}
\caption{Industry-tier real-wage gradient: adjusted hourly wage versus UA hosting count.}
\label{fig:jsic_hosting_vs_bss_wage}
\par\smallskip
\begin{minipage}{0.92\textwidth}
\footnotesize \emph{Notes:} Each point is a JSIC 3-digit industry. The $x$-axis is the number of UAs hosting that industry in the 2021 Economic Census for Business Activity (log scale; the same hosting metric as Figure~\ref{fig:concavity_on_jobs_UA}). The $y$-axis is the industry's adjusted log real hourly wage: the industry fixed effect from a regression of the individual log wage on gender-by-age dummies (single years of age), education (four classes), tenure and its square, employment form (regular or non-regular, with or without a fixed term) and year dummies, evaluated at the sample means of these variables. The wage is the contracted monthly wage (overtime pay and bonuses excluded) divided by the scheduled hours worked in the month, deflated with the same housing-inclusive prefecture-by-year price index as the JPSED wages (Section~\ref{sec:data}). The sample is general workers aged 15--70 in BSS 2015--2019 ($4{,}812{,}937$ records); short-time and temporary workers are excluded, and industries with $<10$ observations are dropped ($N = 490$). Prefecture dummies are not included, because where an industry locates is the object of the figure. The dashed line is the OLS fit across industries. How this fit changes as the worker controls are added one by one is reported in Online Appendix~\ref{oa:app:data}. The labeled points are example industries, from ones hosted in fewer than ten UAs (petroleum refining, recording-media manufacturing) to ones present in almost every UA (restaurants, bars). Source: BSS microdata 2015--2019; hosted counts from the 2021 Economic Census for Business Activity.
\end{minipage}
\end{figure}

\paragraph{The industry wage gradient (Figure~\ref{fig:jsic_hosting_vs_bss_wage}).} The figure is built in two steps from BSS (\emph{Chingin Kouzou Kihon Tokei Chosa}) individual records pooled over 2015--2019 (general workers; the wage is the contracted wage per scheduled hour, deflated as the JPSED wages are, Section~\ref{sec:data}). First, individual log wages are regressed on worker characteristics (gender-by-age, education, tenure, employment form, and year) together with a fixed effect for each JSIC 3-digit industry; the characteristics are the controls, and each industry's fixed effect, evaluated at their sample means, is its \emph{adjusted wage} --- what the industry pays a worker of average characteristics. Second, these $490$ adjusted wages are regressed on the log of the number of UAs hosting the industry, each industry counting once: the slope is $-0.033$ (HC1 SE $0.006$), and $-0.020$ among workers aged 20--29 ($483$ industries) --- the figures of Table~\ref{tab:bss_framework_primitives}, row (2). Section~\ref{sec:descriptive_concavity} uses the figure for two facts: the wage falls as the hosting count rises, and, because the relation is linear in the \emph{log} of the hosting count, the wage gap between neighbouring industries is widest at the specialized end, so the step down to a neighbouring industry costs more there.

\section{Additional instruments: TFR decline and political capital deepening}\label{app:iv_multi}

Section~\ref{sec:strategy_id} instruments the origin decline rate with three predetermined cross-prefecture shocks. This appendix constructs the two instruments beyond the demographic Bartik of Section~\ref{sec:strategy_id}.

The second instrument is the prefecture-level change in the total fertility rate from 1965 to 1985:
\begin{equation}
\widehat{\text{TFR}}_o^{\,\Delta} = \text{TFR}_{o, 1985} - \text{TFR}_{o, 1965},  \label{eq:tfr_dec}
\end{equation}
constructed from the Vital Statistics (Ministry of Health, Labour and Welfare), Volume I, Table 4.5 prefecture-by-year TFR series. This captures the fall in births per woman --- the fertility transition itself --- as distinct from the size of the birth cohorts that the Bartik counts. The 1965 value is missing for Okinawa (pre-reversion), so 2SLS specifications using $\widehat{\text{TFR}}_o^{\,\Delta}$ drop Okinawa from the rural sample (34 prefectures).

The third instrument is the prefecture-level capital-deepening rate over 1980--85:
\begin{equation}
\widehat{g}_o^{\,K/L} = 100 \cdot \log\bigl((K/L)_{o, 1985} \,/\, (K/L)_{o, 1980}\bigr) \,/\, 5,  \label{eq:kdeep}
\end{equation}
where $K$ is the real capital stock at constant 2000 prices and $L$ total employment, both from the R-JIP 2017 industry totals.

\section{Marginal effects of origin size by cell}\label{app:marginal_wls}

Table~\ref{tab:marginal_wls} reports the marginal effect of origin size on the Tokyo MA move premium, $\widehat{\mathrm{ME}}(S) = \hat\beta_3^{\pm} + \hat\beta_{Sg}^{\pm\pm}\,g$, by cell and at two decline depths ($g \in \{\pm 1.0, \pm 1.5\}\,\sigma_g$), for the weighted (WLS) specifications of Section~\ref{sec:results_main}. The unweighted (OLS) columns are in Online Appendix~\ref{oa:app:marginal_effects} and the diagnostics on cell assignment in Online Appendix~\ref{oa:app:cell_assignment}.

\begin{table}[htbp]
\centering
\caption{Marginal effect of origin size on the Tokyo MA wage premium, by cell and decline depth (WLS).}
\label{tab:marginal_wls}
{\small% Cut from the pipeline's tab_marginal.tex by python/postprocess_main_tables.py: WLS columns (Spec 4/5), ±1.0 and ±1.5 sigma.
\begin{tabular}{lcc}
\toprule
 & \multicolumn{2}{c}{Weighted (WLS)} \\
\cmidrule(lr){2-3}
Cell (g evaluation point) & Full rural (Spec 4) & $\leq 29$ (Spec 5) \\
\midrule
Cell I (slow-decl, g = +1.0$\sigma_g$) & +0.3001 & +1.3063$^{***}$ \\
 & (0.2239) & (0.4989) \\[3pt]
Cell I (slow-decl, g = +1.5$\sigma_g$) & +0.7272$^{*}$ & +2.5064$^{***}$ \\
 & (0.4266) & (0.9436) \\[3pt]
\midrule
Cell II (g = -1.0$\sigma_g$) & +0.1169 & +0.2162 \\
 & (0.0877) & (0.1794) \\[3pt]
Cell II (g = -1.5$\sigma_g$) & +0.4524$^{***}$ & +0.8712$^{***}$ \\
 & (0.0783) & (0.2016) \\[3pt]
\midrule
Cell III (slow-decl, g = +1.0$\sigma_g$) & -0.1015 & +0.1530 \\
 & (0.3432) & (0.5046) \\[3pt]
Cell III (slow-decl, g = +1.5$\sigma_g$) & -0.3902 & -0.1973 \\
 & (0.5316) & (0.7814) \\[3pt]
\midrule
Cell IV (g = -1.0$\sigma_g$) & -0.0928 & -1.1096$^{***}$ \\
 & (0.1977) & (0.2670) \\[3pt]
Cell IV (g = -1.5$\sigma_g$) & -0.3772 & -2.0911$^{***}$ \\
 & (0.2964) & (0.4858) \\[3pt]
\bottomrule
\end{tabular}
}
\par\smallskip
\begin{minipage}{0.9\textwidth}
\footnotesize \textit{Notes:} Each number in the table is $\widehat{\mathrm{ME}}(S) = \hat\beta_3^{\pm} + \hat\beta_{Sg}^{\pm\pm}\,g$ for the cell, evaluated at the decline depth $g$ in the row label (multiples of $\sigma_g$), from the WLS estimates of Table~\ref{tab:main} (Spec 4, all ages; Spec 5, ages $\leq 29$). The framework predicts $\mathrm{ME}(S) > 0$ on the large-origin side (cells I and II, read jointly, at the depths where the large cores sit; Section~\ref{sec:results_main}, finding (ii)), $\mathrm{ME}(S) < 0$ deepening with decline in cell IV, and no significant gradient in cell III. $\sigma_g$ is the standard deviation of the demeaned origin decline rate $\tilde g$ across origins: $3.95$ percentage points in the all-ages sample and $3.90$ in the $\leq 29$ sample. Each column evaluates $\mathrm{ME}(S)$ at $\pm 1.0$ and $\pm 1.5$ times its own $\sigma_g$, that is, at decline depths of $3.95$ and $5.93$ points in the all-ages column and $3.90$ and $5.85$ points in the $\leq 29$ column. Standard errors in parentheses (score-based wild cluster bootstrap, $B = 999$, Rademacher, clustered at the origin prefecture). $^{***}$, $^{**}$, $^{*}$: significance at the 1\%, 5\%, 10\% level.
\end{minipage}
\end{table}

\end{document}